\documentclass[10pt,twocolumn,aps,pra,reprint,amsmath,amssymb,preprintnumbers,superscriptaddress,longbibliography]{revtex4-2}
\newcommand*{\thedoctitle}{Limits for realizing single photons}
\newif\iffootinbib\footinbibtrue
\newif\iffigfrompdf\figfrompdftrue

\usepackage[utf8]{inputenc}                                             

\usepackage{xcolor}                                                     

\usepackage{amsmath, amssymb}                                           
\usepackage[italicdiff]{physics}                                        
\usepackage{dsfont}                                                     
\usepackage{bm}                                                         

\usepackage{graphicx}                                                   
\usepackage{tikz}                                                       
\usepackage{pgfplots}                                                   
\pgfplotsset{compat=1.16}

\usepackage{hyperref}                                                   

\newcommand*{\autofootnote}[2]{%
    \iffootinbib%
        { }%
    \fi%
    \footnote{\label{#1}#2}%
    \expandafter\edef\csname citefootnote@#1\endcsname{\thempfn}%
}

\newcommand*{\autofootnotetext}[2]{%
    \footnotetext{\label{#1}#2}%
    \expandafter\edef\csname citefootnote@#1\endcsname{\thempfn}%
}

\newcommand*{\footnotebibid}[1]{%
    \csname citefootnote@#1\endcsname%
}

\newcommand*{\footnoteref}[1]{%
    \iffootinbib%
        { }\cite{\footnotebibid{#1}}%
    \else%
        \textsuperscript{\ref{#1}}%
    \fi%
}

\makeatletter

\renewcommand*{\eqref}[1]{%
    \hyperref[{#1}]{\textup{\tagform@{\protect\ref*{#1}}}}%
}

\newcommand*{\instring}[2]{%
    TT\fi\begingroup\edef\x{\endgroup\noexpand\in@{#1}{#2}}\x\ifin@%
}
\let\ocite\cite
\renewcommand*{\cite}[1]{%
    \if\instring{,}{#1}\textcolor{\@citecolor}{\ocite{#1}}\else\hyperlink{cite.#1}{\protect\NoHyper\ocite{#1}\protect\endNoHyper}\fi%
}

\makeatother

\newcommand*{\vect}[1]{\bm{\mathrm{#1}}}                                
\newcommand*{\reals}{\mathbb{R}}                                        
\newcommand*{\complexs}{\mathbb{C}}                                     
\newcommand*{\lp}{\mathopen{}\left}                                     
\newcommand*{\rp}{\right}                                               
\newcommand*{\hc}{\text{H.c.}}                                          
\newcommand*{\cc}{\text{c.c.}}                                          

\renewcommand*{\ket}[1]{\left\protect\vert #1 \right\protect\rangle}    
\renewcommand*{\bra}[1]{\left\protect\langle #1 \right\protect\vert}    
\renewcommand*{\braket}[2]{
    \left\protect\langle #1 \middle\protect\vert #2 %
    \right\protect\rangle%
}
\newcommand*{\qexv}[2]{\bra{#1} #2 \ket{#1}}                            
\newcommand*{\norder}[1]{:\nbuffb\mathrel{#1}\nbuffc:}                  
\newcommand*{\norderop}[1]{\nbuffa\norder{#1}\nbuffd}                   
\newcommand*{\identity}{\mathds{1}}                                     

\newcommand*{\tbuff}{\mathchoice{\quad}{\>}{\>}{\>}}                    
\newcommand*{\kbuff}{\:}                                                
\newcommand*{\eqbuff}{\hspace{1.2em}}                                   
\newcommand*{\wbuff}{\hspace{.1em}}                                     
\newcommand*{\ibuffa}{\mathchoice
    {\hspace{-.18em}}{\hspace{-.06em}}%
    {\hspace{-.06em}}{\hspace{-.05em}}}
\newcommand*{\ibuffi}{\mathchoice
    {\hspace{-.62em}}{\hspace{-.2em}}{}{}}
\newcommand*{\ibuffn}{\mathchoice
    {\hspace{-.8em}}{}{}{}}
\newcommand*{\ibuffb}{\mathchoice
    {\hspace{.2em}}{\hspace{.2em}}{\hspace{.1em}}{\hspace{.06em}}}
\newcommand*{\nbuffeq}{\mathchoice
    {\mspace{5mu plus 3mu minus 1mu}}%
    {\mspace{5mu plus 3mu minus 1mu}}%
    {\mspace{5mu plus 3mu minus 1mu}}%
    {\mspace{5mu plus 3mu minus 1mu}}}
\newcommand*{\nbuffa}{\mathchoice
    {\hspace{-.15em}}{\hspace{-.14em}}{\hspace{.05em}}{}}
\newcommand*{\nbuffb}{\mathchoice
    {\hspace{.1em}}{\hspace{.09em}}{\hspace{.03em}}{}}
\newcommand*{\nbuffc}{\mathchoice
    {\hspace{.09em}}{\hspace{.06em}}{\hspace{.02em}}{}}
\newcommand*{\nbuffd}{\mathchoice
    {\hspace{-.15em}}{\hspace{-.14em}}{\hspace{.05em}}{}}

\newcommand*{\secref}[1]{\hyperref[#1]{Section~\ref*{#1}}}              
\newcommand*{\figref}[1]{\hyperref[#1]{Fig.~\ref*{#1}}}                 
\newcommand*{\tabref}[1]{\hyperref[#1]{Tab.~\ref*{#1}}}                 
\newcommand*{\thmref}[1]{\hyperref[#1]{Theorem~\ref*{#1}}}              
\newcommand*{\stepref}[1]{\hyperref[#1]{step~\ref*{#1}}}                
\newcommand*{\appref}[1]{\hyperref[#1]{Appendix~\ref*{#1}}}             
\newcommand*{\supplementary}{Appendices}                                

\hypersetup{%
    pdftitle  = {\thedoctitle},%
    pdfauthor = {Jan Gulla, Kai Ryen, Johannes Skaar},%
    colorlinks,%
    linkcolor={blue!50!black},%
    citecolor={blue!50!black},%
    urlcolor={blue!50!black},%
}

\newcommand*{\figone}{
    \begin{figure}[tb]
        \iffigfrompdf
            \includegraphics{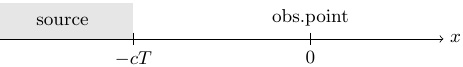}
        \else
            \begin{tikzpicture}[scale=1.5]
                \fill [black!10] (0, 0.4) rectangle (1.5, 0);
                \draw (0.7, 0.07) node[above] {source};
                \draw[->] (0, 0) -- (5, 0) node[right] {$x$};
                \draw (1.5, 0.07) -- (1.5, -0.07) node[below] {$-cT$};
                \draw (3.5, 0.07) -- (3.5, -0.07) node[below] {$0$};
                \draw (3.5, 0.07) node[above] {obs.point};
            \end{tikzpicture}
        \fi
        \caption{\label{fig:setup}A source is located in the region $x < -cT$, and we consider field observables at the observation point $x = 0$. The source is turned on at $t = -T$, which means that any measurement before $t < 0$ must give the same value as for vacuum.}
    \end{figure}
}

\newcommand*{\figtwo}{
    \begin{figure}[tb]
        \iffigfrompdf
            \includegraphics{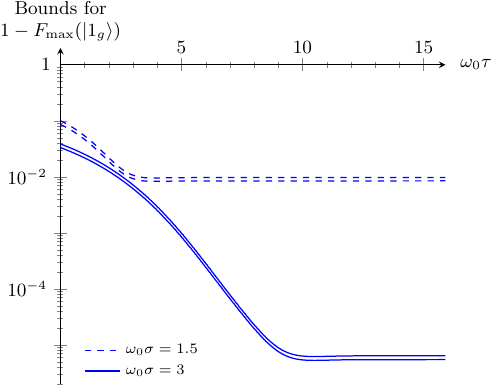}
        \else
            \begin{tikzpicture}[scale=1]
                \begin{semilogyaxis}[%
                    axis lines=center,%
                    x=0.410cm,%
                    xmin=0,%
                    xmax=15.9,%
                    ymax=2,%
                    ymin=2e-6,%
                    xlabel={$\omega_0 \tau$},%
                    x label style={right, xshift=1.0ex},%
                    ylabel={Bounds for \\$1 - F_{\text{max}}\lp(\ket{1_g}\rp)$},%
                    y label style={above, align=center},%
                    major tick length={1.6ex},%
                    xtick={0, 5, 10, 15},%
                    minor x tick num=4,%
                    x tick label style={above, yshift=1.4ex},%
                    yticklabels={, , $10^{-4}$, , $10^{-2}$, , },%
                    extra y ticks={1},%
                    extra y tick labels={$1$},%
                    extra y tick style={%
                        major tick length={0ex},%
                        tick label style={xshift=-0.35ex}%
                    },%
                    legend style={at={(-0.08, 0.07)}, anchor=west, xshift=6ex, font=\scriptsize, draw=none},%
                    legend cell align=left%
                ]
                    \addplot[color=blue, dashed, semithick]%
                        table [x index=0, y index=1, col sep=comma]%
                        {F_max(1_g)_exact_bounds_vs._w0_tau_for_w0_sigma=1.5.csv};
                    \addlegendentry{$\omega_0 \sigma = 1.5$}
                    \addplot[color=blue, semithick]%
                        table [x index=0, y index=1, col sep=comma]%
                        {F_max(1_g)_exact_bounds_vs._w0_tau_for_w0_sigma=3.csv};
                    \addlegendentry{$\omega_0 \sigma = 3$}
                    \addplot[color=blue, dashed, semithick]%
                        table [x index=0, y index=2, col sep=comma]%
                        {F_max(1_g)_exact_bounds_vs._w0_tau_for_w0_sigma=1.5.csv};
                    \addplot[color=blue, semithick]%
                        table [x index=0, y index=2, col sep=comma]%
                        {F_max(1_g)_exact_bounds_vs._w0_tau_for_w0_sigma=3.csv};
                \end{semilogyaxis}
            \end{tikzpicture}
        \fi
        \caption{\label{fig:F_max_bounds_1g_vs_tau}Upper and lower bound for the maximum fidelity between a state strictly localized to $t \geq 0$ and a positive-time single photon $\ket{1_g}$ in a Gaussian-modulated pulse $g(t)$ [see \eqref{g_t_gauss_pos_time}]. The photon has fixed carrier $\omega_0$ and width $\sigma$, while the photon delay $\tau$ is changed along the $x$-axis. The bounds are calculated using the exact expressions \eqref{F_max_1g_upper} and \eqref{F_max_1g_lower}.}
    \end{figure}
}

\newcommand*{\figthree}{
    \begin{figure}[tb]
        \iffigfrompdf
            \includegraphics{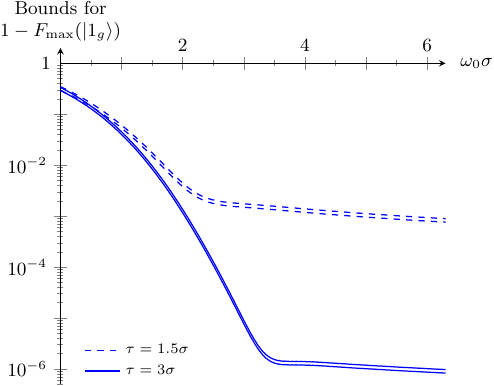}
        \else
            \begin{tikzpicture}[scale=1]
                \begin{semilogyaxis}[%
                    axis lines=center,%
                    x=1.035cm,%
                    xmin=0,%
                    xmax=6.3,%
                    ymax=2,%
                    ymin=5e-7,%
                    xlabel={$\omega_0 \sigma$},%
                    x label style={right, xshift=1.0ex},%
                    ylabel={Bounds for \\$1 - F_{\text{max}}\lp(\ket{1_g}\rp)$},%
                    y label style={above, align=center},%
                    major tick length={1.6ex},%
                    xtick={0, 1, 2, 3, 4, 5, 6},%
                    minor x tick num=1,%
                    xticklabels={0, , 2, , 4, , 6},%
                    x tick label style={above, yshift=1.4ex},%
                    yticklabels={, $10^{-6}$, , $10^{-4}$, , $10^{-2}$, , },%
                    extra y ticks={1},%
                    extra y tick labels={$1$},%
                    extra y tick style={%
                        major tick length={0ex},%
                        tick label style={xshift=-0.35ex}%
                    },%
                    legend style={at={(-0.08, 0.07)}, anchor=west, xshift=6ex, font=\scriptsize, draw=none},%
                    legend cell align=left%
                ]
                    \addplot[color=blue, dashed, semithick]%
                        table [x index=0, y index=1, col sep=comma]%
                        {F_max(1_g)_exact_bounds_vs._w0_sigma_for_tau=1.5sigma.csv};
                    \addlegendentry{$\tau = 1.5 \sigma$}
                    \addplot[color=blue, semithick]%
                        table [x index=0, y index=1, col sep=comma]%
                        {F_max(1_g)_exact_bounds_vs._w0_sigma_for_tau=3sigma.csv};
                    \addlegendentry{$\tau = 3 \sigma$}
                    \addplot[color=blue, dashed, semithick]%
                        table [x index=0, y index=2, col sep=comma]%
                        {F_max(1_g)_exact_bounds_vs._w0_sigma_for_tau=1.5sigma.csv};
                    \addplot[color=blue, semithick]%
                        table [x index=0, y index=2, col sep=comma]%
                        {F_max(1_g)_exact_bounds_vs._w0_sigma_for_tau=3sigma.csv};
                \end{semilogyaxis}
            \end{tikzpicture}
        \fi
        \caption{\label{fig:F_max_bounds_1g_vs_sigma}Same as \figref{fig:F_max_bounds_1g_vs_tau}, except here the fidelity is plotted against the photon width $\sigma$, and the photon delay $\tau$ is kept proportional to $\sigma$. This means that as $\sigma$ is changed, the truncation at $t = 0$ is fixed at some percentage of the pulse width.}
    \end{figure}
}

\newcommand*{\figfour}{
    \begin{figure}[tb]
        \iffigfrompdf
            \includegraphics{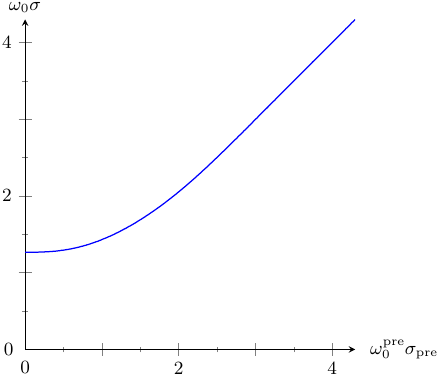}
        \else
            \begin{tikzpicture}[scale=1]
                \begin{axis}[%
                    axis lines=center,%
                    x=1.3cm,%
                    y=1.3cm,%
                    xmin=0,%
                    xmax=4.3,%
                    ymin=0,%
                    ymax=4.3,%
                    xlabel={$\omega_{0}^{\text{pre}} \sigma_{\text{pre}}$},%
                    x label style={right, xshift=1.0ex},%
                    ylabel={$\omega_0 \sigma$},%
                    y label style={above},%
                    major tick length={1.6ex},%
                    xtick={0, 1, 2, 3, 4},%
                    minor x tick num=1,%
                    xticklabels={0, , 2, , 4},%
                    extra x ticks={0},%
                    extra x tick labels={0},%
                    extra x tick style={%
                        major tick length={0ex},%
                        tick label style={yshift=-0.6ex}%
                    },%
                    ytick={0, 1, 2, 3, 4},%
                    minor y tick num=1,%
                    yticklabels={0, , 2, , 4},%
                    extra y ticks={0},%
                    extra y tick labels={0},%
                    extra y tick style={major tick length=0ex,%
                        tick label style={xshift=-0.6ex}%
                    }%
                    ]
                    \addplot[color=blue, semithick]%
                    table [x index=0, y index=1, col sep=comma]%
                    {w0^eff_sigma_eff_vs._w0^pre_sigma_pre_for_tau_eff=1.5sigma_eff.csv};
                \end{axis} 
            \end{tikzpicture}
        \fi
        \caption{\label{fig:w0_eff_sigma_eff_vs_w0_sigma}Mean frequency $\omega_0$ times width $\sigma$ of a pulse $\xi(t)$ containing only positive frequencies [see \eqref{xi_omega_G_pre}]. The spectrum of the pulse $\xi(t)$ is the positive-frequency part of $g_{\text{pre}}(t)$, which is a Gaussian envelope of width $\sigma_{\text{pre}}$ and delay $\tau_{\text{pre}}$ around a center frequency $\omega_0^{\text{pre}}$. As the pulse width of $g_{\text{pre}}(t)$ is reduced, the width of $\xi(t)$ asymptotes to $\approx 1.3$, meaning that a pulse containing only positive frequencies cannot be narrower than that. The plot is independent of $\tau_{\text{pre}}$, which is a desirable trait of the method for calculating the properties of $\xi(t)$ [see \eqref{w0_eff_tau_eff} and below].}
    \end{figure}
}

\newcommand*{\figfive}{
    \begin{figure}[tb]
        \iffigfrompdf
            \includegraphics{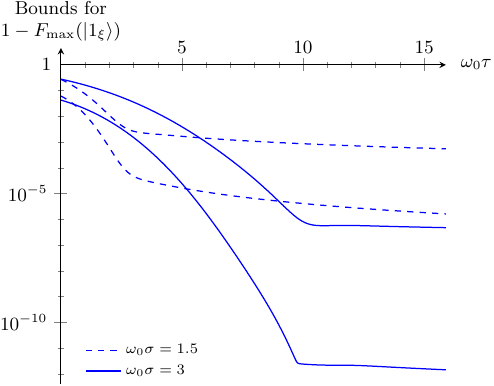}
        \else
            \begin{tikzpicture}[scale=1]
                \begin{semilogyaxis}[%
                    axis lines=center,%
                    x=0.410cm,%
                    xmin=0,%
                    xmax=15.9,%
                    ymax=4.5,%
                    ymin=4e-13,%
                    xlabel={$\omega_0 \tau$},%
                    x label style={right, xshift=1.0ex},%
                    ylabel={Bounds for \\$1 - F_{\text{max}}\lp(\ket{1_\xi}\rp)$},%
                    y label style={above, align=center},%
                    major tick length={1.6ex},%
                    xtick={0, 5, 10, 15},%
                    minor x tick num=4,%
                    x tick label style={above, yshift=1.4ex},%
                    ytick={1e-10, 1e-5},%
                    minor ytick={1e-13, 1e-12, 1e-11, 1e-9, 1e-8, 1e-7, 1e-6, 1e-4, 1e-3, 1e-2, 1e-1},%
                    yticklabels={$10^{-10}$, $10^{-5}$},%
                    extra y ticks={1},%
                    extra y tick labels={$1$},%
                    extra y tick style={%
                        major tick length={0ex},%
                        tick label style={xshift=-0.35ex}%
                    },%
                    legend style={at={(-0.08, 0.07)}, anchor=west, xshift=6ex, font=\scriptsize, draw=none},%
                    legend cell align=left%
                    ]
                    \addplot[color=blue, dashed, semithick]%
                    table [x index=0, y index=1, col sep=comma]%
                    {F_max(1_xi)_exact_bounds_vs._w0^eff_tau_eff_for_w0^eff_sigma_eff=1.5.csv};
                    \addlegendentry{$\omega_0 \sigma = 1.5$}
                    \addplot[color=blue, semithick]%
                    table [x index=0, y index=1, col sep=comma]%
                    {F_max(1_xi)_exact_bounds_vs._w0^eff_tau_eff_for_w0^eff_sigma_eff=3.csv};
                    \addlegendentry{$\omega_0 \sigma = 3$}
                    \addplot[color=blue, dashed, semithick]%
                    table [x index=0, y index=2, col sep=comma]%
                    {F_max(1_xi)_exact_bounds_vs._w0^eff_tau_eff_for_w0^eff_sigma_eff=1.5.csv};
                    \addplot[color=blue, semithick]%
                    table [x index=0, y index=2, col sep=comma]%
                    {F_max(1_xi)_exact_bounds_vs._w0^eff_tau_eff_for_w0^eff_sigma_eff=3.csv};
                \end{semilogyaxis} 
            \end{tikzpicture}
        \fi
        \caption{\label{fig:F_max_bounds_1xi_vs_tau}Upper and lower bound for the maximum fidelity between a state strictly localized to $t \geq 0$ and a physical single photon $\ket{1_\xi}$ with pulse form $\xi(t)$. The photon spectrum is the positive-frequency part of a Gaussian-modulated carrier $g_{\text{pre}}(t)$ [see \eqref{xi_omega_G_pre}]. The mean frequency $\omega_0$, width $\sigma$, and delay $\tau$ of the photon pulse form $\xi(t)$ are computed using the method described in \eqref{w0_eff_tau_eff} and below, and are related to $g_{\text{pre}}(t)$ according to \figref{fig:w0_eff_sigma_eff_vs_w0_sigma}. In the plot, the duration of the single-photon $\sigma$ is fixed, and the photon delay $\tau$ is changed along the $x$-axis. The bounds are calculated using the exact expressions \eqref{F_max_1xi_upper} and \eqref{F_max_1xi_lower}.}
    \end{figure}
}

\newcommand*{\figsix}{
    \begin{figure}[tb]
        \iffigfrompdf
            \includegraphics{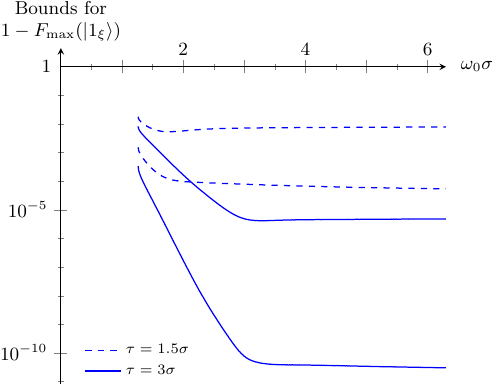}
        \else
            \begin{tikzpicture}[scale=1]
                \begin{semilogyaxis}[%
                    axis lines=center,%
                    x=1.035cm,%
                    xmin=0,%
                    xmax=6.3,%
                    ymax=4.5,%
                    ymin=8e-12,%
                    xlabel={$\omega_0 \sigma$},%
                    x label style={right, xshift=1.0ex},%
                    ylabel={Bounds for \\$1 - F_{\text{max}}\lp(\ket{1_\xi}\rp)$},%
                    y label style={above, align=center},%
                    major tick length={1.6ex},%
                    xtick={0, 1, 2, 3, 4, 5, 6},%
                    minor x tick num=1,%
                    xticklabels={0, , 2, , 4, , 6},%
                    x tick label style={above, yshift=1.4ex},%
                    ytick = {1e-10, 1e-5},%
                    minor ytick={1e-11, 1e-9, 1e-8, 1e-7, 1e-6, 1e-4, 1e-3, 1e-2, 1e-1},%
                    yticklabels={$10^{-10}$, $10^{-5}$},%
                    extra y ticks={1},%
                    extra y tick labels={$1$},%
                    extra y tick style={%
                        major tick length={0ex},%
                        tick label style={xshift=-0.35ex}%
                    },%
                    legend style={at={(-0.08, 0.07)}, anchor=west, xshift=6ex, font=\scriptsize, draw=none},%
                    legend cell align=left%
                    ]
                    \addplot[color=blue, dashed, semithick]%
                    table [x index=0, y index=1, col sep=comma]%
                    {F_max(1_xi)_exact_bounds_vs._w0^eff_sigma_eff_for_tau_eff=1.5sigma_eff.csv};
                    \addlegendentry{$\tau = 1.5 \sigma$}
                    \addplot[color=blue, semithick]%
                    table [x index=0, y index=1, col sep=comma]%
                    {F_max(1_xi)_exact_bounds_vs._w0^eff_sigma_eff_for_tau_eff=3sigma_eff.csv};
                    \addlegendentry{$\tau = 3 \sigma$}
                    \addplot[color=blue, dashed, semithick]%
                    table [x index=0, y index=2, col sep=comma]%
                    {F_max(1_xi)_exact_bounds_vs._w0^eff_sigma_eff_for_tau_eff=1.5sigma_eff.csv};
                    \addplot[color=blue, semithick]%
                    table [x index=0, y index=2, col sep=comma]%
                    {F_max(1_xi)_exact_bounds_vs._w0^eff_sigma_eff_for_tau_eff=3sigma_eff.csv};
                \end{semilogyaxis}
            \end{tikzpicture}
        \fi
        \caption{\label{fig:F_max_bounds_1xi_vs_sigma}Same as \figref{fig:F_max_bounds_1xi_vs_tau}, except here the fidelity is plotted against the photon width $\sigma$, and the photon delay $\tau$ is kept proportional to $\sigma$. Narrower pulse widths than $\omega_0\sigma \approx 1.3$ are not possible, as seen in \figref{fig:w0_eff_sigma_eff_vs_w0_sigma}.}
    \end{figure}
}

\begin{document}


\title{\thedoctitle}

\author{Jan Gulla}
\affiliation{Department of Technology Systems, University of Oslo, NO-0316 Oslo, Norway}

\author{Kai Ryen}
\affiliation{Department of Physics, University of Oslo, NO-0316 Oslo, Norway}

\author{Johannes Skaar}
\email{johannes.skaar@fys.uio.no}
\affiliation{Department of Physics, University of Oslo, NO-0316 Oslo, Norway}

\date{\today}

\begin{abstract}
Exact single photons cannot be generated on demand due to their infinite tails. To quantify how close realizable optical states can be to some target single photon in one dimension, we argue that there are two natural but incompatible ways to specify the target state. Either it can be expressed as a photon with a chosen positive-frequency spectrum, or it can be described as an (unphysical) photon in a chosen positive-time pulse. The results show that for sufficiently short target pulses, the closest realizable states contain substantial multiphoton components. Upper and lower bounds for the maximum fidelity are derived and are expressed as functions of the size of the target state's tail, for negative time or negative frequency, respectively. We also generalize the bounds to arbitrary photon-number states. 
\end{abstract}


\maketitle

\section{Introduction}\label{sec:introduction}
Given an arbitrary electromagnetic source, what optical signals can be generated on demand, i.e., produced reliably by a free, local decision? Although it is not immediately obvious there should be any restrictions, the answer is tied to the limits of photon localization \cite{knight1961,bialynicki-birula1998}, a connection that was first formulated in \cite{gulla2021}.

The question of how well single photons can be localized has a long history. Early efforts were concentrated on direct measures of localization resembling those in nonrelativistic quantum mechanics, such as photon wave functions \cite{landau1930}, photon position operators \cite{newton1949}, and spatial photon number operators \cite{mandel1966}. All these concepts suffered from various difficulties \cite{bohm1951,power1964,pike1995}, which was usually taken as evidence for that photons might not be possible to localize \cite{jauch1967,mandel1995} (see also review articles \cite{keller2005,saari2012}). Although these methods occasionally still receive some interest \cite{keller2005}, it was with time accepted that the only meaningful characterization of particle localization in quantum field theory is through measurements of local observables \cite{haag1996}. For instance for photons, a common local observable is the electromagnetic energy density \cite{bialynicki-birula1998,gulla2021a}. An early calculation of this quantity \cite{amrein1969} produced a long-standing belief that the energy density of a maximally localized photon is spread out in space $\vect{r}$ with an asymptotic fall off of $\abs{\vect{r}}^{-7}$ \cite{mandel1995}. However, this bound was eventually disproved, as solutions with higher-inverse-power fall offs were identified \cite{adlard1997}, before Bialynicki-Birula finally in 1998 discovered the almost-exponential limit of photon localization \cite{bialynicki-birula1998}.

The reason that single photons cannot be localized is the absence of negative frequencies in the quantum field's annihilation operator time dependence $e^{- i \omega t} = e^{- i \abs{\vect{k}} t}$ \cite{knight1961}, where $\omega$ is the frequency, $\vect{k}$ the wavevector, and $t$ the time. To see why, consider a 1D source of a single polarization as in \figref{fig:setup}. For an emitted photon traveling in the $+x$ direction from the source, we can calculate the expected electromagnetic, normal-ordered energy density as a function of position at a fixed time. By expressing this quantity as an integral over the photon frequency (energy), which must be positive, we obtain a Fourier integral over only positive arguments \cite{gulla2021}. Such a function with a purely positive spectrum satisfies quite strict analytic properties in terms of position and time \cite{titchmarsh1948}. In particular by the Paley-Wiener criterion \cite{paley1934} (see also \appref{sec:proof_that_single_photons_in_1d_cannot_be_localized}), this function must either be identically 0, in which case there is no photon, or it is nonzero (almost) everywhere with a asymptotic fall off slower than $e^{-A x}$, for $A > 0$ \cite{bialynicki-birula1998}. In other words, any fall off slower than exponential is possible, but an exponential or faster fall off is impossible. This applies irrespective of the spectrum of the photon, meaning that any single photon in 1D has a nonvanishing energy density everywhere, and is therefore not localizable to any region in space or time (for a detailed derivation see \appref{sec:proof_that_single_photons_in_1d_cannot_be_localized}). 

In contrast, optical states that are distinguishable from vacuum only inside some spacetime region are said to be \emph{strictly localized} to that region \cite{knight1961}. A signal generated on demand is by definition triggered locally (in some freely chosen region in space and time) and reliably (no option of postselection) by some external action (e.g., the experimentalist). The generated pulse then propagates outwards at maximally the speed of light $c$ and should therefore be strictly localized to the light cone of the trigger region. Since this is impossible for single photons, we conclude that exact photons cannot be generated on demand. 

\figone

A natural question is then what optical states can be generated in this manner? Letting some source be located in the region $x < - cT$ for some constant $T > 0$ as in \figref{fig:setup}, we operate it on demand by switching it on at time $t = - T$ and off at $t = - T/2$. The source emits some optical pulse during the time interval $-T \leq t \leq -T/2$. Assuming electromagnetic vacuum initially, the resulting optical state $\ket{\psi}$ is then strictly localized to $t \geq 0$ at the observation point $x = 0$. One class of states that can satisfy this condition was discovered by Knight \cite{knight1961} and later in \cite{bialynicki-birula1998,saari2005}, namely the coherent states. The normal-ordered energy density for a coherent state has a different form where both positive and negative frequencies appear in the Fourier integral \cite{gulla2021} (see also \appref{sec:localization_of_classical_fields_and_coherent_states}). This means that coherent states can be strictly localized to $t \geq 0$ and therefore also generated on demand. 

However, coherent states are very different from single photons, which leads to the next question: is it possible to generate optical states on demand that are close to single photons, and if so, how close can we get? More precisely, we formulate the question as what the maximum fidelity 
\begin{equation}\label{F_MAX_1XI}
    F_{\text{max}}\big(\ket{1_\xi}\big) \equiv \max_{\ket{\psi}\text{ strict. loc.}} \abs{\braket{\psi}{1_\xi}}
\end{equation}
is between any state $\ket{\psi}$ strictly localized to $t \geq 0$ and a single photon in some given spectrum $\xi(\omega)$,
\begin{equation}\label{1xi}
    \ket{1_\xi} = \int_0^\infty \ibuffi \dd\omega \ibuffb \xi(\omega) a^\dagger(\omega) \ket{0}, \tbuff \int_0^\infty \ibuffi \dd\omega \ibuffb \abs{\xi(\omega)}^2 = 1,
\end{equation}
with $\xi(\omega) = 0$ for $\omega < 0$. Here $a^\dagger(\omega)$ is a frequency-mode creation operator satisfying $\comm*{a(\omega)}{a^\dagger(\omega')} = \delta(\omega - \omega')$ \cite{loudon2001} (see also \secref{sec:1D_single_polarization}) and $\ket{0}$ is the electromagnetic vacuum state. Eq. \eqref{F_MAX_1XI} is useful as a measure of how close we can come to a single photon as it clearly captures the size of the multiphoton components of $\ket{\psi}$ necessary to make it a localized state. 

On the other hand, one drawback of this quantity is that the state $\ket{1_\xi}$ has, as all single photons, tails stretching off to infinity. It is therefore not always the best representation of what we in an experiment would intuitively consider the target state: an ``ideal'' single photon in some specified, causal pulse form $g(t)$. Working with $\ket{1_\xi}$ does not allow choosing an arbitrary time-pulse form $g(t)$, since not all such functions can be represented by a spectrum $\xi(\omega)$ with only positive frequencies. For example, if we want the target state to be a single photon in an ultrashort pulse, there is no corresponding, valid target spectrum $\xi(\omega)$ since a short pulse contains a significant amount of negative frequencies. 

We therefore consider an alternative quantity for determining how close we can come to a single photon: the maximum fidelity
\begin{equation}\label{F_MAX_1G}
    F_{\text{max}}\big(\ket{1_g}\big) \equiv \max_{\ket{\psi}\text{ strict. loc.}} \abs{\braket{\psi}{1_g}}
\end{equation}
between a state $\ket{\psi}$ strictly localized to $t \geq 0$ and a single-photon state in some given positive-time pulse $g(t)$,
\begin{equation}\label{1g}
    \ket{1_g} = \int_0^\infty \ibuffi \dd t \ibuffb g(t) a^\dagger (t) \ket{0}, \tbuff \int_0^\infty \ibuffi \dd t \ibuffb \abs{g(t)}^2 = 1,
\end{equation}
with $g(t) = 0$ for $t < 0$. Here $a^\dagger(t)$ is a time-domain creation operator satisfying $\comm*{a(t)}{a^\dagger(t')} = \delta(t - t')$ \cite{loudon2001}. At the same time, $a^\dagger(t)$ is the Fourier transform of $a^\dagger(\omega)$, and the required negative-frequency modes are an artificial construction only used to be able to express the state $\ket{1_g}$. Such a state is of course unphysical, as negative frequencies are not real, but provided we extend the Hilbert space with these negative frequencies, $\ket{1_g}$ is an artificial single-photon state that is localized to $t \geq 0$. 

The advantage of this construction is that the state $\ket{1_g}$ has a causal leading edge and is therefore a more intuitive representation of an ideal target state. The disadvantage is that whereas the state $\ket{1_g}$ has no negative-time content, it instead has negative-frequency content. This means that \eqref{F_MAX_1G} inadvertently measures the amount of the negative frequencies that must be truncated to get a physical state $\ket{\psi}$, which can only have positive frequencies. 

There is thus a trade-off in the choice of target state. To determine the closeness of realizable states to single photons, we want a quantity that captures the size of the necessary multiphoton components of the realizable state. Eqs. \eqref{F_MAX_1XI} and \eqref{F_MAX_1G} both achieve this, but we must choose between either having a target state that is acausal or a target state containing negative frequencies. The state $\ket{\psi}$ we maximize over is of course physical and causal, and the goal is to determine how close to a single photon, either in the form \eqref{1xi} or \eqref{1g}, it can be. 

The exact values for the maximum fidelities \eqref{F_MAX_1XI} and \eqref{F_MAX_1G} will depend on the specific spectra of the target states. Still, we can constrain them by upper and lower bounds expressed by some general properties of the target spectra. For the fidelity with a physical single photon \eqref{F_MAX_1XI}, a useful, key property will turn out to be the weight of the negative-time tail
\begin{equation}\label{mu}
    \mu = \int_{-\infty}^0 \ibuffn \dd t \ibuffb \abs*{\xi(t)}^2,
\end{equation}
where $\xi(t)$ is the inverse Fourier transform of $\xi(\omega)$. Similarly, for the fidelity with a causal single photon \eqref{F_MAX_1G}, the key property is the weight of the negative frequencies
\begin{equation}\label{eta}
    \eta = \int_{-\infty}^0 \ibuffn \dd\omega \ibuffb \abs{G(\omega)}^2,
\end{equation}
where $G(\omega)$ is the Fourier transform of $g(t)$. 

In this work we provide upper and lower bounds for both fidelities \eqref{F_MAX_1XI} and \eqref{F_MAX_1G}, relying on 4 different arguments. The upper bound for \eqref{F_MAX_1XI} is found by constraining the maximum probability of distinguishing the single photon from vacuum by a measurement local to $t < 0$. The corresponding lower bound is found by providing an example using the strictly localized near-single-photon state from \cite{gulla2021}. Providing such an example state clearly constitutes a lower bound for the fidelity, which is a maximum over all strictly localized states. The bounds for \eqref{F_MAX_1G} are already found in \cite{gulla2021}, but the derivation is repeated here in order to coherently present and compare all 4 bounds.

It is worth noting that there is a potential, reasonable objection to the claims presented so far. After all, how can it be that photons cannot be generated on demand when there are numerous proposals and reported experiments for doing so \cite{scheel2009,eisaman2011,senellart2017,wang2019,sinha2019}? The answer is that the limitations imposed by the Paley-Wiener criterion only applies to \emph{exact} single photons. If the optical state is a superposition of different photon numbers, the theorem cannot be applied directly, and such states might be possible to generate on demand. In fact, the reported experiments are a good indication that there exist states realizable on demand that are \emph{very close} to single photons, much closer than for instance coherent states. 

Another important question is whether the Paley-Wiener limitation could have a different interpretation than the one set forth here. For example, is it possible that our notion of sharp causality is incorrect in the sense that generating a signal on demand by a local, free choice is impossible to begin with? Indeed, what if the quantum state governing the external trigger (i.e., the experimentalist) is itself not localized and has its own exponential tails? In that case there would be no sharply defined cause and effect anymore, and the source setup in our analysis would never occur. Instead, all events and interactions would be smeared out in time, with small probability tails stretching off to the infinite past and into the infinite future. Such a deterministic world model would seemingly lead to no inconsistencies, a possibility discussed in \cite{hegerfeldt1998}. 

Yet there are good arguments against such a viewpoint. First, the Paley-Wiener limitation does not apply to all quantum states, as we know that for instance coherent states can be strictly localized \cite{knight1961,bialynicki-birula1998} and therefore generated on demand. In light of this, it is somewhat arbitrary to abandon sharp causality purely because one type of state cannot be localized. Second, it seems counterintuitive that if the world is nonlocal, why does it appear to be local? Even if the fundamental interactions are nonlocal and deterministically predetermined, our own experience of free choice means that these microscopic interactions must somehow conspire in a way to produce at least an illusion of free choice in macroscopic settings. 

Ultimately though, the discussion of whether sharp causality and free will exist is no longer a topic of physics, and we therefore leave it aside. It appears that it is \emph{possible} to formulate a theory where these concepts are present, also for quantum field theory. We therefore include as an assumption in our analysis that on-demand sources exist. 

The paper is organized as follows. \secref{sec:the_fermi_problem_and_causality_in_quantum_field_theory} is a historical overview of analyses similar to our setup, mainly focused on investigations of the Fermi problem. In \secref{sec:local_measurements_and_strictly_localized_states} we look to algebraic quantum field theory to characterize the set of strictly localized states through local measurements and vacuum expectation values. We then find how these states can be generated by Licht or unitary operators. In \secref{sec:strictly_localized_state_eta12_near_single_photon} we construct a specific example of a state $\ket{\eta_{1,2}}$ that is strictly localized yet close to a single photon. The fidelity bounds for \eqref{F_MAX_1G} and \eqref{F_MAX_1XI} are then derived in \secref{sec:bounds_for_F_max_1g} and \secref{sec:bounds_for_F_max_1xi}, respectively. In \secref{sec:numerical_examples} we look at some concrete examples of target states $\ket{1_\xi}$ and $\ket{1_g}$, and plot numerical values for the corresponding fidelity bounds. \secref{sec:fidelity_bounds_for_arbitrary_number_states} generalizes the analysis from single photons to states of arbitrary photon number $\ket{n_\xi}$ and $\ket{n_g}$. A discussion of the results and concluding remarks are given in \secref{sec:discussion_and_conclusion}, where we also indicate connections to experimental results. \appref{sec:proof_that_single_photons_in_1d_cannot_be_localized} gives a short, self-contained proof that single photons in 1D cannot be strictly localized, and we recap the Paley-Wiener criterion. We then show in \appref{sec:localization_of_classical_fields_and_coherent_states} why coherent states avoid this limitation and can be localized. Finally, \appref{sec:eigenvectors_of_quantum_fields} discusses spectral decomposition and eigenvectors of quantum field observables in connection with local measurements.

\section{The Fermi problem and causality in quantum field theory}\label{sec:the_fermi_problem_and_causality_in_quantum_field_theory}
The key insight linking analytic properties implied by energy positivity to particle localization and causality has been rediscovered in several forms and subfields over the years. In 1974 Hegerfeldt made an observation about the impossibility of localizing quantum relativistic particles based on a quite general argument \cite{hegerfeldt1974}. Although the analysis had some weaknesses \cite{kaloyerou1988}, such as the usage of a particle localization operator, the key argument was based on analyticity and energy positivity. Hegerfeldt also subsequently connected his analysis to something called \emph{the Fermi problem} \cite{hegerfeldt1994}, and arrived at the paradoxical conclusion that quantum field theory is in violation with relativistic causality. This conclusion was later refuted \cite{buchholz1994}; however, the arguments used to resolve the paradox, and their connection to particle localization, are somewhat subtle. 

The usual way of demonstrating that quantum field theory is in accordance with relativistic causality is through the field commutator \cite{peskin1995}. E.g., for some real, scalar field $\phi(\vect{r}, t)$, it is verified that $\comm{\phi(\vect{r}, t)}{\phi(\vect{r}', t')} = 0$ for spacelike separations. Since local measurements for bosonic fields are made out of the field evaluated in the measurement region \cite{knight1961} (see \secref{sec:local_measurements_and_strictly_localized_states}), this means that measurements that are spacelike separated cannot influence one another. However, it would also be of interest to verify causality explicitly for the fundamental dynamical process in quantum field theory: particle production, evolution, and detection. 

The first attempt at such a calculation was made already in 1930, immediately following the inception of quantum field theory \cite{heisenberg1929,heisenberg1930}. Here \cite{kikuchi1930} one considered an initially excited atom decaying to its ground state under the production of electromagnetic radiation, and the time development of the radiated energy density was analyzed. Fermi subsequently refined this setup in what has become known as the Fermi problem \cite{fermi1932} by adding a second, spatially separated atom initially in its ground state acting as a detector. Assuming the electromagnetic field starts out in vacuum, Fermi considered the timing of the energy propagation from the first to the second atom by calculating the second atom's excitation probability as a function of time. In both \cite{kikuchi1930} and \cite{fermi1932} it was shown that the emitted radiation propagates at the speed of light, and their solutions demonstrated that quantum field theory is in agreement with relativistic causality \cite{heitler1954}. 

However, in 1964 Shirokov \cite{shirokov1964} pointed out that the causal solutions to the Fermi problem were a result of an approximation, where a certain integral was extended from only positive frequencies to include negative frequencies as well. Without this approximation there would be some nonzero probability for the second atom becoming excited before the signal has had time to propagate there \cite{shirokov1967}. Today we understand that this integral restriction is in fact the same as the positive-energy restriction encountered in the photon localization problem. The contradiction with causality came from another approximation inadvertently done in \cite{kikuchi1930} and \cite{fermi1932}, namely of including only energy-conserving terms in the interaction Hamiltonian. Under this assumption, the excited atom decays under the production of a single photon. Since single photons cannot be localized, this clearly leads to acausal, observable influences on the second atom, meaning that this cannot be an accurate description of the process. 

The correct resolution to this issue came in 1968 by Ferretti \cite{ferretti1968}. First, he pointed out that a proper analysis of atom radiation dynamics can only be done while keeping all interaction terms, meaning that the causally propagating, emitted light state must contain multiple photons. Second, he realized that it is necessary to specify an observable that is local to the measurement region. In this case for measurements at the second atom, an appropriate observable would for example be given by the projector for the second atom in the excited state, averaged over all possible photon states and first-atom states. On the other hand, if we were to include, e.g., the electromagnetic vacuum state in the measurement projector, this would correspond to measuring the photon number everywhere in space at the observation time. Since this is not an observable local to the second atom, the associated measurement probability could show a time variation before the arrival time of the signal at the second atom. 

Third, Ferretti made another important observation, namely that even with a proper, local observable, a nonzero excitation probability will occur for the second atom instantaneously. The reason for this is the now-familiar effect of vacuum fluctuations in quantum field theory, where the interaction Hamiltonian allows an atom to be found in an excited state even if no photons are initially present. Ferretti suggested that the time-independent, nonzero excitation probability for the second atom when no other sources are present constitutes a ``background signal''. By introducing the first, excited atom into the system again, we can obtain its induced ``pure signal'' by calculating the second atom's excitation probability subtracted the background value. The result obtained by Ferretti was that while keeping the exact spectral integral over only positive energies, the second atom experiences a \emph{change} in its excitation probability precisely when the signal has had time to propagate to it. 

Interpreting measurement probabilities in quantum field theory is in general somewhat delicate and has been a recurring source of misunderstanding \cite{buchholz1994} in the interpretation of several theorems and claimed paradoxes \cite{hegerfeldt1974,hegerfeldt1994,hegerfeldt1998,malament1996,halvorson2002} related to particle localization. The question is when the presence of a quantum state can be detected by some local measurement, and naively any nonzero measurement probability seems to be a good detection threshold. However, it turns out that no matter the measurement, some nonzero probability will occur even if there is just vacuum. Formally stated, it follows from the Reeh-Schlieder theorem \cite{reeh1961} that: any possible outcome of any possible local measurement will occur with nonvanishing probability in vacuum \cite{buchholz1994,redhead1995}. Therefore states in quantum field theory are observable, in the sense that they are \emph{distinguishable from vacuum}, when their probability for some measurement is \emph{different compared to} that for the vacuum state. 

Unfortunately, Ferretti's insights seem to have gone mostly unnoticed, and textbooks on quantum field theory repeated Fermi's approximation years later \cite{louisell1973}. The issues and solutions concerning the energy positivity, local observables, and measurement probabilities in vacuum were subsequently forgotten and rediscovered a number of times \cite{hegerfeldt1974,rubin1987,biswas1990,valentini1991,hegerfeldt1994,maddox1994,milonni1994a,hegerfeldt1998} (other discussions of the history of the Fermi problem can be found in \cite{shirokov1978,dickinson2016}). Today, Fermi's original approximation of keeping only the energy-conserving terms, now known as the rotating-wave approximation, and its limitations for precise dynamics is well understood \cite{power1997}. The causal nature of the Fermi problem is easily demonstrated by using the Heisenberg picture \cite{milonni1995}, where the operators' equations of motion are the same as classically \cite{milonni1994}. 

\autofootnotetext{foot:phi_x_0}{Note that for some real, scalar quantum field given by $\phi(\vect{r}, t) = \int_{-\infty}^{\infty} \ibuffn \frac{\dd^3 k}{\lp(2\pi\rp)^3} \ibuffb \frac{1}{\sqrt{2 \omega_{\vect{k}}}} a_{\vect{k}}^{} e^{i \vect{k} \cdot \vect{r} - i \omega t} + \hc$, the state $\ket{\psi_{\vect{r}, t}} = \phi(\vect{r}, t) \ket{0}$ gives a nonzero expectation value $\qexv{\psi_{\vect{r}, t}}{\norder{\phi\lp(\vect{r}', t\rp)^2}}$ for all space $\vect{r}'$ due to the factor $1 / \sqrt{2 \omega_{\vect{k}}}$. Hence this state is not localized to any region. It turns out that this problem is actually avoidable, as we can instead construct a single-particle state $\ket{\psi_{\vect{r}, t}} = \int_{-\infty}^{\infty} \ibuffn \frac{\dd^3 k}{\lp(2\pi\rp)^3} \ibuffb \sqrt{2 \omega_{\vect{k}}} a^\dagger_{\vect{k}} e^{- i \vect{k} \cdot \vect{r} + i \omega t} \ket{0}$, which at time $t$ gives an expectation value for $\norder{\phi(\vect{r}', t)^2}$ that is proportional to $\delta^{(3)}\lp(\vect{r} - \vect{r}'\rp)$ as required. However, the real localization problem is the absence of negative frequencies in the time dependence $e^{i \omega t} = e^{i \sqrt{\vect{k}^2 + m^2} t}$, which makes any single-particle state that is localized at time $t$ instantaneously become spread out everywhere in space at all times $t' > t$.}

On the other hand, the related topic of particle localization appears somewhat confusing in current literature. The majority of established sources today claim that the state given by $\phi(\vect{r}) \ket{0}$ is a single particle localized at $\vect{r}$ \cite{peskin1995,schwartz2014,\footnotebibid{foot:phi_x_0}} or use time-domain ladder operators $\comm{a(t)}{a^\dagger(t')} = \delta(t - t')$ \cite{loudon2001,karpinski2021}. Although these claims are usually approximately valid, the regime where they do not hold are not studied. A second category of works are aware that single particles cannot be localized but do not consider localization of superposition states \cite{hegerfeldt1974,halvorson2002,benincasa2014}. Finally, there are works that deal with causality and localization from the perspective of mode transformations through, e.g., the Heisenberg picture \cite{milonni1995} or quantization in bounded regions of spacetime \cite{su2016a,foo2020}. These analyses generally find that quantum field theory supports localized modes, however, they forgo information about how the states of such localized excitations look, which is needed to evaluate closeness to single particles as in our analysis. 

Lastly, through the circuitous history of this topic, there is one work that stands out as particularly important but overlooked: Knight's 1961 treatment of strictly localized states \cite{knight1961}. Long before Bialynicki-Birula and Hegerfeldt, Knight made the explicit connection between analytic properties implied by the energy positivity and particle localization. Before Ferretti, he realized the need for local observables, and he provided a precise definition of local operators in quantum field theory. He also correctly analyzed the measurement probabilities in terms of a rigorous treatment of vacuum expectation values. On top of that he provided a proof that any bosonic quantum state only containing terms of finite particle-number, such as single-particle states, cannot be localized.

\section{Local measurements and \\strictly localized states}\label{sec:local_measurements_and_strictly_localized_states}

\subsection{General, \texorpdfstring{$\bm{3+1}$}{3+1}D}\label{sec:general_3p1D}
This section introduces the theory of local measurements in the general case of $3 + 1$ dimensions, as established by Knight \cite{knight1961}. To analyze questions of localization in quantum field theory, we need a concept of what, fundamentally, is locally measurable within some spacetime region. Modern methods in particle physics however, mostly deal with asymptotic scattering experiments. In quantum optics, local measurements are treated by Glauber's correlation functions, but these are not suitable either as they are only an approximation to local observables and are not strictly causal \cite{bykov1989,plimak2011}. We therefore go back to first principles, where Bohr and Rosenfeld \cite{bohr1933} argued that the basic measurable quantity is the field itself, averaged over the spacetime region of the measurement. Knight formalized this with the concept of local observables \cite{knight1961}, which was further refined as local algebras with the formation of algebraic quantum field theory \cite{haag1959a}. 

It should be noted that there is still an ongoing discussion of exactly what observables in quantum field theory really are locally measurable \cite{beckman2001,borsten2021,bostelmann2021,ramon2021,jubb2022,albertini2023}. Several potential issues  with causality have been pointed out for local observables of extended spacetime regions \cite{sorkin1993,beckman2002}, and more involved measurement models have been suggested, such as Unruh-DeWitt detectors \cite{unruh1976,dewitt1979,martinmartinez2015} and the FV measurement framework \cite{fewster2018,bostelmann2021}. Still, these topics are unlikely to be fully resolved until the quantum measurement problem is. The concrete observable we will use, the smeared electric field, seems to have some evidence supporting that it is measurable \cite{jubb2022}. In any case, we will assume the simple definition of local observables from Knight in this work; if this assumption is later found to not reflect reality, our results would have to be updated accordingly. 

In this subsection, and here only, we let $x$ denote spacetime coordinates $x^\mu = (x^0, x^1, x^2, x^3)^\mu$. Knight \cite{knight1961} defines local operators for real, scalar quantum fields $\phi(x)$ as sums and products of the field operator, smeared out in the measurement region. Intuitively, the local observables for $\phi(x)$ at point $x$ are
\begin{align}
    &\; C_0, \, \phi(x), \, \phi(x)^2, \, \dotsc \label{local_observables} \\
    &= \text{constant}, \, \text{field strength}, \, \text{(parts of) energy density}, \, \dotsc \nonumber
\end{align}
as well as sums of such quantities. We may also include normal-ordered expressions such as $\,\norder{\phi(x)^2}$, since $\norder{\phi(x)^2} \nbuffeq = \phi(x)^2 - \text{const.}$ To get observables local to some spacetime region $G$, we then smear (integrate) such pointwise observables against some arbitrary function with support in $G$. 

For electrodynamics, the quantum field we consider is the electromagnetic 4-vector potential $A^\mu(x)$. The generalization of Knight's definition to (bosonic) vector fields such as $A^\mu(x)$ is given by Haag in \cite{haag1996}. Formally, \emph{local operators} for $A^\mu(x)$ for some region $G$ are given by
\begin{equation}
\begin{aligned}\label{Q_G}
    Q(G) = \sum_n \int_G \ibuffa &\dd^4 x_1 \dotsm \dd^4 x_n \ibuffb \zeta^{\mu_1 \dotsc \mu_n}_n(x_1, \dotsc, x_n) \\
    &A_{\mu_1}(x_1) \dotsm A_{\mu_n}(x_n).
\end{aligned}
\end{equation}
Here $\zeta_n(\cdot)$ are a sequence of smearing functions of $n$ spacetime arguments, each a tensor that can be contracted with $n$ field operators. \emph{Local observables} are then defined as 
\begin{equation}\label{L_G}
    L(G) = \text{Hermitian, gauge-invariant } Q(G),
\end{equation}
and they represent all measurable quantities in the region $G$. Note that in definition \eqref{Q_G} it is unnecessary to include other fields such as derivatives $\partial_\nu A^\mu(x)$ since they are already covered, as can be seen with integration by parts. 

The smearing of the operator is done because the quantum field $A^\mu(x)$ is strictly speaking not an operator but rather an operator-valued distribution. This distinction can be important; for instance when we wish to use the spectral theorem \cite{hall2013}, the smearing is necessary. Other times we will use the usual physics shorthand of considering simpler, unsmeared operators such as $\norder{A^2(x)} \nbuffeq = \nbuffeq \norder{A^\mu(x) A_\mu(x)}$. Such unsmeared operators are used with the understanding that expressions such as $\qexv{\varphi}{{\norder{A^2(x)}}}$ for some state $\ket{\varphi}$ are to be smeared out with a function $\zeta(x)$ in the end:
\begin{equation}\label{exv_A2_smeared}
    \qexv{\varphi}{\int \ibuffa \dd^4 x \ibuffb \zeta(x) \norder{A^2(x)}} = \int \ibuffa \dd^4 x \ibuffb \zeta(x) \qexv{\varphi}{\norder{A^2(x)}}.
\end{equation}

There are two\autofootnote{foot:self_adjoint}{There is also a potential complication regarding whether $L$ defined in \eqref{L_G} is self-adjoint, and not just Hermitian, which is needed to use the spectral theorem \cite{hall2013} in \eqref{E_zeta_spectral}. Here we will not treat this question. Instead we will simply assume that $A^\mu(x)$ (and its derivatives) is self-adjoint \cite{haag1996}, and that Hermitian superpositions of smeared fields defining $L$ \eqref{L_G} are self-adjoint as well.} subtleties regarding definitions \eqref{Q_G} and \eqref{L_G}. First, the summation range in \eqref{Q_G} is unspecified. For finite sums the meaning is unambiguous, but we must in general consider also infinite sums of the form $Q = \lim_{N \rightarrow \infty} Q_N$, where $Q_N$ are finite sums. However, formalizing this by a specific type of operator convergence is difficult since the operators $Q_N$ are typically unbounded\autofootnote{foot:algebra}{Our definition of local operators \eqref{Q_G} matches eq. (1) in \cite{knight1961} and eq. (II.4.1) in \cite{haag1996}, and it generates the \emph{polynomial algebra} of the field. However, since the (smeared) field is an unbounded operator, technically the domains of the operators $Q$ have to be carefully considered. To avoid this issue, and to handle the convergence of infinite operator sequences, it is convenient to go over to algebras of bounded operators instead through the spectral theorem, as discussed in \cite{haag1996}. For our purposes however, we will simply use the more straight-forward polynomial algebra \eqref{Q_G} and ignore the question of the domains of the operators $Q$.}. For our purposes we will see how this difficulty can be avoided when considering arguments that must hold for all possible operators $Q$. On the other hand, when choosing some specific operator $Q$, we will only use finite sums in \eqref{Q_G}.

The second subtlety concerns the definition of gauge-invariance in \eqref{L_G}. Characterizing the full set of observables that are invariant under gauge transformations of some quantized field $A^\mu(x)$ is not necessarily straight-forward \cite{strocchi1974,beckman2002}. Instead we will again adopt a pragmatic point of view: When we need specific examples of observables $L$, we can use the electric or magnetic fields derived from $A^\mu(x)$, which we know are gauge-invariant \cite{strocchi1974}. These fields are covered by \eqref{Q_G} since they are related to derivatives of $A^\mu(x)$. On the other hand, for arguments concerning all possible observables $L$, it will be easier to consider all possible $Q$ instead, which must include all $L$. 

Knight then introduces the class of strictly localized states. The idea is that such states are indistinguishable from vacuum outside the region they are localized to, meaning that the expectation value of any outside measurement should give the same value as for vacuum. Formalizing this idea, we call a state $\ket{\psi}$ \emph{strictly localized} to $G$ if
\begin{equation}\label{strictly_localized_state}
    \qexv{\psi}{L\lp(G^C\rp)} = \qexv{0}{L\lp(G^C\rp)}, \tbuff \forall L(G^C),
\end{equation}
where $G^C$ is the complement region of $G$. Note that the physical content here is the postulate that all possible measurements in some spacetime region are given by some local observable according to \eqref{L_G}.

An important continuation of Knight's analysis came by Licht in 1963 \cite{licht1963}. Licht was able to show that for every state $\ket{\psi}$ strictly localized to $G$, there exists a \emph{unique} operator $W$ such that $\ket{\psi} = W \ket{0}$, which satisfies
\begin{equation}\label{W_L}
    \comm{W}{L(G^C)} = 0, \tbuff \forall L(G^C)
\end{equation}
and
\begin{equation}\label{W_isometry}
    W^\dagger W = \identity,
\end{equation}
where $\identity$ is the identity operator. We label $W$ the \emph{Licht operator} corresponding to the state $\ket{\psi}$. Eq. \eqref{W_L} says that $W$ commutes with all observables \eqref{L_G} local to the outside of the localization region of $\ket{\psi}$. Note also that since we have an infinite-dimensional space, $W$ may fail to be unitary, $W W^\dagger \neq \identity$, even though it satisfies \eqref{W_isometry}. 

These two conditions for localized states are equivalent in the sense that for every $\ket{\psi}$ satisfying \eqref{strictly_localized_state}, there exists a (unique) operator $W$ satisfying \eqref{W_L} and \eqref{W_isometry}, and for every $W$ satisfying these conditions, the state $\ket{\psi} = W \ket{0}$ satisfies \eqref{strictly_localized_state}. However, although \eqref{W_L} contains a lot of information if we are given a valid Licht operator $W$, it is not as useful for checking whether a particular $W$ satisfies the condition; it must in principle be checked for every possible local observable satisfying \eqref{L_G}, which is not very practical. 

Here we introduce a simpler condition, which is sufficient for guaranteeing \eqref{W_L}:
\begin{equation}\label{W_A}
    \comm{W}{A^\mu(x)} = 0, \tbuff \forall x \in G^C,
\end{equation}
namely that $W$ commutes with every component of $A^\mu(x)$ outside the localization region $G$. This implies that $W$ also commutes with any sums and products of $A^\mu(x)$ evaluated outside $G$, meaning that it commutes with all $Q_N(G^C)$ consisting of finite sums of the form \eqref{Q_G}. Since $W$ is bounded, it must also commute with infinite sums, $\comm{W}{\lim_{N \rightarrow \infty} Q_N} = \lim_{N \rightarrow \infty} \comm{W}{Q_N}$, irrespective of the concrete choice of how to define the limit. Hence \eqref{W_A} is a sufficient condition for \eqref{W_L}. 

It can also be useful to consider unitary operators $U$ instead of the isometric Licht operators $W$. This can always be done by extending the Hilbert space through introducing a source space consisting of ground and excited states $\ket{\text{g}}$ and $\ket{\text{e}}$, and raising and lowering operators $\sigma_+ = \ket{\text{e}} \bra{\text{g}}$ and $\sigma_- = \ket{\text{g}} \bra{\text{e}}$. We can then construct a unitary operator $U$ from $W$ as
\begin{equation}\label{U_W}
    U = W \otimes \sigma_- + W^\dagger \otimes \sigma_+ + \lp(\identity - W W^\dagger\rp) \otimes \ket{\text{g}}\bra{\text{g}},
\end{equation}
which acts on the total, electromagnetic and source Hilbert space. The strictly localized state $\ket{\psi}$ associated with the Licht operator $W$ is then found from $U$ as the electromagnetic reduced state\autofootnote{foot:partial_trace}{Usually the partial trace results in a mixed state given by a density matrix. Here it happens to work out so that the reduced state is pure, hence we can write it directly as in \eqref{psi_from_U}.}:
\begin{equation}\label{psi_from_U}
    \ket{\psi}\bra{\psi} = \tr_{\text{src}} \Big[U \big(\ket{0} \otimes \ket{\text{e}}\big) \big(\bra{0} \otimes \bra{\text{e}}\big) U^\dagger\Big],
\end{equation}
where the partial trace is over the source space. The commutation condition \eqref{W_A} carries over to $U$, and since it is unitary, we may rewrite it as
\begin{equation}\label{U_A}
    U^\dagger A^\mu(x) U = A^\mu(x), \tbuff \forall x \in G^C.
\end{equation}
This condition was also the one considered in \cite{gulla2021} (and in \cite{gulla2021a}), where it was motivated directly from the action of the source. 

In summary, all measurements that can be done locally in some spacetime region $G$ are characterized by a local observable $L$ according to \eqref{L_G}. States $\ket{\psi}$ strictly localized to $G$, i.e., states that are indistinguishable from vacuum outside $G$, are the states that satisfy \eqref{strictly_localized_state}. If we are given a source described by a Licht operator $W$ satisfying \eqref{W_isometry} and \eqref{W_A} or a unitary operator $U$ satisfying \eqref{U_A}, then the state given by $W \ket{0}$ or \eqref{psi_from_U}, respectively, is strictly localized to $G$. 

Note that the negated implications do not work; if we are for instance given an operator $U$ that does not satisfy \eqref{U_A}, we do not automatically know whether the state $\ket{\psi}$ given by \eqref{psi_from_U} is localized or not. This is because Licht's theorem only guarantees the existence of one operator $W$ with $\ket{\psi} = W \ket{0}$ such that \eqref{W_L} and \eqref{W_isometry} hold. There could very well be other operators $W$ with $\ket{\psi} = W \ket{0}$ that do not satisfy \eqref{W_L} and \eqref{W_isometry}. The most straight-forward way of showing that some state is not strictly localized is to pick some specific observable $L$ and showing that \eqref{strictly_localized_state} does not hold for that particular observable.

\subsection{1D, single polarization}\label{sec:1D_single_polarization}
We now specialize the theory of the previous section to the situation in \figref{fig:setup}. We assume the source produces plane-wave modes, so we formulate the problem as one-dimensional by considering measurements along one coordinate direction $x^\mu = (ct, x, 0, 0)^\mu$, and we keep only wavevectors $k^\mu = (\omega/c, k, 0, 0)^\mu$ along this direction. Hence we again let $x$ denote the position along this axis and $k$ denote the 1D wavevector. The frequency is given by the vacuum dispersion relation $\omega = \abs{k}c$. Further, we simplify by assuming the source produces only one, transverse polarization, say in the $z$ direction. We can then treat the electromagnetic potential as a scalar $\vect{A}(x, t) = A(x, t) \hat{z}$, with
\begin{equation}\label{A_x_t}
    A(x, t) = \int_{-\infty}^\infty \ibuffn \dd k \ibuffb \mathcal{A}(\omega) a(k) e^{ikx - i \omega t} + \hc
\end{equation}
Here $a(k)$ is the usual annihilation operator satisfying $\comm*{a(k)}{a^\dagger(k')} = \delta(k - k')$, and $\mathcal{A}(\omega)$ is some function that depends only on $\omega$. For later convenience we note that we may write $\mathcal{A}(\omega) = K / \sqrt{- i \omega}$ for some constant $K > 0$, by absorbing any additional phase factor into $a(k)$. 

The source is located in the region $x < - cT$ and is switched on at $t = -T$, before which we let there be electromagnetic vacuum $\ket{0}$. We assume that the electromagnetic reduced state produced by the source is a pure state $\ket{\psi}$ (see \secref{sec:discussion_and_conclusion} for comments about mixed states). Relativistic causality then dictates that $\ket{\psi}$ must be strictly localized to the region $x \leq ct$, meaning that it satisfies \eqref{strictly_localized_state} for all $x > ct$. From \secref{sec:general_3p1D} we know that this condition is ensured if we for instance can find a unitary source operator $U$ such that \eqref{U_A} holds for $x > ct$. This is an appealing result because we have started with characterizing the source by the types of states it can produce, and we are led back to a requirement on the source operator $U$.

We are interested in how close we can get with this source to target states $\ket{1_\xi}$ and $\ket{1_g}$ consisting of rightward-moving modes, i.e., modes with $k > 0$ only. In this case there is a one-to-one correspondence between $k$ and $\omega = \abs{k}c$, which is why we were able to define $\ket{1_\xi}$ and $\ket{1_g}$ in \eqref{1xi} and \eqref{1g} in terms of their spectra; the creation operator $a^\dagger(\omega)$ is the operator $\frac{1}{\sqrt{c}} a^\dagger(k = \omega/c)$ corresponding to modes propagating in the $+x$ direction. 

We cannot immediately make the same restriction in $k$ for the state produced by the source $\ket{\psi}$, since the maxima in \eqref{F_MAX_1XI} and \eqref{F_MAX_1G} are to be taken over \emph{all} strictly localized states, thus also over states containing $k < 0$ modes. However, we will divide the analysis into two parts: When considering all possible source states $\ket{\psi}$, as we will do in the derivations of the upper bounds for the maximum fidelities, we have to account for states containing modes with $k < 0$. On the other hand, when picking specific examples of source states $\ket{\psi}$, as we will do for the lower bounds, we will restrict our attention to $k > 0$ (this is of course assuming that having a strictly localized state $\ket{\psi}$ with only $k > 0$ modes is possible in the first place, which will be shown to be the case). 

For source states $\ket{\psi}$ only containing $k > 0$ modes, the corresponding source operator $U$ also contains only $k > 0$ and can thus be expressed in terms of $a(\omega)$ and $a^\dagger(\omega)$. In this case, we can also simplify requirement \eqref{U_A}, since we can split $A(x, t)$ from \eqref{A_x_t} into one integral over $k > 0$ and another over $k < 0$. For sources $U$ only containing $k > 0$ modes, the latter integral automatically commutes with $U$, meaning that it is sufficient to check that
\begin{equation}\label{U_A_pos}
    U^\dagger A_{k>0}^{}(x, t) U = A_{k>0}^{}(x, t), \tbuff \forall x > ct,
\end{equation}
for
\begin{equation}\label{A_k_pos}
    A_{k>0}^{}(x,t) = \int_0^\infty \ibuffi \dd \omega \ibuffb \mathcal{A}(\omega) a(\omega) e^{i \omega (x - ct)/c} + \hc,
\end{equation}
where we have rewritten all quantities in terms of frequency $\omega$. To ensure that $U$ is not time-dependent, we consider only $t > -T/2$, i.e., after the source is switched off again, and assume no interactions are present after this point. That is, we assume a free field theory. Since $A_{k>0}^{}(x,t)$ is a function of $x - ct$, checking \eqref{U_A_pos} for all $x$ and $t$ such that $x > ct$ (and $t > - T/2$) amounts to the same as checking for $x = 0$ and all $t < 0$. This also justifies the situation indicated in \figref{fig:setup}, where we have picked a fixed observation point. 

Thus in summary, for finding examples of strictly localized states, we are looking for sources described by unitary operators $U$ satisfying
\begin{equation}\label{U_A_t}
    U^\dagger A(t) U = A(t), \tbuff \forall t < 0,
\end{equation}
or, equivalently, for sources described by Licht operators $W$ satisfying
\begin{equation}\label{W_A_t}
    \comm{W}{A(t)} = 0, \tbuff \forall t < 0,
\end{equation}
where we have defined
\begin{equation}\label{A_t}
    A(t) = A_{k>0}^{}(x = 0, t) = \int_0^\infty \ibuffi \dd \omega \ibuffb \mathcal{A}(\omega) a(\omega) e^{-i \omega t} + \hc
\end{equation}

As a final ingredient for our analysis, in the derivation of the upper bound for \eqref{F_MAX_1XI}, we need an observable $L$ local to $x = 0$ and $t < 0$. It is beneficial for this local observable to consist of only $k > 0$ modes in the same way \eqref{A_t} does, since a measurement is better at distinguishing $\ket{1_\xi}$ and $\ket{0}$ when it only contains modes present in $\ket{1_\xi}$. One such observable could be a smeared version of the electric field with $k > 0$:
\begin{equation}\label{E_t}
    E(t) = E_{k>0}^{}(x = 0, t) = \int_{0}^\infty \ibuffi \dd \omega \ibuffb \mathcal{E}(\omega) a(\omega) e^{- i \omega t} + \hc
\end{equation}
Here $\mathcal{E}(\omega) = i\omega\mathcal{A}(\omega)$. To show that $E(t)$ in fact is a local observable, calculate 
\begin{align}\label{A_t_dx_A_t_dt}
    \frac{c\partial_x A(0, t) - \partial_t A(0, t)}{2} = \int_0^\infty \ibuffi \dd k \ibuffb i \omega \mathcal{A}(\omega) a(k) e^{- i \omega t} + \hc
\end{align}
Since the right-hand side matches that of \eqref{E_t}, and since $\partial_x A(0, t)$ and $\partial_t A(0, t)$ are covered by the definition of $Q$ in \eqref{Q_G}, we see that $E(t)$ is a local operator. Further, with only one polarization, the left-hand-side terms in \eqref{A_t_dx_A_t_dt} are simply the magnetic and electric field, which we know are gauge-invariant. Finally, $E(t)$ is clearly Hermitian. Thus $E(t)$ is a local observable to $x = 0$ and the time $t$ of the measurement. 

We will make use of the operator $E(t)$ in both smeared and unsmeared form. For showing that some state is \emph{not} strictly localized, we will typically use the unsmeared observable $\norder{E^2(t)} \nbuffeq = E^2(t) - \text{const.}$ For any state $\ket{\varphi}$, a nonzero expectation value $\qexv{\varphi}{\norderop{E^2(t)}}$ for any negative $t$ is enough to conclude that $\ket{\varphi}$ is not strictly localized to $t \geq 0$, since $\qexv{0}{\norderop{E^2(t)}} = 0$. For finding the upper bound for \eqref{F_MAX_1G}, we will need the spectral decomposition of an operator local to $t < 0$. In this case we will use a smeared observable of the form $E_\zeta = \int \ibuffa \dd t \ibuffb \zeta(t) E(t)$ for some real function $\zeta(t)$, which is local to the support of $\zeta(t)$.

\section{Strictly localized state \texorpdfstring{$\bm{\ket{\eta_{1,2}}}$}{|eta1,2>} \\near single photon}\label{sec:strictly_localized_state_eta12_near_single_photon}

\subsection{Pulse modes}\label{sec:pulse_modes}
In the following analysis it will be convenient to decompose the Fock space into a countable basis rather than the usual (uncountable) frequency decomposition. We do this by introducing a set of pulse modes $\xi_n(\omega)$ forming a (countable) basis for the function space $L^2(0, \infty)$:
\begin{subequations}\label{xi_n_basis}
\begin{align}
    \int_0^\infty \ibuffi \dd \omega \ibuffb \xi_n^*(\omega) \xi_m^{\vphantom{*}}(\omega) &= \delta_{nm}, \label{xi_orthog} \\
    \sum_n \xi_n^*(\omega) \xi_n^{\vphantom{*}}(\omega') &= \delta(\omega - \omega').
\end{align}
\end{subequations}
To each pulse mode we then define corresponding ladder operators
\begin{equation}\label{a_n}
    a_n^\dagger = \int_0^\infty \ibuffi \dd\omega \ibuffb \xi_n(\omega) a^\dagger(\omega),
\end{equation}
which then satisfy $\comm*{a_n^{\vphantom{\dagger}}}{a_m^\dagger} = \delta_{nm}$. Analogous to $a^\dagger(\omega)$ creating a photon with frequency $\omega$, the operator $a_n^\dagger$ creates a photon in pulse mode $\xi_n(\omega)$. Since the states generated by $a_n^\dagger$ for different $n$ must be orthogonal, we can write the Fock space as a tensor product of a state space for each pulse mode; for example we can write the total identity operator as
\begin{equation}\label{identity_n}
    \identity = \Big(\sum_n \ket{n_1}\bra{n_1}\Big) \otimes \Big(\sum_n \ket{n_2}\bra{n_2}\Big) \otimes \dotsb,
\end{equation}
with pulse-mode Fock states
\begin{equation}\label{Fock_n_m}
    \ket{n_m} = \frac{1}{\sqrt{n!}} a_m^{\dagger^n} \ket{0_m},
\end{equation}
where $\ket{0_m}$ is the vacuum state of pulse mode $m$. 

We can also rewrite \eqref{A_t} and \eqref{E_t} in the pulse-mode basis as
\begin{equation}\label{A_t_an}
    A(t) = \sum_n A_n(t) a_n + \hc
\end{equation}
and
\begin{equation}\label{E_t_an}
    E(t) = \sum_n E_n(t) a_n + \hc,
\end{equation}
with associated functions
\begin{equation}\label{A_n_t}
    A_n(t) = \int_0^\infty \ibuffi \dd\omega \ibuffb \mathcal{A}(\omega) \xi_n(\omega) e^{-i\omega t}
\end{equation}
and
\begin{equation}\label{E_n_t}
    E_n(t) = \int_0^\infty \ibuffi \dd\omega \ibuffb \mathcal{E}(\omega) \xi_n(\omega) e^{-i\omega t}.
\end{equation}
Note that according to the Paley-Wiener criterion \cite{paley1934}, the functions $A_n(t)$ and $E_n(t)$ must be nonzero (almost) everywhere, since they contain only positive frequencies. In particular they have infinite tails for $t < 0$.

\subsection{Algorithm for \texorpdfstring{$\bm{\ket{\eta_{1,2}}}$}{|eta1,2>}}\label{sec:algorithm_for_eta12}
Ref. \cite{gulla2021} gives the following algorithm for constructing a state $\ket{\eta_{1,2}}$ that is strictly localized to $t \geq 0$ while also being close to a single photon. For understanding this construction and why it works, it is easiest to start with the description in \cite{gulla2021}. 

\begin{enumerate}
    \item Pick a complex-valued function $g(t)$ with $g(t) = 0$ for $t < 0$, and calculate its Fourier transform $G(\omega)$. We will refer to $g(t)$ as the seed function for the state. We let $g(t)$ be normalized, $\int_0^\infty \ibuffi \dd t \ibuffb \abs{g(t)}^2 = 1$.

    \item\label{itm:G_tilde} Modify $G(\omega) \mapsto \widetilde{G}(\omega)$ as follows:
    \begin{equation}\label{G_tilde_omega}
        \widetilde{G}(\omega) = G(\omega) - \beta G^*(-\omega),
    \end{equation}
    where
    \begin{subequations}\label{beta_I}
    \begin{align}
        \beta &= \frac{1}{2I^*} \lp(1 - \sqrt{1 - 4 \abs{I}^2}\rp), \\
        I &= \int_0^\infty \ibuffi \dd\omega \ibuffb G(\omega)G(-\omega). \label{I}
    \end{align}
    \end{subequations}
    Note that the inverse Fourier transform $\widetilde{g}(t)$ of $\widetilde{G}(\omega)$ vanishes for $t < 0$ since $g(t)$ does. 

    \item\label{itm:G_xi} Normalize $\widetilde{G}(\omega)$ such that $\int_0^\infty \ibuffi \dd\omega \ibuffb \abs*{\widetilde{G}(\omega)}^2 = 1$. Identify two pulse-mode spectra $\xi_1(\omega)$ and $\xi_2(\omega)$ using
    \begin{subequations}\label{xi1_2_omega}
    \begin{alignat}{3}
        \xi_1(\omega) &= \widetilde{G}(\omega), \tbuff &&\omega > 0, \\
        \xi_2(\omega) &= \sqrt{\frac{1 - \widetilde{\eta}}{\widetilde{\eta}}} {\widetilde{G}\null}^*(-\omega), \tbuff &&\omega > 0.
    \end{alignat}
    \end{subequations}
    The constant $\widetilde{\eta} > 0$ is picked such that $\xi_2(\omega)$ gets normalized. The two pulse modes $\xi_1(\omega)$ and $\xi_2(\omega)$ are normalized and orthogonal because of \stepref{itm:G_tilde}, and can therefore be chosen as two modes in the basis $\xi_n(\omega)$ in \eqref{xi_n_basis}.

    \item Define operators
    \begin{align}
        \widetilde{a}_1^\dagger &= a_1^\dagger \frac{1}{\sqrt{a_1^{\vphantom{\dagger}} a_1^\dagger}} = \sum_n \ket{{n + 1}_1}\bra{n_1}, \label{A1} \\
        S &= e^{\gamma a_1^{\vphantom{\dagger}} a_2^{\vphantom{\dagger}} - \gamma a_1^\dagger a_2^\dagger}, \label{S}
    \end{align}
    where $\tanh\gamma = \sqrt{\widetilde{\eta}/(1 - \widetilde{\eta})}$, which act on the mode space of $\xi_1(\omega)$ and $\xi_2(\omega)$. Our strictly localized state is then
    \begin{equation}\label{eta12}
        \ket{\eta_{1,2}} = W \ket{0},
    \end{equation}
    given by the Licht operator
    \begin{equation}\label{W_eta12}
        W \equiv S^\dagger \widetilde{a}_1^\dagger S.
    \end{equation}
\end{enumerate}

To see that $\ket{\eta_{1,2}}$ indeed is strictly localized to $t \geq 0$, we use condition \eqref{W_isometry} and \eqref{W_A_t}. The operator $S$ from \eqref{S} is a two-mode squeeze operator with the property
\begin{equation}\label{S_a1}
    S a_1^{\vphantom{\dagger}} S^\dagger = a_1^{\vphantom{\dagger}} \cosh \gamma + a_2^\dagger \sinh \gamma, 
\end{equation}
and similar for $S a_2 S^\dagger$. In addition, ref. \cite{schumaker1985} gives a list of possible decompositions of $S$, of which we will use
\begin{equation}\label{S_factorized}
    S = e^{- a_1^\dagger a_2^\dagger \tanh\gamma} \lp(\cosh\gamma\rp)^{-a_1^{\vphantom{\dagger}} a_1^\dagger - a_2^\dagger a_2^{\vphantom{\dagger}}} e^{a_1^{\vphantom{\dagger}} a_2^{\vphantom{\dagger}} \tanh\gamma}.
\end{equation}
Using that $S$ is unitary, \eqref{W_isometry} is in this case equivalent to showing that $\widetilde{a}_1 \widetilde{a}_1^\dagger = \identity$. Writing it out,
\begin{equation}\label{A1_A1d}
    \widetilde{a}_1 \widetilde{a}_1^\dagger = \frac{1}{\sqrt{a_1 a_1^\dagger}} a_1 a_1^\dagger \frac{1}{\sqrt{a_1 a_1^\dagger}},
\end{equation}
and considering how this operator acts on pulse-mode Fock states $\ket{n_m}$ from \eqref{Fock_n_m}, it is easy to see that \eqref{W_isometry} holds. 

To show \eqref{W_A_t}, we again use the unitarity of $S$ to get that \eqref{W_A_t} is equivalent to
\begin{equation}\label{A1d_S_A_Sd}
    \comm{\widetilde{a}_1^\dagger}{S A(t) S^\dagger} = 0, \tbuff t < 0.
\end{equation}
Using \eqref{S_a1}, it follows that
\begin{equation}\label{S_a1_a2_Sd}
    S \lp(a_1^{\vphantom{\dagger}} - \sqrt{\frac{1 - \widetilde{\eta}}{\widetilde{\eta}}} a_2^\dagger\rp) S^\dagger = - a_2^\dagger \sqrt{\frac{1 - 2 \widetilde{\eta}}{\widetilde{\eta}}}.
\end{equation}
At the same time, using that $\mathcal{A}^*(\omega) = \mathcal{A}(-\omega^*)$, we get from \eqref{xi1_2_omega} that
\begin{equation}\label{A2c_t_A1_t}
    \sqrt{\frac{\widetilde{\eta}}{1 - \widetilde{\eta}}} A_2^*(t) + A_1(t) = \int_{-\infty}^\infty \ibuffn \dd \omega \ibuffb \mathcal{A}(\omega) \widetilde{G}(\omega) e^{-i \omega t}.
\end{equation}
Since $\widetilde{g}(t) = 0$ for $t < 0$, it follows that $\widetilde{G}(\omega)$ is analytic in the upper half-plane of complex frequency $\omega$. The function $\mathcal{A}(\omega) \sim 1/\sqrt{- i \omega}$ is also analytic there with the conventional branch cut of the complex square root. Therefore the product $\mathcal{A}(\omega) \widetilde{G}(\omega)$ is analytic in this half-plane, and since it decays sufficiently fast\autofootnote{foot:G_w_A_w}{For any $g(t)$ in $L^2(0, \infty)$, Titchmarsh' theorem \cite{titchmarsh1948} dictates how quickly $G(\omega)$ falls off for large $\omega$ in the upper half-plane of complex frequency $\omega$. The factor $\mathcal{A}(\omega) \sim 1/\sqrt{\omega}$ only strengthens this convergence, but there is a possibility of divergence at the origin. This only happens for very special choices of seed functions $g(t)$ that have ``almost-diverging'' norm at $\omega = 0$, which we must exclude.} for $\omega \rightarrow \infty$ there, it follows that its inverse Fourier transform vanishes for $t < 0$. Thus the left-hand side of \eqref{A2c_t_A1_t} is zero for negative times. Using this we can rewrite \eqref{A_t_an} for $t < 0$ as
\begin{equation}\label{A_t_neg}
\begin{aligned}
    A(t) &= A_1(t) \lp(a_1^{\vphantom{\dagger}} - \sqrt{\frac{1 - \widetilde{\eta}}{\widetilde{\eta}}} a_2^\dagger\rp) \\
    &\eqbuff + \sum_{n \geq 3} A_n(t) a_n + \hc, \tbuff\tbuff t < 0.
\end{aligned}
\end{equation}
From \eqref{S_a1_a2_Sd} and \eqref{A_t_neg}, it is then clear that \eqref{A1d_S_A_Sd} holds, meaning that $\ket{\eta_{1,2}}$ is strictly localized to $t \geq 0$ as desired.

\subsection{Fidelity of \texorpdfstring{$\bm{\ket{\eta_{1,2}}}$}{|eta1,2>}}\label{sec:fidelity_of_eta12}
By expanding the exponential of $S$ and $S^\dagger$ in \eqref{S}, and inserting into \eqref{W_eta12}, we find that the state $\ket{\eta_{1,2}}$ is of the form
\begin{equation}\label{eta12_expansion}
    \ket{\eta_{1,2}} = c_1 \ket{1_1 \kbuff 0_2} + c_2 \ket{2_1 \kbuff 1_2} + c_3 \ket{3_1 \kbuff 2_2} + \dotsb,
\end{equation}
for coefficients $c_1, c_2, \dotsc$ Its fidelity with the single-photon state $\ket{1_1 \kbuff 0_2}$ is given by the size of the first coefficient, which we can calculate by using \eqref{S_factorized} on \eqref{W_eta12} and some algebra, giving
\begin{equation}\label{F}
    F \equiv \abs{\braket{1_1 \kbuff 0_2}{\eta_{1,2}}} = \sqrt{\frac{\lp(1 - 2\widetilde{\eta}\rp)^3}{{\widetilde{\eta}\wbuff}^2 - {\widetilde{\eta}\wbuff}^3}} \mathrm{Li}_{-\frac{1}{2}}\lp(\frac{\widetilde{\eta}}{1 - \widetilde{\eta}}\rp),
\end{equation}
where $\mathrm{Li_s}(z) = \sum_{k=1}^\infty z^k / k^s$ is the polylogarithm function. 

Let $\eta$ be the negative-frequency fraction of the square norm of $G(\omega)$ as defined in \eqref{eta}. Similarly, it follows from \eqref{xi1_2_omega} that $\widetilde{\eta}$ is the negative-frequency fraction of the square norm of $\widetilde{G}(\omega)$,
\begin{equation}\label{eta_tilde}
    \widetilde{\eta} = \frac{\int_{-\infty}^0 \ibuffn \dd\omega \ibuffb \abs*{\widetilde{G}(\omega)}^2}{\int_{-\infty}^\infty \ibuffn \dd\omega \ibuffb \abs*{\widetilde{G}(\omega)}^2}.
\end{equation}
Using \eqref{G_tilde_omega}, \eqref{beta_I}, and some algebra, we can show that
\begin{equation}\label{eta_tilde_eta_J}
    \widetilde{\eta} - \eta = - \frac{1 - J}{2J} \lp(1 - 2\eta \rp),
\end{equation}
where
\begin{equation}\label{J}
    J = \sqrt{1 - 4 \abs{I}^2}.
\end{equation}
We also get that 
\begin{equation}\label{I_upper}
    \abs{I}^2 \leq \eta (1 - \eta)
\end{equation}
by applying the Cauchy-Schwarz inequality to \eqref{I}.

In the algorithm for constructing $\ket{\eta_{1,2}}$, we assume that $g(t)$ is chosen such that $\eta < 1/2$, meaning that $G(\omega)$ has its main weight for positive frequencies. This ensures that $\widetilde{G}(\omega)$ is nonvanishing. Additionally, it is clear from \eqref{eta_tilde_eta_J} and \eqref{I_upper} that this assumption means that the modification in \stepref{itm:G_tilde} never increases the amount of negative frequencies. In other words,
\begin{equation}\label{eta_tilde_eta}
    \widetilde{\eta} \leq \eta.
\end{equation}

In practical situations we will often consider functions $g(t)$ with a very small amount of negative frequencies, i.e., $\eta \ll 1$. It will be useful to have a lower bound for $F$ in this regime, expressed purely as a function of $\eta$. This is easily obtained by expanding \eqref{F} for small $\widetilde{\eta}$ and using \eqref{eta_tilde_eta}, giving
\begin{equation}\label{F_lower_1order}
    F \geq 1 - \lp(\frac{3}{2} - \sqrt{2}\rp) \eta + \order{\eta^2}.
\end{equation}
Thus we see that the parameter $\eta$, which is determined by the choice of seed function $g(t)$ of the localized state, quantifies the state's similarity with a single photon: as $\eta \to 0$, the state $\ket{\eta_{1,2}}$ tends to a single photon in pulse mode $\xi_1(\omega)$ according to \eqref{F_lower_1order}. Also note that by the Paley-Wiener criterion \cite{paley1934}, it is impossible for a function and its Fourier transform both to be supported for positive arguments only. Since $\widetilde{g}(t) = 0$ for $t < 0$, this means that $\eta = 0$ (a strictly localized single photon) is impossible.

\section{Bounds for \texorpdfstring{$\bm{F_{\text{max}}\big(\ket{1_g}\big)}$}{Fmax(|1g>)}}\label{sec:bounds_for_F_max_1g}

\subsection{Upper bound}\label{sec:upper_bound_for_F_max_1g}
We now turn to finding an upper bound for \eqref{F_MAX_1G}. Here we consider the maximum fidelity between any strictly localized state $\ket{\psi}$ and a causal single photon $\ket{1_g}$ in some positive-time pulse $g(t)$, as defined in \eqref{1g}. Any physical state $\ket{\psi}$ contains a superposition of (products of) ladder operators $a^\dagger(\omega)$ only for $\omega > 0$, acting on the vacuum state. Therefore the maximum fidelity satisfies
\begin{align}
    F_{\text{max}}\big(\ket{1_g}\big) &\equiv \max_{\ket{\psi}\text{ strict. loc.}} \abs{\braket{\psi}{1_g}} \label{F_max_1g_G} \\
    &= \max_{\ket{\psi}\text{ strict. loc.}} \abs{\bra{\psi} \int_0^\infty \ibuffi \dd \omega \ibuffb G(\omega) a^\dagger(\omega) \ket{0}}, \nonumber
\end{align}
where $G(\omega)$ is the Fourier transform of $g(t)$. Using the Cauchy-Schwarz inequality, we obtain an upper bound
\begin{equation}\label{F_max_1g_upper}
    F_{\text{max}}\big(\ket{1_g}\big) \leq \lp(\int_{0}^\infty \ibuffi \dd \omega \ibuffb \abs{G(\omega)}^2\rp)^{1/2} = \sqrt{1 - \eta}.
\end{equation}
Similar to in \eqref{F_lower_1order}, it is useful to consider the behavior for small $\eta$, which we find by expanding,
\begin{equation}\label{F_max_1g_upper_1order}
    F_{\text{max}}\big(\ket{1_g}\big) \leq 1 - \eta/2 + \order{\eta^2}.
\end{equation}

\subsection{Lower bound}\label{sec:lower_bound_for_F_max_1g}
A lower bound for \eqref{F_MAX_1G} can be found by using the strictly localized state $\ket{\eta_{1,2}}$ described earlier. Clearly
\begin{equation}\label{F_max_1g_lower_eta12}
    F_{\text{max}}\big(\ket{1_g}\big) \equiv \max_{\ket{\psi}\text{ strict. loc.}} \abs{\braket{\psi}{1_g}} \geq \abs{\braket{\eta_{1,2}}{1_g}}.
\end{equation}
We also see that the weight function $g(t)$ for the state $\ket{1_g}$ is a valid seed function for constructing a localized state $\ket{\eta_{1,2}}$, so we set them equal. Substituting $\ket{\eta_{1,2}}$ in \eqref{F_max_1g_lower_eta12} with its expansion from \eqref{eta12_expansion} leads to
\begin{equation}\label{fid_1g_eta12_F}
    \abs{\braket{1_g}{\eta_{1,2}}} = F \abs{\braket{1_g}{1_1 \kbuff 0_2}},
\end{equation}
where $F$ is given by \eqref{F}. The state $\ket{1_1 \kbuff 0_2}$ is just a single photon in the mode $\xi_1(\omega)$, which is the positive-frequency part of $\widetilde{G}(\omega)$:
\begin{equation}\label{1xi1_G_tilde}
    \ket{1_1 \kbuff 0_2} = \int_0^\infty \ibuffi \dd\omega \ibuffb \widetilde{G}(\omega) a^\dagger(\omega) \ket{0}.
\end{equation}
By using \eqref{G_tilde_omega}, \eqref{beta_I}, and accounting for the required normalization of $\widetilde{G}(\omega)$, we get after some algebra that
\begin{equation}\label{F_max_1g_lower}
    F_{\text{max}}\big(\ket{1_g}\big) \geq \frac{F}{2} \sqrt{\frac{(1 + J)\lp(1 + J - 2 \eta\rp)}{J}}.
\end{equation}

As in \eqref{F_max_1g_upper_1order}, we are interested in examining the regime $\eta \ll 1$. To find this we plug \eqref{J} and \eqref{F_lower_1order} into \eqref{F_max_1g_lower}, use \eqref{I_upper}, and expand to first order in $\eta$, to get
\begin{equation}\label{F_max_1g_lower_1order}
    F_{\text{max}}\big(\ket{1_g}\big) \geq 1 - \lp(2 - \sqrt{2}\rp)\eta + \order{\eta^2}.
\end{equation}

\section{Bounds for \texorpdfstring{$\bm{F_{\text{max}}\big(\ket{1_\xi}\big)}$}{Fmax(|1xi>)}}\label{sec:bounds_for_F_max_1xi}

\subsection{Upper bound}\label{sec:upper_bound_for_F_max_1xi}
Next, we consider the fidelity \eqref{F_MAX_1XI} between a strictly localized state $\ket{\psi}$ and a physical single photon $\ket{1_\xi}$ with some spectrum $\xi(\omega)$, as defined in \eqref{1xi}. The strategy for determining this quantity is to use that the trace distance between two states is related to the probability of distinguishing the states by some measurement:
\begin{align}\label{D_max_P}
    D\big(\ket{\psi}, \ket{1_\xi}\big) &= \max_P \big(\qexv{\psi}{P} - \qexv{1_\xi}{P}\big),
\end{align}
where $P$ is any projector. If we choose a measurement that is local to $t < 0$, the state $\ket{\psi}$ is here indistinguishable from vacuum $\ket{0}$, whereas the state $\ket{1_\xi}$ has some nonzero tail extending to $t \rightarrow - \infty$. Thus for any projector $P$ that is local to $t < 0$, we have
\begin{equation}\label{D_P_0}
    D\big(\ket{\psi}, \ket{1_\xi}\big) \geq \qexv{0}{P} - \qexv{1_\xi}{P}.
\end{equation}

To find a local projector, we begin by choosing a local observable. In accordance with \secref{sec:local_measurements_and_strictly_localized_states}, we select a smeared electric field
\begin{equation}\label{E_zeta}
    E_\zeta = \frac{1}{2 \sqrt{\pi}} \int_{-\infty}^\infty \ibuffn \dd t \ibuffb \zeta(t) E(t),
\end{equation}
for a real smearing function $\zeta(t)$. According to \eqref{Q_G}, the observable $E_\zeta$ is local to the support of $\zeta(t)$, so by setting $\zeta(t) = 0$ for $t \geq 0$, it is then local to $t < 0$. The factor $1/(2\sqrt{\pi})$ is a normalization constant for later convenience. 

Note that there is no normalization requirement for $\zeta(t)$, as we are free to set the scale of our measurement. However, \eqref{D_P_0} depends purely on the projectors of the measurement, which are unaffected by the scale of $E_\zeta$, so for convenience we choose the normalization of $\zeta(t)$ so that
\begin{equation}\label{zeta_normalization}
    \int_0^\infty \ibuffi \dd\omega \ibuffb \abs{\mathcal{E}(\omega)}^2 \abs{\zeta(\omega)}^2 = 1,
\end{equation}
where $\zeta(\omega)$ is the Fourier transform of $\zeta(t)$. Expression \eqref{E_zeta} can be rewritten as
\begin{equation}\label{E_zeta_x}
    E_\zeta = \frac{1}{\sqrt{2}} \lp(a_\zeta + a_\zeta^\dagger\rp),
\end{equation}
with
\begin{equation}\label{a_zeta}
    a_\zeta^\dagger = \int_0^\infty \ibuffi \dd\omega \ibuffb \mathcal{E}^*(\omega) a^\dagger(\omega) \zeta(\omega).
\end{equation}
With the normalization \eqref{zeta_normalization}, we get that $\comm*{a_\zeta}{a_\zeta^\dagger} = 1$, meaning that the smeared field observable $E_\zeta$ has the same form as the position operator $\hat{x}$ in a regular quantum harmonic oscillator. We therefore know that the spectrum of $E_\zeta$ is the real line $\reals$. We also see from \eqref{E_zeta_x} and \eqref{a_zeta} that even though $\zeta(\omega)$ is defined for all $\omega$, it is only its positive frequencies that determine $E_\zeta$.

To find the eigenvectors of $E_\zeta$, we again use the decomposition of the Fock space into a tensor product of spaces \eqref{identity_n} associated with the pulse mode set $\xi_n(\omega)$ in \eqref{xi_n_basis}. This is useful because the smeared field observable $E_\zeta$ clearly operates only on the subspace of $\mathcal{E}^*(\omega) \zeta(\omega)$ (restricted to positive frequencies). We therefore select the first pulse mode
\begin{equation}\label{xi1_omega_zeta}
    \xi_1(\omega) = \mathcal{E}^*(\omega) \zeta(\omega),
\end{equation}
which has the required normalization \eqref{xi_n_basis} because of \eqref{zeta_normalization}. This means that $a_\zeta^\dagger = a_1^\dagger$, where $a_1^\dagger$ is the creation operator on the $\xi_1(\omega)$ mode subspace according to \eqref{a_n}. With this basis choice, we can find the spectral decomposition of the self-adjoint smeared-field observable $E_\zeta$ as
\begin{equation}\label{E_zeta_spectral}
    E_\zeta = \int_{-\infty}^\infty \ibuffn \dd X \ibuffb X P_X,
\end{equation}
where
\begin{equation}\label{P_X}
    P_X = \ket{X_1}\bra{X_1} \otimes \identity_2 \otimes \identity_3 \otimes \dotsb
\end{equation}
is the projector density associated with the eigenvalue $X$ and
\begin{equation}\label{eigen_X}
    \ket{X_1} = \pi^{-1/4} e^{-X^2/2} e^{-a_1^{\dagger^2}/2 + \sqrt{2} X a_1^\dagger} \ket{0_1}
\end{equation}
is the corresponding eigenstate of $E_\zeta$ on the $\xi_1(\omega)$ subspace (we show that this is true in \appref{sec:eigenvectors_of_quantum_fields}). On this subspace, the eigenstates are not degenerate, and they satisfy delta-function normalization as is usual for continuous spectra: $\braket{Y_1}{X_1} = \delta\lp(X - Y\rp)$. 

We then select $\xi_2(\omega)$ as the component of $\xi(\omega)$ orthogonal to $\xi_1(\omega)$ [in the edge case $\xi_1(\omega) = \xi(\omega)$, we let $\xi_2(\omega)$ be arbitrary], also normalized according to \eqref{xi_n_basis}. We can then write
\begin{equation}\label{xi_omega_xi1_xi2}
    \xi(\omega) = c_\xi \xi_1(\omega) + \sqrt{1 - \abs{c_\xi}^2} \xi_2(\omega),
\end{equation}
with
\begin{equation}\label{c_xi}
    c_\xi = \int_0^\infty \ibuffi \dd\omega \ibuffb \xi_1^*(\omega) \xi(\omega).
\end{equation}
This means that for the state $\ket{1_\xi} = a_\xi^\dagger \ket{0}$, we can write
\begin{equation}\label{a_xi_c_xi}
    a_\xi^\dagger = \int_0^\infty \ibuffi \dd \omega \ibuffb a^\dagger(\omega) \xi(\omega) = c_\xi a_1^\dagger + \sqrt{1 - \abs{c_\xi}^2} a_2^\dagger.
\end{equation}

To form a projector $P$ from the projector density $P_X$, we must integrate over an indicator function $1_\chi(X)$,
\begin{equation}\label{P_chi}
    P = \int_{-\infty}^\infty \ibuffn \dd X \ibuffb 1_\chi(X) P_X.
\end{equation}
We now use the fact that the spectral projectors of any local observable are local to the same region as the observable is. To show this, note that by \eqref{W_isometry} the Licht operator $W$ is isometric and therefore bounded. Theorem 13.33 from \cite{rudin1973} then asserts that a bounded operator $W$ commuting with an (unbounded) self-adjoint operator $L$, also commutes with the spectral projectors $P_X$ of $L$ (and $W$ being isometric takes care of the domain condition). Thus $\qexv{\psi}{P} = \qexv{0}{P}$. We can therefore insert \eqref{P_chi} into \eqref{D_P_0} and use \eqref{a_xi_c_xi} to get
\begin{equation}\label{D_chi}
\begin{aligned}
    &D\big(\ket{\psi}, \ket{1_\xi}\big) \\
    &\quad \geq \int_{-\infty}^\infty \ibuffn \dd X \ibuffb 1_\chi(X) \big( \qexv{0}{P_X} - \qexv{1_\xi}{P_X} \big) \\
    &\quad = \frac{\abs{c_\xi}^2}{\sqrt{\pi}} \int_{-\infty}^\infty \ibuffn \dd X \ibuffb 1_\chi(X) e^{-X^2} (1 - 2 X^2).
\end{aligned}
\end{equation}
We are free to maximize over indicator functions $1_\chi(X)$, and selecting $1_\chi(X) = 1$ for $\abs{X} < 1/\sqrt{2}$ and 0 otherwise gives
\begin{equation}\label{D_c_xi}
    D\big(\ket{\psi}, \ket{1_\xi}\big) \geq \sqrt{\frac{2}{\pi e}} \abs{c_\xi}^2.
\end{equation}
Finally, converting to fidelity, we get
\begin{equation}\label{F_max_1xi_upper_c_xi}
    F_{\text{max}}\big(\ket{1_\xi}\big) \equiv \max_{\ket{\psi}\text{ strict. loc.}} \abs{\braket{\psi}{1_\xi}} \leq \sqrt{1 - \frac{2}{\pi e} \abs{c_\xi}^4}.
\end{equation}

The quantity $c_\xi$ is given by the overlap between the spectrum of the target single photon $\xi(\omega)$ and the spectrum of the measurement $\zeta(\omega)$ [weighted by $\mathcal{E}(\omega)$]. To get as good an upper bound \eqref{F_max_1xi_upper_c_xi} as possible, we need to select the real smearing function $\zeta(t)$ so that $\zeta(\omega)$ has a large overlap with $\mathcal{E}(\omega) \xi(\omega)$ while satisfying $\zeta(t) = 0$ for $t \geq 0$ and having normalization \eqref{zeta_normalization}. Letting $\xi(t)$ be the inverse Fourier transform of $\xi(\omega)$, we define
\begin{equation}\label{f_t}
    f(t) = 
    \begin{cases}
        0, \tbuff & t \geq 0, \\
        f_0 e^{i \phi / 2} \xi(t) + \cc, \tbuff & t < 0,
    \end{cases}
\end{equation}
and $F(\omega)$ as its Fourier transform. Here $f_0 > 0$ is a normalization constant and $\phi$ is an arbitrary real phase. One choice for the smearing function is then to let $\zeta(t)$ be the inverse Fourier transform of
\begin{equation}\label{zeta_omega}
    \zeta(\omega) = F(\omega) / \mathcal{E}^*(\omega).
\end{equation}
To verify that this is a valid choice of $\zeta(t)$, first note that $f(t)$ is real, meaning that $\zeta^*(\omega) = \zeta(-\omega^*)$ so that $\zeta(t)$ is also real. Second, since $f(t)$ vanishes for positive times, $F(\omega)$ is analytic in the lower half-plane. The Fourier integral involves only real $\omega$, meaning that we can substitute $1/\mathcal{E}^*(\omega)$ with $1/\mathcal{E}^*(\omega^*) = 1/\mathcal{E}(-\omega)$ in \eqref{zeta_omega}. The latter function is also analytic in the lower half-plane of complex $\omega$ with the conventional branch cut of the complex square root. Therefore $\zeta(\omega)$ is analytic in the lower half-plane, and since it decays sufficiently fast (by a similar argument as in\footnoteref{foot:G_w_A_w}), it follows that $\zeta(t) = 0$ for $t \geq 0$. Finally, we choose the normalization constant $f_0$ so that \eqref{zeta_normalization} is satisfied. 

With the choice \eqref{zeta_omega}, the first basis pulse mode \eqref{xi1_omega_zeta} becomes $\xi_1(\omega) = F(\omega)$. In addition to $\mu$ from \eqref{mu}, we define the complex constant
\begin{equation}\label{nu}
    \nu = \int_{-\infty}^0 \ibuffn \dd t \ibuffb \xi^2(t),
\end{equation}
and write $\nu = \abs{\nu}e^{i \theta_\nu}$. Since $F^*(\omega) = F(-\omega^*)$, we have that
\begin{equation}\label{F_omega_integral}
    \int_0^\infty \ibuffi \dd \omega \ibuffb \abs{F(\omega)}^2 = \frac{1}{2} \int_{-\infty}^\infty \ibuffn \dd \omega \ibuffb \abs{F(\omega)}^2.
\end{equation}
With this relation, some algebra, and using the property that function inner products are preserved under the Fourier transform, we can calculate \eqref{c_xi},
\begin{equation}\label{c_xi_phi}
    \abs{c_\xi}^2 = \frac{\mu^2 + \abs{\nu}^2 + 2 \mu \abs{\nu} \cos(\phi + \theta_\nu)}{\mu + \abs{\nu}\cos(\phi + \theta_\nu)}.
\end{equation}
We are free to maximize with respect to $\phi$. Setting $\cos(\phi + \theta_\nu) = 1$ gives
\begin{equation}\label{c_xi_mu}
    \abs{c_\xi}^2 = \mu + \abs{\nu}. 
\end{equation}
Inserting \eqref{c_xi_mu} into \eqref{F_max_1xi_upper_c_xi}, we finally get
\begin{equation}\label{F_max_1xi_upper}
    F_{\text{max}}\big(\ket{1_\xi}\big) \leq \sqrt{1 - \frac{2}{\pi e}\lp(\mu + \abs{\nu}\rp)^2}.
\end{equation}

Similar to the regime $\eta \ll 1$ for causal single photons, we often have $\mu \ll 1$ for physical single photons in practical situations. We can find an expression for the upper bound \eqref{F_max_1xi_upper} for this case by expanding for small $\mu$ and noting that $\abs{\nu}$ is nonnegative, giving
\begin{equation}\label{F_max_1xi_upper_1order}
    F_{\text{max}}\big(\ket{1_\xi}\big) \leq 1 - \frac{1}{\pi e}\mu^2 + \order{\mu^4}.
\end{equation}

\subsection{Lower bound}\label{sec:lower_bound_for_F_max_1xi}
A lower bound for \eqref{F_MAX_1XI} can be found by providing an example of a state $\ket{\psi}$ that is strictly localized to $t\geq 0$. The state $\ket{\eta_{1,2}}$ is a valid choice, although the single-photon state it approximates, $\ket{1_1 \kbuff 0_2}$, has a particularly chosen spectrum $\xi_1(\omega)$, namely the positive-frequency part of some $\widetilde{G}(\omega)$ coming from \eqref{G_tilde_omega}. We can however use $\ket{\eta_{1,2}}$ and try to find a suitable spectrum $\xi_1(\omega)$ that has a large overlap with $\xi(\omega)$, while having the required properties. From the expanded form of $\ket{\eta_{1,2}}$ \eqref{eta12_expansion}, we can then find the fidelity between $\ket{1_\xi}$ and $\ket{\eta_{1,2}}$ as
\begin{equation}\label{F_max_1xi_lower_1xi1}
\begin{aligned}
    F_{\text{max}}\big(\ket{1_\xi}\big) &\equiv \max_{\ket{\psi}\text{ strict. loc.}} \abs{\braket{\psi}{1_\xi}} \\
    &\geq \abs{\braket{1_\xi}{\eta_{1,2}}} = F\abs{\braket{1_\xi}{1_1 \kbuff 0_2}},
\end{aligned}
\end{equation}
where $F$ is given by \eqref{F}. The quality of the approximation is largely dependent on the procedure for selecting the spectrum $\xi_1(\omega)$. The following method is probably not optimal, but it provides a lower bound.

Start with taking the inverse Fourier transform of $\xi(\omega)$ to obtain $\xi(t)$. Separate the positive and negative times into two functions
\begin{subequations}\label{hp_hm}
\begin{align}
    h_{+}(t) &=
    \begin{cases}
        \xi(t), \tbuff & t \geq 0, \\
        0, \tbuff & t < 0,
    \end{cases} \label{g_t_xi}\\
    h_{-}(t) &=
    \begin{cases}
        0, \tbuff & t \geq 0, \\
        \xi(t), \tbuff & t < 0.
    \end{cases}
\end{align}
\end{subequations}
Label their Fourier transforms by $H_{+}(\omega)$ and $H_{-}(\omega)$, respectively. We have
\begin{subequations}
\begin{align}
    \xi(t) &= h_{+}(t) + h_{-}(t), \\
    \xi(\omega) &= H_{+}(\omega) + H_{-}(\omega). \label{xi_omega_G_H}
\end{align}
\end{subequations}
By normalizing,
\begin{equation}\label{g_t_hp}
    g(t) = \frac{1}{\sqrt{1 - \mu}} h_{+}(t)
\end{equation}
is then clearly a valid seed function for generating $\ket{\eta_{1,2}}$. From \eqref{G_tilde_omega} and \eqref{beta_I}, accounting for the required normalization of $\widetilde{G}(\omega)$, and using that function inner products are preserved under Fourier transforms, we get after some algebra that
\begin{equation}\label{F_max_1xi_lower}
    F_{\text{max}}\big(\ket{1_\xi}\big) \geq F \sqrt{1 - \mu} \sqrt{\frac{J \lp(1 + J\rp)}{1 + J - 2\eta}}.
\end{equation}

In order to expand \eqref{F_max_1xi_lower} for $\mu \ll 1$, we must first find an upper bound for $\eta$ in terms of $\mu$. It follows from \eqref{xi_omega_G_H} that $\int_{-\infty}^0 \ibuffn \dd \omega \ibuffb \abs{H_{+}(\omega)}^2 = \int_{-\infty}^0 \ibuffn \dd \omega \ibuffb \abs{H_{-}(\omega)}^2$. Then Fourier transforming \eqref{g_t_hp} and using \eqref{mu}, \eqref{eta}, and the Plancherel theorem gives
\begin{equation}\label{eta_mu}
    \eta = \frac{1}{1 - \mu} \int_{-\infty}^0 \ibuffn \dd \omega \ibuffb \abs{H_{+}(\omega)}^2 < \frac{\mu}{1 - \mu} = \mu + \order{\mu^2}.
\end{equation}
In other words, $\mu \ll 1$ implies $\eta \ll 1$. An expansion of \eqref{F_max_1xi_lower} for small $\mu$, similar to that leading to \eqref{F_max_1xi_upper_1order}, finally gives
\begin{equation}\label{F_max_1xi_lower_1order}
    F_{\text{max}}\lp( \ket{1_\xi} \rp) \geq 1 - \mu + \order{\mu^2}.
\end{equation}

\section{Numerical examples}\label{sec:numerical_examples}

\figtwo

\figthree

We now use a numerical routine to evaluate the bounds for $F_{\text{max}}\lp( \ket{1_g} \rp)$ and $F_{\text{max}}\lp( \ket{1_\xi} \rp)$ for some specific examples of target states. Starting with $\ket{1_g}$, one possible example for the pulse form $g(t)$ is a Gaussian envelope around a carrier frequency $\omega_0$. Since $g(t)$ must be zero for negative times, we must truncate the Gaussian:
\begin{equation}\label{g_t_gauss_pos_time}
    g(t) \propto \theta(t) e^{- (t - \tau)^2 / 2\sigma^2} e^{- i\omega_0 t},
\end{equation}
where $\sigma$ is the pulse width, $\tau$ is the pulse delay, and $\theta(t)$ is the Heaviside function. The delay $\tau$ controls the amount of truncation at $t = 0$. Using a numerical Fourier transform routine, we perform the steps of the algorithm for constructing $\ket{\eta_{1,2}}$ and determine numerical values of the upper and lower bounds \eqref{F_max_1g_upper} and \eqref{F_max_1g_lower} for various choices of the seed-function parameters. 

In \figref{fig:F_max_bounds_1g_vs_tau} we keep the pulse width $\sigma$ fixed and plot the bounds for $F_{\text{max}}\lp(\ket{1_g}\rp)$ as a function of delay $\tau$. To understand the plot behavior, consider the bound approximations \eqref{F_max_1g_upper_1order} and \eqref{F_max_1g_lower_1order}, which are expressed purely as a function of the amount of negative frequencies $\eta$ in the Fourier transform $G(\omega)$ of $g(t)$. For $\tau = 0$, both the clipping at $t = 0$ and the width of the pulse contribute to the negative-frequency content, and a narrow pulse has more negative frequencies, giving a low fidelity. Increasing $\tau$ decreases the truncation at $t = 0$, giving less negative frequencies and thus higher fidelity. This effect is slower for large $\sigma$ since wider pulses are effectively moved less by the same delay. At some point the fidelity saturates when the main contribution to the negative frequencies comes from the pulse width. Increasing $\tau$ further after this point has no more effect, and the fidelity remains roughly constant at a value that increases with $\sigma$.

In \figref{fig:F_max_bounds_1g_vs_sigma} we plot the bounds for $F_{\text{max}}\lp(\ket{1_g}\rp)$ now as a function of the pulse width $\sigma$. Instead of keeping the delay $\tau$ fixed, we instead let it be proportional to $\sigma$ so that the pulses have a truncation fixed at some percentage of their width. For $\sigma \rightarrow 0$, the pulse is narrow and thus has a large negative-frequency content and low fidelity. Increasing $\sigma$ gives less negative frequencies and a higher fidelity, up to a point where the negative frequencies come mainly from the truncation at $t = 0$. The fidelity at this saturation point is higher for larger $\tau$ since there is less truncation. Increasing $\sigma$ further after that still improves the fidelity, although only very slowly since the $\sim 1 / \omega$ tail in $G(\omega)$ introduced by the truncation at $t = 0$ has a slow fall off. 

In both \figref{fig:F_max_bounds_1g_vs_tau} and \figref{fig:F_max_bounds_1g_vs_sigma}, the upper and lower bounds are quite close together, meaning that the optimal fidelity is constrained tightly. One contributing factor to this is the artificial negative-frequency modes in the construction of $\ket{1_g}$. They allow us to specify an arbitrary causal pulse in time $g(t)$, but removing these artificial modes to achieve a physical state results in a substantial, unavoidable factor $\sqrt{1 - \eta}$ [in \eqref{F_max_1g_upper}] in the fidelity. This factor gives the main contribution in both the upper and lower bound for $F_{\text{max}}\lp(\ket{1_g}\rp)$, making the bounds relatively close. 

Note also that $F_{\text{max}}\big(\ket{1_g}\big)$ being less than 1 is not just a consequence of the truncation at $t = 0$. To show this explicitly, we can consider an experiment where only the shape, and not the timing, of the target pulse matters, meaning that we can delay the target pulse infinitely. This means that for, e.g., the Gaussian pulse \eqref{g_t_gauss_pos_time}, we may take $\omega_0 \tau \rightarrow \infty$. The truncation at $t = 0$ is then redundant, and we can drop the Heaviside factor $\theta(t)$, so that $g(t)$ is a pure Gaussian:
\begin{equation}\label{g_t_gauss_pos_time_inf}
    g(t) \propto e^{- (t - \tau)^2 / 2\sigma^2} e^{- i\omega_0 t}, \tbuff \omega_0 \tau \rightarrow \infty.
\end{equation}
In this case we can calculate the parameter $\eta$ analytically,
\begin{equation}\label{eta_inf}
    \eta = \frac{1}{2}\lp[1 - \mathrm{erf}(\omega_0 \sigma)\rp],
\end{equation}
where $\mathrm{erf}(\cdot)$ is the error function. Using this we can calculate the first-order approximations for the upper and lower bounds for $F_{\text{max}}\big(\ket{1_g}\big)$ as given by \eqref{F_max_1g_upper_1order} and \eqref{F_max_1g_lower_1order}. The bounds are indistinguishable from the case $\tau = 3 \sigma$ in \figref{fig:F_max_bounds_1g_vs_sigma} until $\omega_0 \sigma \approx 3$, after which they continue downwards instead of flattening out. Thus for $0 \leq \omega_0 \sigma \lessapprox 3$, the plot for $\tau = 3 \sigma$ in \figref{fig:F_max_bounds_1g_vs_sigma} also represents the fidelity bounds for a Gaussian target pulse without truncation \eqref{g_t_gauss_pos_time_inf}. 

Next, we consider the upper and lower bounds \eqref{F_max_1xi_upper} and \eqref{F_max_1xi_lower} of $F_{\text{max}}\lp(\ket{1_\xi}\rp)$. We would like to specify the pulse of the target state $\ket{1_\xi}$ in time domain similar to \eqref{g_t_gauss_pos_time}. However, we must choose a spectrum $\xi(\omega)$ defined on $\omega > 0$, meaning that not every time-domain pulse is possible. This fundamental issue was also the main motivation for constructing the target state $\ket{1_g}$, which is unphysical but more convenient in this regard. We construct the spectrum $\xi(\omega)$ in two steps: First choose some time-domain Gaussian similar to before,
\begin{equation}\label{g_pre_t}
    g_{\text{pre}}(t) \propto e^{- (t - \tau_{\text{pre}})^2 / 2 \sigma_{\text{pre}}^2} e^{- i \omega_{0}^{\text{pre}} t},
\end{equation}
where $\omega_{0}^{\text{pre}}$, $\sigma_{\text{pre}}$, and $\tau_{\text{pre}}$ are free parameters as in \eqref{g_t_gauss_pos_time}. Note that there is no requirement of truncating for negative times. We then choose $\xi(\omega)$ as the (normalized) positive-frequency part of the Fourier transform $G_{\text{pre}}(\omega)$ of $g_{\text{pre}}(t)$,
\begin{equation}\label{xi_omega_G_pre}
    \xi(\omega) = \frac{G_{\text{pre}}(\omega)}{\sqrt{ \int_0^\infty \ibuffi \dd \omega \ibuffb \lp| G_{\text{pre}}(\omega) \rp|^2 }}, \tbuff \omega > 0,
\end{equation}
and label its inverse Fourier transform $\xi(t)$.

\figfour

When $G_{\text{pre}}(\omega)$ has its main weight for positive frequencies, the target pulse $\xi(t)$ is close to $g_{\text{pre}}(t)$ and it therefore has parameters close to those chosen in \eqref{g_pre_t}. In general however, $\xi(t)$ will be significantly different from $g_{\text{pre}}(t)$, meaning that its carrier frequency, width, and delay must be computed. Picking a procedure for how to do this will always involve some amount of choice, especially since $\xi(t)$ may be significantly different from a Gaussian in some cases. We use the method:
\begin{subequations}\label{w0_eff_tau_eff}
\begin{align}
    \omega_0 &= \int_0^\infty \ibuffi \dd \omega \ibuffb \omega \abs{\xi(\omega)}^2, \\
    \tau &= \int_0^\infty \ibuffi \dd t \ibuffb t \abs{\xi(t)}^2.
\end{align}
\end{subequations}
Finally, $\sigma$ is chosen so that the square of a Gaussian with standard deviation $\sigma$ has a width at 5\% of its peak equal to the width at 5\% of the peak of $\abs{\xi(t)}^2$.

The behavior of this method is checked in \figref{fig:w0_eff_sigma_eff_vs_w0_sigma}. When $g_{\text{pre}}(t)$ has a wide pulse form, the width of $\xi(t)$ is about the same. As the pulse is made narrower, the mean frequency of $\xi(\omega)$ is pushed up compared to in $G_{\text{pre}}(\omega)$, meaning that the width of $\xi(t)$ relative to its carrier decreases slower and finally reaches a constant value of $\approx 1.3$. Achieving a pulse with only positive frequencies that is narrower than that is impossible. 

In \figref{fig:F_max_bounds_1xi_vs_tau} we keep the pulse width $\sigma$ fixed and plot the bounds for $F_{\text{max}}\lp(\ket{1_\xi}\rp)$ as a function of delay $\tau$. Again we can use the first-order bound approximations \eqref{F_max_1xi_upper_1order} and \eqref{F_max_1xi_lower_1order} to understand the plot, since they depend only on the negative-time tail $\mu$ of the target pulse $\xi(t)$. When $\tau = 0$, half of $g_{\text{pre}}(t)$ is located for negative $t$, and therefore $\xi(t)$ also has a significant portion there which gives a large negative-time tail and low fidelity. As $\tau$ increases, the tail is reduced, increasing the fidelity. At some point the fidelity saturates when the Gaussian tail of $g_{\text{pre}}(t)$ is insignificant, and it is instead the $\sim 1/t$ tail introduced by the truncation at $\omega = 0$ in $\xi(\omega)$ that dominates. This tail is smaller for larger values of $\sigma$ since wide pulses have less negative frequencies. Increasing $\tau$ further after this point still improves the fidelity, but only very slowly since the $1/t$ tail is so slowly decreasing. 

\figfive

In \figref{fig:F_max_bounds_1xi_vs_sigma} we plot $F_{\text{max}}\lp(\ket{1_\xi}\rp)$ as a function of pulse width $\sigma$. Similar to before, we keep $\tau$ at a constant proportionality with $\sigma$. As $\sigma_{\text{pre}} \rightarrow 0$, $\sigma$ reaches the minimum, nonzero possible width of a pulse with only positive frequencies, as seen in \figref{fig:w0_eff_sigma_eff_vs_w0_sigma}. At this point the negative-time tail comes mainly from the severe truncation at $\omega = 0$, and the fidelity is low. As $\sigma$ is increased, there is less frequency truncation, meaning that the fidelity increases and $\xi(t)$ approaches $g_{\text{pre}}(t)$. At some point the truncation for $\omega = 0$ becomes insignificant compared to the tail of $g_{\text{pre}}(t)$ itself, and the fidelity saturates depending on the delay $\tau$.

\figsix

\section{Fidelity bounds for arbitrary number states}\label{sec:fidelity_bounds_for_arbitrary_number_states}
So far we have considered the maximum fidelity between states strictly localized to $t \geq 0$ and single photons, either in the form $\ket{1_\xi}$ or $\ket{1_g}$. Here we consider the generalization of these bounds to states of arbitrary photon number. Concretely, we wish to find upper and lower bounds for the maximum fidelity
\begin{equation}\label{F_max_n_xi}
    F_{\text{max}}\big(\ket{n_\xi}\big) \equiv \max_{\ket{\psi}\text{ strict. loc.}} \abs{\braket{\psi}{n_\xi}},
\end{equation}
between any state $\ket{\psi}$ strictly localized to $t \geq 0$ and a physical $n$-photon state
\begin{equation}\label{n_xi}
    \ket{n_\xi} = \frac{1}{\sqrt{n!}} {a_\xi^\dagger}^n \ket{0}
\end{equation}
in some spectrum $\xi(\omega)$,
\begin{equation}\label{ad_xi}
    a_\xi^\dagger = \int_0^\infty \ibuffi \dd \omega \ibuffb \xi(\omega) a^\dagger(\omega), \tbuff \int_0^\infty \ibuffi \dd \omega \ibuffb \abs{\xi(\omega)}^2 = 1,
\end{equation}
with $\xi(\omega) = 0$ for $\omega < 0$. Similarly, we wish to find upper and lower bounds for the maximum fidelity
\begin{equation}\label{F_max_n_g}
    F_{\text{max}}\big(\ket{n_g}\big) \equiv \max_{\ket{\psi}\text{ strict. loc.}} \abs{\braket{\psi}{n_g}},
\end{equation}
between any state $\ket{\psi}$ strictly localized to $t \geq 0$ and a causal $n$-photon state
\begin{equation}\label{n_g}
    \ket{n_g} = \frac{1}{\sqrt{n!}} {a_g^\dagger}^n \ket{0}
\end{equation}
in some pulse form $g(t)$,
\begin{equation}\label{ad_g}
    a_g^\dagger = \int_0^\infty \ibuffi \dd t \ibuffb g(t) a^\dagger(t), \tbuff \int_0^\infty \ibuffi \dd t \ibuffb \abs{g(t)}^2 = 1,
\end{equation}
with $g(t) = 0$ for $t < 0$. 

Similar to $\ket{\eta_{1,2}}$, we also construct an example of a strictly localized state $\ket{\eta^n_{1,2}}$ that is close to an $n$-photon state, which we obtain by iterated application of Licht operators \eqref{W_eta12}:
\begin{align}
    \ket{\eta_{1,2}^n} &= W^n \ket{0} = S^\dagger {\widetilde{a}_1^\dagger{}}^n S \ket{0} \label{eta12n} \\
    &= c_1 \ket{n_1 \kbuff 0_2} + c_2 \ket{{n+1}_1 \kbuff 1_2} + c_3 \ket{{n+2}_1 \kbuff 2_2} + \dotsb, \nonumber
\end{align}
for some coefficients $c_1, c_2, \dotsc$ [different from those in \eqref{eta12_expansion}]. In \secref{sec:algorithm_for_eta12} we showed that the Licht operator $W$ satisfies $W^\dagger W = \identity$ and $\comm{W}{A(t)} = 0$ for $t < 0$. It then follows immediately that $W^n$ is another Licht operator since it satisfies the same conditions, and thus $\ket{\eta_{1,2}^n}$ is a state strictly localized to $t \geq 0$. 

We can calculate the fidelity $F_n$ between this state and its $n$-photon component in the same way as was done in \eqref{F}:
\begin{equation}\label{Fn}
\begin{aligned}
    F_n &\equiv \abs{\braket{n_1 \kbuff 0_2}{\eta_{1,2}^n}} \\
    &= \lp(\frac{1 - 2 \widetilde{\eta}}{1 - \widetilde{\eta}} \rp)^{1 + \frac{n}{2}} \sum_{k=0}^\infty \lp(\frac{\widetilde{\eta}}{1 - \widetilde{\eta}} \rp)^k \sqrt{\binom{n+k}{n}},
\end{aligned}
\end{equation}
and as before it is useful to obtain a lower bound for $F_n$ in terms of $\eta$ for the regime $\eta \ll 1$,
\begin{equation}\label{Fn_lower_1order}
    F_n \geq 1 - \lp(1 + \frac{n}{2} - \sqrt{n + 1} \rp)\eta + \order{\eta^2}.
\end{equation}

The calculations for the upper and lower bounds for the fidelities follow the same lines as for single photons. We make repeated use of that for any operators $a$ and $b$ such that $a \ket{0} = 0$ and $\comm{a}{b} = c$ for some constant $c$, we have that
\begin{equation}\label{abc}
    \bra{0} a^n {b^\dagger}^n \ket{0} = n! c^n.
\end{equation}

In the calculation for the upper bound for \eqref{F_max_n_xi}, we encounter an expression for the probability density of obtaining result $X$ when measuring the smeared field $E_\zeta$ \eqref{E_zeta} for the state $\ket{n_\xi}$:
\begin{equation}\label{n_xi_P_X}
    \qexv{n_\xi}{P_X} = \sum_{k = 0}^n \binom{n}{k} \abs{c_\xi}^{2k} \lp(1 - \abs{c_\xi}^2 \rp)^{n - k} \psi_k^2(X),
\end{equation}
where $\psi_k(X)$ is the $k$-th Hermite function
\begin{equation}\label{psi_k_X}
    \psi_k(X) = \lp(\sqrt{\pi} 2^k k!\rp)^{-1/2} e^{-X^2/2} H_k(X)
\end{equation}
and $H_k(X)$ is the $k$-th (physicist) Hermite polynomial. To continue with this expression, we must find an upper bound for the integral of $\psi^2_k(X)$ over some interval. Using the same technique as in \cite{indritz1961}, we can show that the following bound holds for the Hermite functions: Let $L \geq 0$ and $k$ be a positive integer, so that $\psi_k(X)$ has a local maximum at $X_0 \geq L$. Then $\abs{\psi_l(X)} \leq \psi_k(X_0)$ for all $X \in \lp[-L, L \rp]$ and $l \geq k$. 

For simplicity we pick the same projector as for $n = 1$,
\begin{equation}\label{P_P_X}
    P = \int_{-1/\sqrt{2}}^{1/\sqrt{2}} \hspace{0em} \dd X \ibuffb P_X,
\end{equation}
and we then want to show that
\begin{equation}\label{psi_k_X_bound}
    \int_{-1/\sqrt{2}}^{1/\sqrt{2}} \hspace{0em} \dd X \ibuffb \psi_k^2(X) \leq \mathrm{erf}\lp(\frac{1}{\sqrt{2}} \rp) - \sqrt{\frac{2}{\pi e}}, \tbuff \forall k \geq 1.
\end{equation}
This can be done by first a brute-force calculation for $k = 1, \dotsc, 14$, and then noticing that there is a local maximum $\psi_{15}(X_0) < 0.35$ at $X_0 \approx 0.85 \geq 1 / \sqrt{2}$. Using the above result with $L = 1 / \sqrt{2}$, we can then verify \eqref{psi_k_X_bound} for all $k \geq 15$. 

In the end we obtain the following upper and lower bounds for \eqref{F_max_n_xi} and \eqref{F_max_n_g}:
\begin{equation}\label{F_max_n_xi_bounds}
\begin{aligned}
    1 - n \mu &\lessapprox F_n \lp(1 - \mu\rp)^{n/2} \lp[\frac{J (1 + J)}{1 + J - 2 \eta}\rp]^{n/2} \leq F_{\text{max}}\big(\ket{n_\xi}\big) \\
    &\leq \sqrt{1 - \frac{2}{\pi e} \lp(1 - \lp(1 - \mu - \abs{\nu} \rp)^n \rp)^2} \lessapprox 1 - \frac{n^2}{\pi e} \mu^2
\end{aligned}
\end{equation}
and
\begin{align}
    &1 - \lp(n + 1 - \sqrt{n + 1} \rp) \eta \lessapprox F_n \lp[ \frac{\lp(1 + J \rp)\lp(1 + J - 2\eta \rp)}{4J} \rp]^{n/2} \nonumber \\
    &\qquad \leq F_{\text{max}}\big(\ket{n_g}\big) \leq \lp(1 - \eta \rp)^{n/2} \approx 1 - \frac{n}{2} \eta, \label{F_max_n_g_bounds}
\end{align}
where the approximations are valid for small $\mu$ and small $\eta$, respectively.

\section{Discussion and conclusion}\label{sec:discussion_and_conclusion}
We have considered the question of how close a state $\ket{\psi}$ produced by an on-demand, 1D, photonic source can be to a single photon, or to an $n$-photon state. By causality, the state $\ket{\psi}$ generated on demand must be strictly localized to $t \geq 0$ at the observation point $x = 0$. We argue that there are two natural but incompatible ways to specify the target photon state. The most obvious is a photon $\ket{1_\xi}$ with a given positive-frequency spectrum $\xi(\omega)$ as defined in \eqref{1xi}. On the other hand, as discussed in the introduction, sometimes a better representation can be an (unphysical) photon $\ket{1_g}$ in a given positive-time pulse $g(t)$ as defined in \eqref{1g}. 

We answer the question by constraining the maximum possible fidelity between $\ket{\psi}$ and the target states $\ket{1_\xi}$ and $\ket{1_g}$. We also find it convenient to obtain first-order approximations for the fidelity bounds expressed purely as a function of the negative-time tail $\mu$ of $\ket{1_\xi}$, and of the negative-frequency tail $\eta$ of $\ket{1_g}$. The results are that the maximum fidelity between any state $\ket{\psi}$ strictly localized to $t \geq 0$ and a physical (acausal) single photon $\ket{1_\xi}$ satisfies
\begin{equation}\label{F_max_1xi_bounds}
\begin{aligned}
    1 - \mu &\lessapprox F \sqrt{1 - \mu} \sqrt{\frac{J \lp(1 + J\rp)}{1 + J - 2\eta}} \\
    &\leq F_{\text{max}}\big(\ket{1_\xi}\big) \\ 
    &\leq \sqrt{1 - \frac{2}{\pi e}\lp(\mu + \abs{\nu}\rp)^2} \lessapprox 1 - 0.12 \mu^2.
\end{aligned}
\end{equation}
Here $\mu$ is given by \eqref{mu}, $\eta$ by \eqref{eta_mu}, $F$ by \eqref{F}, $J$ by \eqref{J}, and $\nu$ by \eqref{nu}. The approximations are valid for the regime $\mu \ll 1$. On the other hand, the maximum fidelity between any state $\ket{\psi}$ strictly localized to $t \geq 0$ and a causal (unphysical) single photon $\ket{1_g}$ satisfies
\begin{equation}\label{F_max_1g_bounds}
\begin{aligned}
    1 - 0.59 \eta &\lessapprox \frac{F}{2} \sqrt{\frac{(1 + J) \lp(1 + J - 2 \eta\rp)}{J}} \\
    &\leq F_{\text{max}}\big(\ket{1_g}\big) \\
    &\leq \sqrt{1 - \eta} \approx 1 - 0.5 \eta,
\end{aligned}
\end{equation}
where $\eta$ is given by \eqref{eta}. The approximations are valid for $\eta \ll 1$. The generalizations of these bounds to arbitrary number states are given in \eqref{F_max_n_xi_bounds} and \eqref{F_max_n_g_bounds}, respectively. 

The fidelity \eqref{F_max_1xi_bounds} is limited by the size of the negative-time tail $\mu$ associated with the target state's spectrum $\xi(\omega)$. Thus it can always be improved by delaying the target state in time, corresponding to a linear phase factor in $\xi(\omega)$, as seen by the forever-decreasing curves in \figref{fig:F_max_bounds_1xi_vs_tau}. Despite this improvement, it is important to note that there is already a limitation inherent in the requirement of only positive frequencies in the target state; a spectrum $\xi(\omega)$ for $\omega > 0$ can never accurately describe, for example, an ultrashort, few-cycle pulse, even if it is infinitely delayed in time. 

The fidelity \eqref{F_max_1g_bounds} is related to the target pulse's negative-frequency content $\eta$. The bounds are severe for ultrashort pulses of the order of a few cycles. Unlike \eqref{F_max_1xi_bounds}, the fidelity \eqref{F_max_1g_bounds} is at some point not improved by delaying the pulse more, because few-cycle pulses necessarily contain a significant amount of negative frequencies regardless of the delay. This is discussed around \eqref{g_t_gauss_pos_time_inf} and seen in \figref{fig:F_max_bounds_1g_vs_tau}, where the curves flatten out. Nevertheless, the fidelity tends to 1 rapidly as the pulse envelope becomes slowly varying over an optical cycle.

Note that we have so far assumed that the photonic state produced by the source $\ket{\psi}$ is pure. However, the source can very well entangle electromagnetic and internal degrees of freedom, making the reduced photonic state mixed. Interestingly, the 4 bounds derived in this paper (including the generalizations to arbitrary number states) all apply also when the maximizations in \eqref{F_MAX_1XI} and \eqref{F_MAX_1G} are over all strictly localized mixed states $\rho$ instead. The definition of strict localization for mixed states is the straight-forward generalization of \eqref{strictly_localized_state}. The validity of the 2 lower bounds to mixed states follows trivially since they are proven by example. The derivation of the upper bound for \eqref{F_MAX_1G} is easily generalized since it relies only on the source state being physical, which applies also to the pure states of an ensemble expansion of $\rho$ (even if these states may fail to be strictly localized). Finally, the derivation of the upper bound of \eqref{F_MAX_1XI} uses an operator local to a region complementary to the localization region of the source state, which works equally well for mixed states. 

Also note that the regime in which the effects discussed in this work become appreciable is quite far away from current technology. For typical on-demand single-photon sources today \cite{wang2019,scheel2009,eisaman2011,senellart2017,sinha2019}, the bandwidth is orders of magnitude smaller than the carrier frequency \cite{eisaman2011}, making the effect of truncating for $\omega < 0$ vanishingly small. Thus it seems plausible that we might with high accuracy replace the true state produced by such sources with a single photon. This is true even though there have been demonstrations of pulsed lasers with pulse lengths comparable to a single cycle \cite{sansone2006}; it remains to achieve similarly short pulses for sources of (near) single photons, which is much more difficult. 

Yet these are purely technological limitations that will surely improve over time. There are already suggestions for how one might create single-photon sources with a pulse length on the order of a single cycle \cite{su2016}. For on-demand sources in this regime, the theoretical maximum single-photon fidelity will be significantly less than one, as shown by the plots in \secref{sec:numerical_examples}. Given the importance of single-photon sources for quantum information and communication \cite{eisaman2011,scheel2009}, we can expect the results presented here to be relevant for describing potentials and limitations of future quantum technologies. 

Additionally, it is not a priori given that replacing the true source state with a single photon is actually a valid approximation, even though the amount of negative frequencies is very low. The states $\ket{\eta_{1,2}}$ provide a specific mechanism for justifying and analyzing this approximation. The fidelity bounds \eqref{F_max_1xi_bounds} and \eqref{F_max_1g_bounds} give an exact range for which regime such an approximation can be warranted, and for which it cannot. The states $\ket{\eta_{1,2}}$ also allow a manifestly causal description of propagating signals in quantum field theory, potentially opening up new methods for analyzing such processes. 

The main limitation of our analysis is the assumption of a free theory. We let the source produce some state and then be switched off, assuming that the field is subsequently free of any interactions, which is of course an unphysical idealization. It would be interesting to generalize our analysis to a full, interacting theory including the effects of renormalization. This work is also limited to analyzing photon localization along one dimension. Some of the results are generalized to 3 dimensions in \cite{ryen2022}.

For future work, it would also be interesting to further explore connections between our results and measurement theories for quantum fields in curved spacetime. Specifically, there is some similarity between the mixing of annihilation and creation operators in \eqref{S_a1} and the Bogoliubov transformations relating operators associated with modes of global spacetime to modes of bounded regions \cite{bruschi2010,su2016a}. Another possible direction could be to check how the states $\ket{\eta_{1,2}}$ look in other measurement models, such as Unruh-DeWitt detectors, and see whether they are still strictly localized. A full generalization of our analysis to curved spacetime would also be interesting, for instance investigating the limitations imposed by energy positivity when there is no longer an identification of positive-frequency modes \cite{birrell1984}. 

Looking back at the discussion of the Fermi problem and causality in quantum field theory in \secref{sec:the_fermi_problem_and_causality_in_quantum_field_theory}, we hope our results might bridge some gaps in the understanding. There is an apparent disconnect between experimental experience of atoms emitting single photons \cite{scheel2009}, the theorems showing that single photons are infinitely delocalized \cite{knight1961}, and the abstract analyses showing that causality is manifest but revealing little about the actual quantum states \cite{buchholz1994,milonni1995}. Our results show that a possible resolution is that the emitted states can be strictly localized and propagate causally, while for quasi-monochromatic pulses being extremely close to single photons. Indeed, perhaps it is a general feature that real particles in quantum field theory are not exact single-particle states?


\section*{\supplementary{}}
\appendix

\section{Proof that single photons in 1D\\cannot be localized}\label{sec:proof_that_single_photons_in_1d_cannot_be_localized}
In the introduction we indicate why single photons traveling to the right cannot be localized in space or time, using an argument that neglects the leftward-moving modes ($k < 0$). If we include leftward-moving modes $k < 0$, we get an apparent possibility of localization of some observables at a single point in time. For example, the electric energy density of a single photon at $t = 0$ can be nonzero in a finite spatial interval. However, this localization is only apparent, and the photon becomes infinitely spread-out in space instantaneously for any $t > 0$. Here we show these claims in detail. For a proof that single photons in general (in 3D) cannot be localized, see \cite{knight1961}.

Let $\ket{1}$ be an arbitrary single-photon state containing any 1D wavevectors $k$ (both rightward- and leftward-moving modes):
\begin{equation}\label{1_photon_k}
    \ket{1} = \int_{-\infty}^{\infty} \ibuffn \dd k \ibuffb G(k) a^\dagger(k) \ket{0},
\end{equation}
where $G(k)$ is an arbitrary function. The electric field at position $x$ and $t$ is given by
\begin{equation}\label{E_x_t_neg_k}
    E(x, t) = \int_{-\infty}^{\infty} \ibuffn \dd k \ibuffb \mathcal{E}(\omega) a(k) e^{i k x - i \omega t} + \hc
\end{equation}
Note that unlike in \eqref{E_t}, we are here including both positive and negative $k$. Consider the expectation value of, e.g., the normal-ordered observable $\norder{E^2(x,t)}$,
\begin{equation}\label{exv_1_photon_nE2}
    \qexv{1}{\norder{E^2(x,t)}} = 2 \abs{s(x,t)}^2,
\end{equation}
where
\begin{equation}\label{1_photon_nE2_s}
\begin{aligned}
    s(x, t) &= \int_{-\infty}^\infty \ibuffn \dd k \ibuffb \mathcal{E}(\omega) G(k) e^{ikx - i \omega t} \\
    &= \int_{0}^\infty \ibuffi \dd k \ibuffb \mathcal{E}(kc) G(k) e^{ikx - ikct} \\
    &+ \int_{0}^\infty \ibuffi \dd k \ibuffb \mathcal{E}(kc) G(-k) e^{-ikx - ikct} \\
    &= u(x - ct) + v(x + ct),
\end{aligned}
\end{equation}
for some functions $u(\cdot)$ and $v(\cdot)$. 

For a fixed time, say $t = 0$, we can choose a function $G(k)$ in the first integral in \eqref{1_photon_nE2_s} such that the inverse Fourier transform of $\mathcal{E}(\abs{k}c) G(k)$ vanishes for any desired spatial region. Thus a single photon may appear to be localized instantaneously, e.g., so that $s(x, 0) = 0$ for $\abs{x} > l/2$, where $l$ is a localization width. 

However, this localization disappears instantaneously since it is impossible to have $s(x, t) = 0$ for an interval in space and time. Indeed, in the interior of the region where $s(x, t)$ is zero, we can differentiate the last line in \eqref{1_photon_nE2_s} wrt. $x$ and $t$ separately. Assuming for simplicity that the time interval contains $t = 0$, we get that $u'(x)$ and $v'(x)$ must both vanish in the space interval. Yet we see from \eqref{1_photon_nE2_s} that $u'(x)$ and $v'(x)$ are (inverse) Fourier transforms of only positive frequencies.

Fourier transforms over only positive arguments are limited by the Paley-Wiener criterion (Theorem XII in \cite{paley1934}; see also \cite{bialynicki-birula1998}): Let $F(\omega)$ be in a nonzero function in $L^2(0, \infty)$, meaning that it vanishes for $\omega < 0$. Then its Fourier transform $f(t)$ satisfies
\begin{equation}\label{paley_wiener}
    \int_{-\infty}^\infty \ibuffn dt \ibuffb \frac{\abs{\log\abs{f(t)}}}{1 + t^2} < \infty.
\end{equation}
I.e., $f(t)$ is nonzero (almost) everywhere on $\reals$ and has an asymptotic fall off that is slower than $e^{- A t}$, for some constant $A > 0$. For example, it may have an asymptotic fall off of $e^{- A t^\gamma}$ for $\gamma < 1$. 

Since $u'(x)$ and $v'(x)$ in \eqref{1_photon_nE2_s} are (inverse) Fourier transforms of only positive frequencies, they cannot vanish in any finite space interval as is required. Thus 1D single photons cannot be localized to any interval in space and time. This includes for instance the light cone region $x < ct$ dictated by causality for the source analyzed in the main text.

\section{Localization of classical fields and coherent states}\label{sec:localization_of_classical_fields_and_coherent_states}
We have seen in \appref{sec:proof_that_single_photons_in_1d_cannot_be_localized} that single photons $\ket{1}$ cannot be strictly localized. Here we show that classical fields as well as coherent states $\ket{\alpha_\xi}$ can be localized, for instance to $t \geq 0$. 

In classical electrodynamics we are free to specify the fields as any functions of space and time. For instance, letting $A(t)$ be a 1D, single-polarization classical electromagnetic potential at the point $x = 0$, we can always choose it to satisfy
\begin{equation}\label{causality_A_t_classical}
    A(t) = 0, \tbuff t < 0,
\end{equation}
making all observable fields localized to $t \geq 0$. To compare to the quantum case, we can Fourier transform the function $A(t)$ as
\begin{equation}\label{A_t_classical}
    A(t) = \int_0^\infty \ibuffi \dd \omega \ibuffb c(\omega) e^{-i\omega t} + \cc
\end{equation}
Requirement \eqref{causality_A_t_classical}, along with $A(t)$ being real, can then be formulated as requirements on the Fourier coefficients $c(\omega)$. 

For a quantum field $A(t)$ from \eqref{A_t}, the localization condition \eqref{U_A_t} for $t < 0$ cannot be satisfied for single photons $\ket{1}$, but it can be satisfied for, e.g., coherent states. For $\alpha \in \complexs$, define the coherent state $\ket{\alpha_\xi} = D_\xi(\alpha) \ket{0}$ in an arbitrary spectrum $\xi(\omega)$:
\begin{equation}\label{D_xi}
    D_\xi(\alpha) = e^{\alpha a_\xi^\dagger - \alpha^* a_\xi^{\vphantom{\dagger}}},
\end{equation}
with
\begin{equation}\label{a_xi_app}
    a_\xi^\dagger = \int_0^\infty \ibuffi \dd \omega \ibuffb \xi(\omega) a^\dagger(\omega).
\end{equation}
To show that $\ket{\alpha_\xi}$ is strictly localized, we can find an operator $U$ so that \eqref{U_A_t} is satisfied. In this case we do not need a source space as we can take $D_\xi(\alpha)$ directly as our $U$ and calculate
\begin{equation}\label{Dd_A_t_D}
    D_\xi^\dagger(\alpha) A(t) D_\xi(\alpha) = A(t) + \int_{-\infty}^\infty \ibuffn \dd \omega \ibuffb Z(\omega) e^{- i \omega t},
\end{equation}
where
\begin{equation}\label{Z_omega}
    Z(\omega) =
    \begin{cases}
        \alpha \mathcal{A}(\omega) \xi(\omega), \tbuff & \omega > 0, \\
        \alpha^* \mathcal{A}^*(-\omega) \xi^*(-\omega), \tbuff & \omega < 0.
    \end{cases}
\end{equation}
The second term in \eqref{Dd_A_t_D} is a Fourier integral that contains both positive and negative frequencies, and it is therefore not limited by the Paley-Wiener criterion. In particular we can select a spectrum $\xi(\omega)$ so that the inverse Fourier transform of $Z(\omega)$ vanishes for $t < 0$ and that $Z(\omega) = Z^*(-\omega)$. Thus coherent states $\ket{\alpha_\xi}$ can be strictly localized to $t \geq 0$.

\section{Eigenvectors of quantum fields}\label{sec:eigenvectors_of_quantum_fields}
We are interested in measuring a (smeared) quantum field observable, e.g., the electric field smeared with some real function $\zeta(t)$,
\begin{equation}\label{L_E_zeta}
    E_\zeta = \int_{-\infty}^\infty \ibuffn \dd t \ibuffb \zeta(t) E(t).
\end{equation}
As usual in quantum mechanics, the possible measurement outcomes and corresponding probabilities are found by the spectral decomposition of $E_\zeta$. Since smeared quantum fields are generally unbounded, we must use the spectral theorem for unbounded operators \cite{hall2013} on $E_\zeta$,
\begin{equation}\label{E_zeta_spectral_app}
    E_\zeta = \int_{\sigma(E_\zeta)} \hspace{0em} \lambda \hspace{.3em} \dd \mu_{E_\zeta}(\lambda).
\end{equation}
Here $\sigma(E_\zeta)$ is the spectrum of $E_\zeta$ and $\mu_{E_\zeta}(\lambda)$ are the spectral projectors. When measuring the observable $E_\zeta$ for some state $\ket{\varphi}$, the possible outcomes $\lambda$ are given by the set $\sigma(E_\zeta)$, and the probability (density) for outcome $\lambda$ is given by $\qexv{\varphi}{\mu_{E_\zeta}(\lambda)}$. 

However, \eqref{E_zeta_spectral_app} is not very constructive; it only asserts that such a decomposition is possible and not how to actually find it. In \eqref{E_zeta} -- \eqref{eigen_X} we showed how to perform the spectral decomposition of $E_\zeta$ by writing
\begin{equation}\label{E_zeta_a1}
    E_\zeta = \frac{1}{\sqrt{2}} \lp(a_1 + a_1^\dagger\rp)
\end{equation}
as an operator acting on the $\xi_1(\omega) = \mathcal{E}^*(\omega) \zeta(\omega)$ subspace in the pulse mode decomposition \eqref{identity_n} of the Hilbert space. Further, with an appropriate normalization of $\zeta(t)$, we showed that the creation operator $a_1^\dagger$ satisfies $\comm*{a_1}{a_1^\dagger} = 1$. This means that $E_\zeta$ is isomorphic to the position operator in a quantum harmonic oscillator, and thus the possible outcomes when measuring $E_\zeta$ are the real numbers $\sigma(E_\zeta) = \reals$. For any outcome $X \in \reals$, we then claimed that the corresponding spectral projector of $E_\zeta$ is given by the outer product $\mu_{E_\zeta}(X) = \ket{X_1}\bra{X_1} \otimes \identity_2 \otimes \dotsb$ of the (generalized) eigenvectors
\begin{equation}\label{eigen_X_app}
    \ket{X_1} = \pi^{-1/4} e^{-X^2/2} e^{-{a_1^\dagger}^2/2 + \sqrt{2}Xa_1^\dagger} \ket{0_1}
\end{equation}
of $E_\zeta$ (see also \cite{barnett2002}). Here $\ket{0_1}$ is the vacuum state of mode $\xi_1(\omega)$. In this section we demonstrate the required properties of $\ket{X_1}$. 

To see that $\ket{X_1}$ is an eigenvector of $E_\zeta$, note that the commutator of $a_1$ and $a_1^\dagger$ implies that
\begin{align}
    \comm{a_1}{e^{r a_1^\dagger}} &= r e^{r a_1^\dagger}, \\
    \comm{a_1}{e^{r {a_1^\dagger}^2 / 2}} &= r a_1^\dagger e^{r {a_1^\dagger}^2 / 2},
\end{align}
for any constant $r$. This can in turn be used to show that
\begin{equation}\label{a1_eigen_X}
    \frac{a_1}{\sqrt{2}} \ket{X_1} = X \ket{X_1} - \frac{a_1^\dagger}{\sqrt{2}} \ket{X_1},
\end{equation}
which means that $\ket{X_1}$ is an eigenvector of $E_\zeta$ corresponding to the eigenvalue $X$.

Next, we demonstrate that $\ket{X_1}$ has the correct delta-function normalization: $\braket{Y_1}{X_1} = \delta\lp(X - Y\rp)$. On the $\xi_1(\omega)$ subspace, define corresponding number states $\ket{n_1} = {a_1^\dagger}^n \ket{0_1} / \sqrt{n!}$, as in \eqref{Fock_n_m}, as well as coherent states $\ket{\alpha_1} = e^{\alpha a_1^\dagger - \alpha^* a_1} \ket{0_1}$. We can then calculate the inner product
\begin{equation}\label{fid_alpha1_eigen_X}
    \braket{\alpha_1}{X_1} = \pi^{-1/4} e^{-X^2/2} e^{-\abs{\alpha}^2/2} \sum_{l = 0}^\infty \frac{{\alpha^*}^l}{2^{l/2} l!} H_l(X),
\end{equation}
using the generating function
\begin{equation}\label{Hermite_generating}
    e^{2Xt - t^2} = \sum_{l=0}^\infty H_l(X) \frac{t^l}{l!}
\end{equation}
for the Hermite polynomials $H_l(X)$ for any constant $t$. By expanding in the states $\ket{\alpha_1}$, we can calculate the inner product
\begin{equation}\label{fid_n1_eigen_X}
\begin{aligned}
    \braket{n_1}{X_1} &= \frac{1}{\pi} \bra{n_1} \int \ibuffa \dd^2 \alpha \ibuffb \ket{\alpha_1} \bra{\alpha_1} \ket{X_1} \\
    &= \lp(\sqrt{\pi} 2^n n!\rp)^{-1/2} e^{-X^2/2} H_n(X),
\end{aligned}
\end{equation}
which is simply the $n$-th Hermite function $\psi_n(X)$. Using the completeness of the states $\ket{n_1}$ and of the Hermite polynomials, we then get that
\begin{equation}\label{fid_eigen_Y_eigen_X}
    \braket{Y_1}{X_1} = \delta \lp(X - Y\rp),
\end{equation}
as required.


\begin{thebibliography}{93}%
\makeatletter
\providecommand \@ifxundefined [1]{%
 \@ifx{#1\undefined}
}%
\providecommand \@ifnum [1]{%
 \ifnum #1\expandafter \@firstoftwo
 \else \expandafter \@secondoftwo
 \fi
}%
\providecommand \@ifx [1]{%
 \ifx #1\expandafter \@firstoftwo
 \else \expandafter \@secondoftwo
 \fi
}%
\providecommand \natexlab [1]{#1}%
\providecommand \enquote  [1]{``#1''}%
\providecommand \bibnamefont  [1]{#1}%
\providecommand \bibfnamefont [1]{#1}%
\providecommand \citenamefont [1]{#1}%
\providecommand \href@noop [0]{\@secondoftwo}%
\providecommand \href [0]{\begingroup \@sanitize@url \@href}%
\providecommand \@href[1]{\@@startlink{#1}\@@href}%
\providecommand \@@href[1]{\endgroup#1\@@endlink}%
\providecommand \@sanitize@url [0]{\catcode `\\12\catcode `\$12\catcode
  `\&12\catcode `\#12\catcode `\^12\catcode `\_12\catcode `\%12\relax}%
\providecommand \@@startlink[1]{}%
\providecommand \@@endlink[0]{}%
\providecommand \url  [0]{\begingroup\@sanitize@url \@url }%
\providecommand \@url [1]{\endgroup\@href {#1}{\urlprefix }}%
\providecommand \urlprefix  [0]{URL }%
\providecommand \Eprint [0]{\href }%
\providecommand \doibase [0]{https://doi.org/}%
\providecommand \selectlanguage [0]{\@gobble}%
\providecommand \bibinfo  [0]{\@secondoftwo}%
\providecommand \bibfield  [0]{\@secondoftwo}%
\providecommand \translation [1]{[#1]}%
\providecommand \BibitemOpen [0]{}%
\providecommand \bibitemStop [0]{}%
\providecommand \bibitemNoStop [0]{.\EOS\space}%
\providecommand \EOS [0]{\spacefactor3000\relax}%
\providecommand \BibitemShut  [1]{\csname bibitem#1\endcsname}%
\let\auto@bib@innerbib\@empty
\bibitem [{\citenamefont {Knight}(1961)}]{knight1961}%
  \BibitemOpen
  \bibfield  {author} {\bibinfo {author} {\bibfnamefont {J.~M.}\ \bibnamefont
  {Knight}},\ }\bibfield  {title} {\bibinfo {title} {Strict localization in
  quantum field theory},\ }\href {https://doi.org/10.1063/1.1703731} {\bibfield
   {journal} {\bibinfo  {journal} {J. Math. Phys.}\ }\textbf {\bibinfo {volume}
  {2}},\ \bibinfo {pages} {459} (\bibinfo {year} {1961})}\BibitemShut {NoStop}%
\bibitem [{\citenamefont {Bialynicki-Birula}(1998)}]{bialynicki-birula1998}%
  \BibitemOpen
  \bibfield  {author} {\bibinfo {author} {\bibfnamefont {I.}~\bibnamefont
  {Bialynicki-Birula}},\ }\bibfield  {title} {\bibinfo {title} {Exponential
  localization of photons},\ }\href
  {https://doi.org/10.1103/PhysRevLett.80.5247} {\bibfield  {journal} {\bibinfo
   {journal} {Phys. Rev. Lett.}\ }\textbf {\bibinfo {volume} {80}},\ \bibinfo
  {pages} {5247} (\bibinfo {year} {1998})}\BibitemShut {NoStop}%
\bibitem [{\citenamefont {Gulla}\ and\ \citenamefont
  {Skaar}(2021{\natexlab{a}})}]{gulla2021}%
  \BibitemOpen
  \bibfield  {author} {\bibinfo {author} {\bibfnamefont {J.}~\bibnamefont
  {Gulla}}\ and\ \bibinfo {author} {\bibfnamefont {J.}~\bibnamefont {Skaar}},\
  }\bibfield  {title} {\bibinfo {title} {Approaching single-photon pulses},\
  }\href {https://doi.org/10.1103/PhysRevLett.126.073601} {\bibfield  {journal}
  {\bibinfo  {journal} {Phys. Rev. Lett.}\ }\textbf {\bibinfo {volume} {126}},\
  \bibinfo {pages} {073601} (\bibinfo {year} {2021}{\natexlab{a}})},\ \Eprint
  {https://arxiv.org/abs/2008.07483} {arXiv:2008.07483 [quant-ph]} \BibitemShut
  {NoStop}%
\bibitem [{\citenamefont {Landau}\ and\ \citenamefont
  {Peierls}(1930)}]{landau1930}%
  \BibitemOpen
  \bibfield  {author} {\bibinfo {author} {\bibfnamefont {L.}~\bibnamefont
  {Landau}}\ and\ \bibinfo {author} {\bibfnamefont {R.}~\bibnamefont
  {Peierls}},\ }\bibfield  {title} {\bibinfo {title} {{Quantenelektrodynamik im
  Konfigurationsraum} [{Q}uantum electrodynamics in configuration space]},\
  }\href {https://doi.org/10.1007/BF01339793} {\bibfield  {journal} {\bibinfo
  {journal} {Z. Phys.}\ }\textbf {\bibinfo {volume} {62}},\ \bibinfo {pages}
  {188} (\bibinfo {year} {1930})},\ \bibinfo {note} {(in German)}\BibitemShut
  {NoStop}%
\bibitem [{\citenamefont {Newton}\ and\ \citenamefont
  {Wigner}(1949)}]{newton1949}%
  \BibitemOpen
  \bibfield  {author} {\bibinfo {author} {\bibfnamefont {T.~D.}\ \bibnamefont
  {Newton}}\ and\ \bibinfo {author} {\bibfnamefont {E.~P.}\ \bibnamefont
  {Wigner}},\ }\bibfield  {title} {\bibinfo {title} {Localized states for
  elementary systems},\ }\href {https://doi.org/10.1103/revmodphys.21.400}
  {\bibfield  {journal} {\bibinfo  {journal} {Rev. Mod. Phys.}\ }\textbf
  {\bibinfo {volume} {21}},\ \bibinfo {pages} {400} (\bibinfo {year}
  {1949})}\BibitemShut {NoStop}%
\bibitem [{\citenamefont {Mandel}(1966)}]{mandel1966}%
  \BibitemOpen
  \bibfield  {author} {\bibinfo {author} {\bibfnamefont {L.}~\bibnamefont
  {Mandel}},\ }\bibfield  {title} {\bibinfo {title} {Configuration-space photon
  number operators in quantum optics},\ }\href
  {https://doi.org/10.1103/physrev.144.1071} {\bibfield  {journal} {\bibinfo
  {journal} {Phys. Rev.}\ }\textbf {\bibinfo {volume} {144}},\ \bibinfo {pages}
  {1071} (\bibinfo {year} {1966})}\BibitemShut {NoStop}%
\bibitem [{\citenamefont {Bohm}(1951)}]{bohm1951}%
  \BibitemOpen
  \bibfield  {author} {\bibinfo {author} {\bibfnamefont {D.}~\bibnamefont
  {Bohm}},\ }\href@noop {} {\emph {\bibinfo {title} {Quantum Theory}}}\
  (\bibinfo  {publisher} {Prentice-Hall},\ \bibinfo {address} {Englewood
  Cliffs},\ \bibinfo {year} {1951})\ Chap.\ \bibinfo {chapter}
  {4.6}\BibitemShut {NoStop}%
\bibitem [{\citenamefont {Power}(1964)}]{power1964}%
  \BibitemOpen
  \bibfield  {author} {\bibinfo {author} {\bibfnamefont {E.~A.}\ \bibnamefont
  {Power}},\ }\href@noop {} {\emph {\bibinfo {title} {Introductory Quantum
  Electrodynamics}}},\ Mathematical Physics Series 4\ (\bibinfo  {publisher}
  {Longmans},\ \bibinfo {address} {London},\ \bibinfo {year} {1964})\ Chap.\
  \bibinfo {chapter} {5.1}\BibitemShut {NoStop}%
\bibitem [{\citenamefont {Pike}\ and\ \citenamefont {Sarkar}(1995)}]{pike1995}%
  \BibitemOpen
  \bibfield  {author} {\bibinfo {author} {\bibfnamefont {E.~R.}\ \bibnamefont
  {Pike}}\ and\ \bibinfo {author} {\bibfnamefont {S.}~\bibnamefont {Sarkar}},\
  }\href@noop {} {\emph {\bibinfo {title} {The Quantum Theory of Radiation}}},\
  International Series of Monographs on Physics 86\ (\bibinfo  {publisher}
  {Clarendon Press},\ \bibinfo {address} {Oxford},\ \bibinfo {year} {1995})\
  Chap.~\bibinfo {chapter} {2}\BibitemShut {NoStop}%
\bibitem [{\citenamefont {Jauch}\ and\ \citenamefont
  {Piron}(1967)}]{jauch1967}%
  \BibitemOpen
  \bibfield  {author} {\bibinfo {author} {\bibfnamefont {J.~M.}\ \bibnamefont
  {Jauch}}\ and\ \bibinfo {author} {\bibfnamefont {C.}~\bibnamefont {Piron}},\
  }\bibfield  {title} {\bibinfo {title} {Generalized localizability},\ }\href
  {https://doi.org/10.5169/SEALS-113783} {\bibfield  {journal} {\bibinfo
  {journal} {{H}elv. Phys. Acta}\ }\textbf {\bibinfo {volume} {40}},\ \bibinfo
  {pages} {559} (\bibinfo {year} {1967})}\BibitemShut {NoStop}%
\bibitem [{\citenamefont {Mandel}\ and\ \citenamefont
  {Wolf}(1995)}]{mandel1995}%
  \BibitemOpen
  \bibfield  {author} {\bibinfo {author} {\bibfnamefont {L.}~\bibnamefont
  {Mandel}}\ and\ \bibinfo {author} {\bibfnamefont {E.}~\bibnamefont {Wolf}},\
  }\href {https://doi.org/10.1017/cbo9781139644105} {\emph {\bibinfo {title}
  {Optical Coherence and Quantum Optics}}}\ (\bibinfo  {publisher} {Cambridge
  University Press},\ \bibinfo {address} {Cambridge},\ \bibinfo {year} {1995})\
  Chap.\ \bibinfo {chapter} {12.11}\BibitemShut {NoStop}%
\bibitem [{\citenamefont {Keller}(2005)}]{keller2005}%
  \BibitemOpen
  \bibfield  {author} {\bibinfo {author} {\bibfnamefont {O.}~\bibnamefont
  {Keller}},\ }\bibfield  {title} {\bibinfo {title} {On the theory of spatial
  localization of photons},\ }\href
  {https://doi.org/10.1016/j.physrep.2005.01.002} {\bibfield  {journal}
  {\bibinfo  {journal} {Phys. Rep.}\ }\textbf {\bibinfo {volume} {411}},\
  \bibinfo {pages} {1} (\bibinfo {year} {2005})}\BibitemShut {NoStop}%
\bibitem [{\citenamefont {Saari}(2012)}]{saari2012}%
  \BibitemOpen
  \bibfield  {author} {\bibinfo {author} {\bibfnamefont {P.}~\bibnamefont
  {Saari}},\ }\bibfield  {title} {\bibinfo {title} {Photon localization
  revisited},\ }in\ \href {https://doi.org/10.5772/1394} {\emph {\bibinfo
  {booktitle} {Quantum Optics and Laser Experiments}}},\ \bibinfo {editor}
  {edited by\ \bibinfo {editor} {\bibfnamefont {S.}~\bibnamefont {Lyagushyn}}}\
  (\bibinfo  {publisher} {InTech},\ \bibinfo {address} {Rijeka},\ \bibinfo
  {year} {2012})\ Chap.~\bibinfo {chapter} {3}, pp.\ \bibinfo {pages}
  {49--66}\BibitemShut {NoStop}%
\bibitem [{\citenamefont {Haag}(1996)}]{haag1996}%
  \BibitemOpen
  \bibfield  {author} {\bibinfo {author} {\bibfnamefont {R.}~\bibnamefont
  {Haag}},\ }\href {https://doi.org/10.1007/978-3-642-61458-3} {\emph {\bibinfo
  {title} {Local Quantum Physics: Fields, Particles, Algebras}}},\ \bibinfo
  {edition} {2nd}\ ed.,\ Texts and Monographs in Physics\ (\bibinfo
  {publisher} {Springer},\ \bibinfo {address} {Berlin},\ \bibinfo {year}
  {1996})\ Chap.\ \bibinfo {chapter} {II.4 and III.1}\BibitemShut {NoStop}%
\bibitem [{\citenamefont {Gulla}\ and\ \citenamefont
  {Skaar}(2021{\natexlab{b}})}]{gulla2021a}%
  \BibitemOpen
  \bibfield  {author} {\bibinfo {author} {\bibfnamefont {J.}~\bibnamefont
  {Gulla}}\ and\ \bibinfo {author} {\bibfnamefont {J.}~\bibnamefont {Skaar}},\
  }\bibfield  {title} {\bibinfo {title} {Tunneling times of single photons},\
  }\href {https://doi.org/10.1364/JOSAB.437386} {\bibfield  {journal} {\bibinfo
   {journal} {J. Opt. Soc. {A}m. B}\ }\textbf {\bibinfo {volume} {38}},\
  \bibinfo {pages} {3457} (\bibinfo {year} {2021}{\natexlab{b}})},\ \Eprint
  {https://arxiv.org/abs/2110.01483} {arXiv:2110.01483 [quant-ph]} \BibitemShut
  {NoStop}%
\bibitem [{\citenamefont {Amrein}(1969)}]{amrein1969}%
  \BibitemOpen
  \bibfield  {author} {\bibinfo {author} {\bibfnamefont {W.~O.}\ \bibnamefont
  {Amrein}},\ }\bibfield  {title} {\bibinfo {title} {Localizability for
  particles of mass zero},\ }\href@noop {} {\bibfield  {journal} {\bibinfo
  {journal} {{H}elv. Phys. Acta}\ }\textbf {\bibinfo {volume} {42}},\ \bibinfo
  {pages} {149} (\bibinfo {year} {1969})}\BibitemShut {NoStop}%
\bibitem [{\citenamefont {Adlard}\ \emph {et~al.}(1997)\citenamefont {Adlard},
  \citenamefont {Pike},\ and\ \citenamefont {Sarkar}}]{adlard1997}%
  \BibitemOpen
  \bibfield  {author} {\bibinfo {author} {\bibfnamefont {C.}~\bibnamefont
  {Adlard}}, \bibinfo {author} {\bibfnamefont {E.~R.}\ \bibnamefont {Pike}},\
  and\ \bibinfo {author} {\bibfnamefont {S.}~\bibnamefont {Sarkar}},\
  }\bibfield  {title} {\bibinfo {title} {Localization of one-photon states},\
  }\href {https://doi.org/10.1103/physrevlett.79.1585} {\bibfield  {journal}
  {\bibinfo  {journal} {Phys. Rev. Lett.}\ }\textbf {\bibinfo {volume} {79}},\
  \bibinfo {pages} {1585} (\bibinfo {year} {1997})}\BibitemShut {NoStop}%
\bibitem [{\citenamefont {Titchmarsh}(1948)}]{titchmarsh1948}%
  \BibitemOpen
  \bibfield  {author} {\bibinfo {author} {\bibfnamefont {E.~C.}\ \bibnamefont
  {Titchmarsh}},\ }\href@noop {} {\emph {\bibinfo {title} {Introduction to the
  Theory of {F}ourier Integrals}}},\ \bibinfo {edition} {2nd}\ ed.\ (\bibinfo
  {publisher} {Clarendon Press},\ \bibinfo {address} {Oxford},\ \bibinfo {year}
  {1948})\ Chap.\ \bibinfo {chapter} {5.5}\BibitemShut {NoStop}%
\bibitem [{\citenamefont {Paley}\ and\ \citenamefont
  {Wiener}(1934)}]{paley1934}%
  \BibitemOpen
  \bibfield  {author} {\bibinfo {author} {\bibfnamefont {R.~E. A.~C.}\
  \bibnamefont {Paley}}\ and\ \bibinfo {author} {\bibfnamefont
  {N.}~\bibnamefont {Wiener}},\ }\href {https://doi.org/10.1090/coll/019}
  {\emph {\bibinfo {title} {{F}ourier Transforms in the Complex Domain}}},\
  {A}merican Mathematical Society Colloquium Publications Vol. {XIX}\ (\bibinfo
   {publisher} {American Mathematical Society},\ \bibinfo {address} {New
  York},\ \bibinfo {year} {1934})\ Chap.\ \bibinfo {chapter} {I.7}\BibitemShut
  {NoStop}%
\bibitem [{\citenamefont {Saari}\ \emph {et~al.}(2005)\citenamefont {Saari},
  \citenamefont {Menert},\ and\ \citenamefont {Valtna}}]{saari2005}%
  \BibitemOpen
  \bibfield  {author} {\bibinfo {author} {\bibfnamefont {P.}~\bibnamefont
  {Saari}}, \bibinfo {author} {\bibfnamefont {M.}~\bibnamefont {Menert}},\ and\
  \bibinfo {author} {\bibfnamefont {H.}~\bibnamefont {Valtna}},\ }\bibfield
  {title} {\bibinfo {title} {Photon localization barrier can be overcome},\
  }\href {https://doi.org/10.1016/j.optcom.2004.11.020} {\bibfield  {journal}
  {\bibinfo  {journal} {Opt. Commun.}\ }\textbf {\bibinfo {volume} {246}},\
  \bibinfo {pages} {445} (\bibinfo {year} {2005})},\ \Eprint
  {https://arxiv.org/abs/quant-ph/0409034} {arXiv:quant-ph/0409034}
  \BibitemShut {NoStop}%
\bibitem [{\citenamefont {Loudon}(2001)}]{loudon2001}%
  \BibitemOpen
  \bibfield  {author} {\bibinfo {author} {\bibfnamefont {R.}~\bibnamefont
  {Loudon}},\ }\href@noop {} {\emph {\bibinfo {title} {The Quantum Theory of
  Light}}},\ \bibinfo {edition} {3rd}\ ed.\ (\bibinfo  {publisher} {Oxford
  University Press},\ \bibinfo {address} {Oxford},\ \bibinfo {year} {2001})\
  Chap.\ \bibinfo {chapter} {6.2}\BibitemShut {NoStop}%
\bibitem [{\citenamefont {Scheel}(2009)}]{scheel2009}%
  \BibitemOpen
  \bibfield  {author} {\bibinfo {author} {\bibfnamefont {S.}~\bibnamefont
  {Scheel}},\ }\bibfield  {title} {\bibinfo {title} {Single-photon sources --
  an introduction},\ }\href {https://doi.org/10.1080/09500340802331849}
  {\bibfield  {journal} {\bibinfo  {journal} {J. Mod. Opt.}\ }\textbf {\bibinfo
  {volume} {56}},\ \bibinfo {pages} {141} (\bibinfo {year} {2009})}\BibitemShut
  {NoStop}%
\bibitem [{\citenamefont {Eisaman}\ \emph {et~al.}(2011)\citenamefont
  {Eisaman}, \citenamefont {Fan}, \citenamefont {Migdall},\ and\ \citenamefont
  {Polyakov}}]{eisaman2011}%
  \BibitemOpen
  \bibfield  {author} {\bibinfo {author} {\bibfnamefont {M.~D.}\ \bibnamefont
  {Eisaman}}, \bibinfo {author} {\bibfnamefont {J.}~\bibnamefont {Fan}},
  \bibinfo {author} {\bibfnamefont {A.}~\bibnamefont {Migdall}},\ and\ \bibinfo
  {author} {\bibfnamefont {S.~V.}\ \bibnamefont {Polyakov}},\ }\bibfield
  {title} {\bibinfo {title} {Invited review article: Single-photon sources and
  detectors},\ }\href {https://doi.org/10.1063/1.3610677} {\bibfield  {journal}
  {\bibinfo  {journal} {Rev. Sci. Instrum.}\ }\textbf {\bibinfo {volume}
  {82}},\ \bibinfo {pages} {071101} (\bibinfo {year} {2011})}\BibitemShut
  {NoStop}%
\bibitem [{\citenamefont {Senellart}\ \emph {et~al.}(2017)\citenamefont
  {Senellart}, \citenamefont {Solomon},\ and\ \citenamefont
  {White}}]{senellart2017}%
  \BibitemOpen
  \bibfield  {author} {\bibinfo {author} {\bibfnamefont {P.}~\bibnamefont
  {Senellart}}, \bibinfo {author} {\bibfnamefont {G.}~\bibnamefont {Solomon}},\
  and\ \bibinfo {author} {\bibfnamefont {A.}~\bibnamefont {White}},\ }\bibfield
   {title} {\bibinfo {title} {High-performance semiconductor quantum-dot
  single-photon sources},\ }\href {https://doi.org/10.1038/nnano.2017.218}
  {\bibfield  {journal} {\bibinfo  {journal} {Nat. Nanotechnol.}\ }\textbf
  {\bibinfo {volume} {12}},\ \bibinfo {pages} {1026} (\bibinfo {year}
  {2017})}\BibitemShut {NoStop}%
\bibitem [{\citenamefont {Wang}\ \emph {et~al.}(2019)\citenamefont {Wang},
  \citenamefont {He}, \citenamefont {Chung}, \citenamefont {Hu}, \citenamefont
  {Yu}, \citenamefont {Chen}, \citenamefont {Ding}, \citenamefont {Chen},
  \citenamefont {Qin}, \citenamefont {Yang}, \citenamefont {Liu}, \citenamefont
  {Duan}, \citenamefont {Li}, \citenamefont {Gerhardt}, \citenamefont
  {Winkler}, \citenamefont {Jurkat}, \citenamefont {Wang}, \citenamefont
  {Gregersen}, \citenamefont {Huo}, \citenamefont {Dai}, \citenamefont {Yu},
  \citenamefont {H{\"o}fling}, \citenamefont {Lu},\ and\ \citenamefont
  {Pan}}]{wang2019}%
  \BibitemOpen
  \bibfield  {author} {\bibinfo {author} {\bibfnamefont {H.}~\bibnamefont
  {Wang}}, \bibinfo {author} {\bibfnamefont {Y.-M.}\ \bibnamefont {He}},
  \bibinfo {author} {\bibfnamefont {T.-H.}\ \bibnamefont {Chung}}, \bibinfo
  {author} {\bibfnamefont {H.}~\bibnamefont {Hu}}, \bibinfo {author}
  {\bibfnamefont {Y.}~\bibnamefont {Yu}}, \bibinfo {author} {\bibfnamefont
  {S.}~\bibnamefont {Chen}}, \bibinfo {author} {\bibfnamefont {X.}~\bibnamefont
  {Ding}}, \bibinfo {author} {\bibfnamefont {M.-C.}\ \bibnamefont {Chen}},
  \bibinfo {author} {\bibfnamefont {J.}~\bibnamefont {Qin}}, \bibinfo {author}
  {\bibfnamefont {X.}~\bibnamefont {Yang}}, \bibinfo {author} {\bibfnamefont
  {R.-Z.}\ \bibnamefont {Liu}}, \bibinfo {author} {\bibfnamefont {Z.-C.}\
  \bibnamefont {Duan}}, \bibinfo {author} {\bibfnamefont {J.-P.}\ \bibnamefont
  {Li}}, \bibinfo {author} {\bibfnamefont {S.}~\bibnamefont {Gerhardt}},
  \bibinfo {author} {\bibfnamefont {K.}~\bibnamefont {Winkler}}, \bibinfo
  {author} {\bibfnamefont {J.}~\bibnamefont {Jurkat}}, \bibinfo {author}
  {\bibfnamefont {L.-J.}\ \bibnamefont {Wang}}, \bibinfo {author}
  {\bibfnamefont {N.}~\bibnamefont {Gregersen}}, \bibinfo {author}
  {\bibfnamefont {Y.-H.}\ \bibnamefont {Huo}}, \bibinfo {author} {\bibfnamefont
  {Q.}~\bibnamefont {Dai}}, \bibinfo {author} {\bibfnamefont {S.}~\bibnamefont
  {Yu}}, \bibinfo {author} {\bibfnamefont {S.}~\bibnamefont {H{\"o}fling}},
  \bibinfo {author} {\bibfnamefont {C.-Y.}\ \bibnamefont {Lu}},\ and\ \bibinfo
  {author} {\bibfnamefont {J.-W.}\ \bibnamefont {Pan}},\ }\bibfield  {title}
  {\bibinfo {title} {Towards optimal single-photon sources from polarized
  microcavities},\ }\href {https://doi.org/10.1038/s41566-019-0494-3}
  {\bibfield  {journal} {\bibinfo  {journal} {Nat. Photonics}\ }\textbf
  {\bibinfo {volume} {13}},\ \bibinfo {pages} {770} (\bibinfo {year} {2019})},\
  \Eprint {https://arxiv.org/abs/1907.06818} {arXiv:1907.06818 [quant-ph]}
  \BibitemShut {NoStop}%
\bibitem [{\citenamefont {Sinha}\ \emph {et~al.}(2019)\citenamefont {Sinha},
  \citenamefont {Sahoo}, \citenamefont {Singh}, \citenamefont {Joarder},
  \citenamefont {Chatterjee},\ and\ \citenamefont {Chakraborti}}]{sinha2019}%
  \BibitemOpen
  \bibfield  {author} {\bibinfo {author} {\bibfnamefont {U.}~\bibnamefont
  {Sinha}}, \bibinfo {author} {\bibfnamefont {S.~N.}\ \bibnamefont {Sahoo}},
  \bibinfo {author} {\bibfnamefont {A.}~\bibnamefont {Singh}}, \bibinfo
  {author} {\bibfnamefont {K.}~\bibnamefont {Joarder}}, \bibinfo {author}
  {\bibfnamefont {R.}~\bibnamefont {Chatterjee}},\ and\ \bibinfo {author}
  {\bibfnamefont {S.}~\bibnamefont {Chakraborti}},\ }\bibfield  {title}
  {\bibinfo {title} {Single-photon sources},\ }\href
  {https://doi.org/10.1364/OPN.30.9.000032} {\bibfield  {journal} {\bibinfo
  {journal} {Opt. Photonics News}\ }\textbf {\bibinfo {volume} {30}},\ \bibinfo
  {pages} {32} (\bibinfo {year} {2019})},\ \Eprint
  {https://arxiv.org/abs/1906.09565} {arXiv:1906.09565 [quant-ph]} \BibitemShut
  {NoStop}%
\bibitem [{\citenamefont {Hegerfeldt}(1998)}]{hegerfeldt1998}%
  \BibitemOpen
  \bibfield  {author} {\bibinfo {author} {\bibfnamefont {G.~C.}\ \bibnamefont
  {Hegerfeldt}},\ }\bibfield  {title} {\bibinfo {title} {Causality, particle
  localization and positivity of the energy},\ }in\ \href
  {https://doi.org/10.1007/BFb0106784} {\emph {\bibinfo {booktitle}
  {Irreversibility and Causality: Semigroups and Rigged {H}ilbert Spaces, Proc.
  21st Int. Colloq. On Group Theoretical Methods in Physics, Goslar, Germany,
  Jul 16--21, 1996}}},\ \bibinfo {series and number} {Lecture Notes in Physics
  504},\ \bibinfo {editor} {edited by\ \bibinfo {editor} {\bibfnamefont
  {A.}~\bibnamefont {Bohm}}, \bibinfo {editor} {\bibfnamefont {H.-D.}\
  \bibnamefont {Doebner}},\ and\ \bibinfo {editor} {\bibfnamefont
  {P.}~\bibnamefont {Kielanowski}}}\ (\bibinfo  {publisher} {Springer},\
  \bibinfo {address} {Berlin},\ \bibinfo {year} {1998})\ Chap.~\bibinfo
  {chapter} {{IV}}, pp.\ \bibinfo {pages} {238--245},\ \Eprint
  {https://arxiv.org/abs/quant-ph/9806036} {arXiv:quant-ph/9806036}
  \BibitemShut {NoStop}%
\bibitem [{\citenamefont {Hegerfeldt}(1974)}]{hegerfeldt1974}%
  \BibitemOpen
  \bibfield  {author} {\bibinfo {author} {\bibfnamefont {G.~C.}\ \bibnamefont
  {Hegerfeldt}},\ }\bibfield  {title} {\bibinfo {title} {Remark on causality
  and particle localization},\ }\href
  {https://doi.org/10.1103/PhysRevD.10.3320} {\bibfield  {journal} {\bibinfo
  {journal} {Phys. Rev. D}\ }\textbf {\bibinfo {volume} {10}},\ \bibinfo
  {pages} {3320} (\bibinfo {year} {1974})}\BibitemShut {NoStop}%
\bibitem [{\citenamefont {Kaloyerou}(1988)}]{kaloyerou1988}%
  \BibitemOpen
  \bibfield  {author} {\bibinfo {author} {\bibfnamefont {P.~N.}\ \bibnamefont
  {Kaloyerou}},\ }\bibfield  {title} {\bibinfo {title} {Comments on the
  {H}egerfeldt \textquotedblleft paradox\textquotedblright},\ }\href
  {https://doi.org/https://doi.org/10.1016/0375-9601(88)90333-7} {\bibfield
  {journal} {\bibinfo  {journal} {Phys. Lett. A}\ }\textbf {\bibinfo {volume}
  {129}},\ \bibinfo {pages} {285} (\bibinfo {year} {1988})}\BibitemShut
  {NoStop}%
\bibitem [{\citenamefont {Hegerfeldt}(1994)}]{hegerfeldt1994}%
  \BibitemOpen
  \bibfield  {author} {\bibinfo {author} {\bibfnamefont {G.~C.}\ \bibnamefont
  {Hegerfeldt}},\ }\bibfield  {title} {\bibinfo {title} {Causality problems for
  {F}ermi's two-atom system},\ }\href
  {https://doi.org/10.1103/PhysRevLett.72.596} {\bibfield  {journal} {\bibinfo
  {journal} {Phys. Rev. Lett.}\ }\textbf {\bibinfo {volume} {72}},\ \bibinfo
  {pages} {596} (\bibinfo {year} {1994})}\BibitemShut {NoStop}%
\bibitem [{\citenamefont {Buchholz}\ and\ \citenamefont
  {Yngvason}(1994)}]{buchholz1994}%
  \BibitemOpen
  \bibfield  {author} {\bibinfo {author} {\bibfnamefont {D.}~\bibnamefont
  {Buchholz}}\ and\ \bibinfo {author} {\bibfnamefont {J.}~\bibnamefont
  {Yngvason}},\ }\bibfield  {title} {\bibinfo {title} {There are no causality
  problems for {F}ermi's two-atom system},\ }\href
  {https://doi.org/10.1103/PhysRevLett.73.613} {\bibfield  {journal} {\bibinfo
  {journal} {Phys. Rev. Lett.}\ }\textbf {\bibinfo {volume} {73}},\ \bibinfo
  {pages} {613} (\bibinfo {year} {1994})},\ \Eprint
  {https://arxiv.org/abs/hep-th/9403027} {arXiv:hep-th/9403027} \BibitemShut
  {NoStop}%
\bibitem [{\citenamefont {Peskin}\ and\ \citenamefont
  {Schroeder}(1995)}]{peskin1995}%
  \BibitemOpen
  \bibfield  {author} {\bibinfo {author} {\bibfnamefont {M.~E.}\ \bibnamefont
  {Peskin}}\ and\ \bibinfo {author} {\bibfnamefont {D.~V.}\ \bibnamefont
  {Schroeder}},\ }\href@noop {} {\emph {\bibinfo {title} {An Introduction to
  Quantum Field Theory}}}\ (\bibinfo  {publisher} {Perseus Books},\ \bibinfo
  {address} {Reading},\ \bibinfo {year} {1995})\ Chap.\ \bibinfo {chapter} {2.3
  and 2.4}\BibitemShut {NoStop}%
\bibitem [{\citenamefont {Heisenberg}\ and\ \citenamefont
  {Pauli}(1929)}]{heisenberg1929}%
  \BibitemOpen
  \bibfield  {author} {\bibinfo {author} {\bibfnamefont {W.}~\bibnamefont
  {Heisenberg}}\ and\ \bibinfo {author} {\bibfnamefont {W.}~\bibnamefont
  {Pauli}},\ }\bibfield  {title} {\bibinfo {title} {{Zur Quantendynamik der
  Wellenfelder} [{O}n the quantum dynamics of wave fields]},\ }\href
  {https://doi.org/10.1007/bf01340129} {\bibfield  {journal} {\bibinfo
  {journal} {Z. Phys.}\ }\textbf {\bibinfo {volume} {56}},\ \bibinfo {pages}
  {1} (\bibinfo {year} {1929})},\ \bibinfo {note} {(in German)}\BibitemShut
  {NoStop}%
\bibitem [{\citenamefont {Heisenberg}\ and\ \citenamefont
  {Pauli}(1930)}]{heisenberg1930}%
  \BibitemOpen
  \bibfield  {author} {\bibinfo {author} {\bibfnamefont {W.}~\bibnamefont
  {Heisenberg}}\ and\ \bibinfo {author} {\bibfnamefont {W.}~\bibnamefont
  {Pauli}},\ }\bibfield  {title} {\bibinfo {title} {{Zur Quantentheorie der
  Wellenfelder. {II}} [{O}n the quantum theory of wave fields. {II}]},\ }\href
  {https://doi.org/10.1007/bf01341423} {\bibfield  {journal} {\bibinfo
  {journal} {Z. Phys.}\ }\textbf {\bibinfo {volume} {59}},\ \bibinfo {pages}
  {168} (\bibinfo {year} {1930})},\ \bibinfo {note} {(in German)}\BibitemShut
  {NoStop}%
\bibitem [{\citenamefont {Kikuchi}(1930)}]{kikuchi1930}%
  \BibitemOpen
  \bibfield  {author} {\bibinfo {author} {\bibfnamefont {S.}~\bibnamefont
  {Kikuchi}},\ }\bibfield  {title} {\bibinfo {title} {{{\"U}ber die
  Fortpflanzung von Lichtwellen in der Heisenberg-Paulischen Formulierung der
  Quantenelektrodynamik} [{A}bout the propagation of light waves in the
  {H}eisenberg-{P}auli formulation of quantum electrodynamics]},\ }\href
  {https://doi.org/10.1007/bf01402038} {\bibfield  {journal} {\bibinfo
  {journal} {Z. Phys.}\ }\textbf {\bibinfo {volume} {66}},\ \bibinfo {pages}
  {558} (\bibinfo {year} {1930})},\ \bibinfo {note} {(in German)}\BibitemShut
  {NoStop}%
\bibitem [{\citenamefont {Fermi}(1932)}]{fermi1932}%
  \BibitemOpen
  \bibfield  {author} {\bibinfo {author} {\bibfnamefont {E.}~\bibnamefont
  {Fermi}},\ }\bibfield  {title} {\bibinfo {title} {Quantum theory of
  radiation},\ }\href {https://doi.org/10.1103/RevModPhys.4.87} {\bibfield
  {journal} {\bibinfo  {journal} {Rev. Mod. Phys.}\ }\textbf {\bibinfo {volume}
  {4}},\ \bibinfo {pages} {87} (\bibinfo {year} {1932})}\BibitemShut {NoStop}%
\bibitem [{\citenamefont {Heitler}(1954)}]{heitler1954}%
  \BibitemOpen
  \bibfield  {author} {\bibinfo {author} {\bibfnamefont {W.}~\bibnamefont
  {Heitler}},\ }\href@noop {} {\emph {\bibinfo {title} {The Quantum Theory of
  Radiation}}},\ \bibinfo {edition} {3rd}\ ed.,\ International Series of
  Monographs on Physics\ (\bibinfo  {publisher} {Clarendon Press},\ \bibinfo
  {address} {Oxford},\ \bibinfo {year} {1954})\ Chap.~\bibinfo {chapter}
  {20}\BibitemShut {NoStop}%
\bibitem [{\citenamefont {Shirokov}(1964)}]{shirokov1964}%
  \BibitemOpen
  \bibfield  {author} {\bibinfo {author} {\bibfnamefont {M.~I.}\ \bibnamefont
  {Shirokov}},\ }\href@noop {} {\emph {\bibinfo {title} {{Skorost' peredachi
  vozbuzhdenija v kvantovoj elektrodinamike} [{T}he velocity of excitation
  transfer in quantum electrodynamics]}}},\ \bibinfo {type} {{\unskip\null}}\
  \bibinfo {number} {P-1719}\ (\bibinfo  {institution} {JINR},\ \bibinfo {year}
  {1964})\ \bibinfo {note} {(in Russian)}\BibitemShut {NoStop}%
\bibitem [{\citenamefont {Shirokov}(1967)}]{shirokov1967}%
  \BibitemOpen
  \bibfield  {author} {\bibinfo {author} {\bibfnamefont {M.~I.}\ \bibnamefont
  {Shirokov}},\ }\bibfield  {title} {\bibinfo {title} {The velocity of
  electromagnetic radiation in quantum electrodynamics},\ }\href@noop {}
  {\bibfield  {journal} {\bibinfo  {journal} {{S}ov. J. Nucl. Phys.}\ }\textbf
  {\bibinfo {volume} {4}},\ \bibinfo {pages} {774} (\bibinfo {year} {1967})},\
  \bibinfo {note} {[\href@noop {} {\bibfield {journal} {\bibinfo {journal}
  {Yad. Fiz.}\ }\textbf {\bibinfo {volume} {4}},\ \bibinfo {pages} {1077}
  (\bibinfo {year} {1966})}]}\BibitemShut {NoStop}%
\bibitem [{\citenamefont {Ferretti}(1968)}]{ferretti1968}%
  \BibitemOpen
  \bibfield  {author} {\bibinfo {author} {\bibfnamefont {B.}~\bibnamefont
  {Ferretti}},\ }\bibfield  {title} {\bibinfo {title} {Propagation of signals
  and particles},\ }in\ \href
  {https://doi.org/10.1016/b978-0-12-395657-6.50011-2} {\emph {\bibinfo
  {booktitle} {Old and New Problems in Elementary Particles: A Volume Dedicated
  to {G}ilberto {B}ernardini in His Sixtieth Birthday}}},\ \bibinfo {editor}
  {edited by\ \bibinfo {editor} {\bibfnamefont {G.}~\bibnamefont {Puppi}}}\
  (\bibinfo  {publisher} {Academic Press},\ \bibinfo {address} {New York},\
  \bibinfo {year} {1968})\ pp.\ \bibinfo {pages} {108--119}\BibitemShut
  {NoStop}%
\bibitem [{\citenamefont {Malament}(1996)}]{malament1996}%
  \BibitemOpen
  \bibfield  {author} {\bibinfo {author} {\bibfnamefont {D.~B.}\ \bibnamefont
  {Malament}},\ }\bibfield  {title} {\bibinfo {title} {In defense of dogma: Why
  there cannot be a relativistic quantum mechanics of (localizable)
  particles},\ }in\ \href {https://doi.org/10.1007/978-94-015-8656-6_1} {\emph
  {\bibinfo {booktitle} {Perspectives on Quantum Reality: Non-Relativistic,
  Relativistic, and Field-Theoretic}}},\ \bibinfo {series and number} {The
  Western Ontario Series in Philosophy of Science Vol. 57},\ \bibinfo {editor}
  {edited by\ \bibinfo {editor} {\bibfnamefont {R.}~\bibnamefont {Clifton}}}\
  (\bibinfo  {publisher} {Springer},\ \bibinfo {address} {Dordrecht},\ \bibinfo
  {year} {1996})\ pp.\ \bibinfo {pages} {1--10}\BibitemShut {NoStop}%
\bibitem [{\citenamefont {Halvorson}\ and\ \citenamefont
  {Clifton}(2002)}]{halvorson2002}%
  \BibitemOpen
  \bibfield  {author} {\bibinfo {author} {\bibfnamefont {H.}~\bibnamefont
  {Halvorson}}\ and\ \bibinfo {author} {\bibfnamefont {R.}~\bibnamefont
  {Clifton}},\ }\bibfield  {title} {\bibinfo {title} {No place for particles in
  relativistic quantum theories?},\ }in\ \href
  {https://doi.org/10.1142/9789812776440_0010} {\emph {\bibinfo {booktitle}
  {Ontological Aspects of Quantum Field Theory}}},\ \bibinfo {editor} {edited
  by\ \bibinfo {editor} {\bibfnamefont {M.}~\bibnamefont {Kuhlmann}}, \bibinfo
  {editor} {\bibfnamefont {H.}~\bibnamefont {Lyre}},\ and\ \bibinfo {editor}
  {\bibfnamefont {A.}~\bibnamefont {Wayne}}}\ (\bibinfo  {publisher} {World
  Scientific},\ \bibinfo {address} {New Jersey},\ \bibinfo {year} {2002})\
  Chap.~\bibinfo {chapter} {10}, pp.\ \bibinfo {pages} {181--213},\ \Eprint
  {https://arxiv.org/abs/quant-ph/0103041} {arXiv:quant-ph/0103041}
  \BibitemShut {NoStop}%
\bibitem [{\citenamefont {Reeh}\ and\ \citenamefont
  {Schlieder}(1961)}]{reeh1961}%
  \BibitemOpen
  \bibfield  {author} {\bibinfo {author} {\bibfnamefont {H.}~\bibnamefont
  {Reeh}}\ and\ \bibinfo {author} {\bibfnamefont {S.}~\bibnamefont
  {Schlieder}},\ }\bibfield  {title} {\bibinfo {title} {{Bemerkungen zur
  Unit{\"a}r{\"a}quivalenz von Lorentzinvarianten Feldern} [{R}emarks on the
  unitary equivalence of {L}orentz-invariant fields]},\ }\href
  {https://doi.org/10.1007/BF02787889} {\bibfield  {journal} {\bibinfo
  {journal} {Nuovo Cimento}\ }\textbf {\bibinfo {volume} {22}},\ \bibinfo
  {pages} {1051} (\bibinfo {year} {1961})},\ \bibinfo {note} {(in
  German)}\BibitemShut {NoStop}%
\bibitem [{\citenamefont {Redhead}(1995)}]{redhead1995}%
  \BibitemOpen
  \bibfield  {author} {\bibinfo {author} {\bibfnamefont {M.}~\bibnamefont
  {Redhead}},\ }\bibfield  {title} {\bibinfo {title} {More ado about nothing},\
  }\href {https://doi.org/10.1007/bf02054660} {\bibfield  {journal} {\bibinfo
  {journal} {Found. Phys.}\ }\textbf {\bibinfo {volume} {25}},\ \bibinfo
  {pages} {123} (\bibinfo {year} {1995})}\BibitemShut {NoStop}%
\bibitem [{\citenamefont {Louisell}(1973)}]{louisell1973}%
  \BibitemOpen
  \bibfield  {author} {\bibinfo {author} {\bibfnamefont {W.~H.}\ \bibnamefont
  {Louisell}},\ }\href@noop {} {\emph {\bibinfo {title} {Quantum Statistical
  Properties of Radiation}}},\ Wiley Series in Pure and Applied Optics\
  (\bibinfo  {publisher} {John Wiley {\&} Sons},\ \bibinfo {address} {New
  York},\ \bibinfo {year} {1973})\ Chap.\ \bibinfo {chapter} {5.2}\BibitemShut
  {NoStop}%
\bibitem [{\citenamefont {Rubin}(1987)}]{rubin1987}%
  \BibitemOpen
  \bibfield  {author} {\bibinfo {author} {\bibfnamefont {M.~H.}\ \bibnamefont
  {Rubin}},\ }\bibfield  {title} {\bibinfo {title} {Violation of {E}instein
  causality in a model quantum system},\ }\href
  {https://doi.org/10.1103/physrevd.35.3836} {\bibfield  {journal} {\bibinfo
  {journal} {Phys. Rev. D}\ }\textbf {\bibinfo {volume} {35}},\ \bibinfo
  {pages} {3836} (\bibinfo {year} {1987})}\BibitemShut {NoStop}%
\bibitem [{\citenamefont {Biswas}\ \emph {et~al.}(1990)\citenamefont {Biswas},
  \citenamefont {Compagno}, \citenamefont {Palma}, \citenamefont {Passante},\
  and\ \citenamefont {Persico}}]{biswas1990}%
  \BibitemOpen
  \bibfield  {author} {\bibinfo {author} {\bibfnamefont {A.~K.}\ \bibnamefont
  {Biswas}}, \bibinfo {author} {\bibfnamefont {G.}~\bibnamefont {Compagno}},
  \bibinfo {author} {\bibfnamefont {G.~M.}\ \bibnamefont {Palma}}, \bibinfo
  {author} {\bibfnamefont {R.}~\bibnamefont {Passante}},\ and\ \bibinfo
  {author} {\bibfnamefont {F.}~\bibnamefont {Persico}},\ }\bibfield  {title}
  {\bibinfo {title} {Virtual photons and causality in the dynamics of a pair of
  two-level atoms},\ }\href {https://doi.org/10.1103/physreva.42.4291}
  {\bibfield  {journal} {\bibinfo  {journal} {Phys. Rev. A}\ }\textbf {\bibinfo
  {volume} {42}},\ \bibinfo {pages} {4291} (\bibinfo {year}
  {1990})}\BibitemShut {NoStop}%
\bibitem [{\citenamefont {Valentini}(1991)}]{valentini1991}%
  \BibitemOpen
  \bibfield  {author} {\bibinfo {author} {\bibfnamefont {A.}~\bibnamefont
  {Valentini}},\ }\bibfield  {title} {\bibinfo {title} {Non-local correlations
  in quantum electrodynamics},\ }\href
  {https://doi.org/10.1016/0375-9601(91)90952-5} {\bibfield  {journal}
  {\bibinfo  {journal} {Phys. Lett. A}\ }\textbf {\bibinfo {volume} {153}},\
  \bibinfo {pages} {321} (\bibinfo {year} {1991})}\BibitemShut {NoStop}%
\bibitem [{\citenamefont {Maddox}(1994)}]{maddox1994}%
  \BibitemOpen
  \bibfield  {author} {\bibinfo {author} {\bibfnamefont {J.}~\bibnamefont
  {Maddox}},\ }\bibfield  {title} {\bibinfo {title} {Time machines still over
  horizon},\ }\href {https://doi.org/10.1038/367509a0} {\bibfield  {journal}
  {\bibinfo  {journal} {Nature}\ }\textbf {\bibinfo {volume} {367}},\ \bibinfo
  {pages} {509} (\bibinfo {year} {1994})}\BibitemShut {NoStop}%
\bibitem [{\citenamefont {Milonni}(1994{\natexlab{a}})}]{milonni1994a}%
  \BibitemOpen
  \bibfield  {author} {\bibinfo {author} {\bibfnamefont {P.~W.}\ \bibnamefont
  {Milonni}},\ }\bibfield  {title} {\bibinfo {title} {Interatomic signalling in
  {QED}},\ }\href {https://doi.org/10.1038/372325b0} {\bibfield  {journal}
  {\bibinfo  {journal} {Nature}\ }\textbf {\bibinfo {volume} {372}},\ \bibinfo
  {pages} {325} (\bibinfo {year} {1994}{\natexlab{a}})}\BibitemShut {NoStop}%
\bibitem [{\citenamefont {Shirokov}(1978)}]{shirokov1978}%
  \BibitemOpen
  \bibfield  {author} {\bibinfo {author} {\bibfnamefont {M.~I.}\ \bibnamefont
  {Shirokov}},\ }\bibfield  {title} {\bibinfo {title} {Signal velocity in
  quantum electrodynamics},\ }\href@noop {} {\bibfield  {journal} {\bibinfo
  {journal} {{S}ov. Phys. Usp.}\ }\textbf {\bibinfo {volume} {21}},\ \bibinfo
  {pages} {345} (\bibinfo {year} {1978})},\ \bibinfo {note} {[\href@noop {}
  {\bibfield {journal} {\bibinfo {journal} {Usp. Fiz. Nauk}\ }\textbf {\bibinfo
  {volume} {124}},\ \bibinfo {pages} {697} (\bibinfo {year}
  {1978})}]}\BibitemShut {NoStop}%
\bibitem [{\citenamefont {Dickinson}\ \emph {et~al.}(2016)\citenamefont
  {Dickinson}, \citenamefont {Forshaw},\ and\ \citenamefont
  {Millington}}]{dickinson2016}%
  \BibitemOpen
  \bibfield  {author} {\bibinfo {author} {\bibfnamefont {R.}~\bibnamefont
  {Dickinson}}, \bibinfo {author} {\bibfnamefont {J.}~\bibnamefont {Forshaw}},\
  and\ \bibinfo {author} {\bibfnamefont {P.}~\bibnamefont {Millington}},\
  }\bibfield  {title} {\bibinfo {title} {Probabilities and signalling in
  quantum field theory},\ }\href {https://doi.org/10.1103/physrevd.93.065054}
  {\bibfield  {journal} {\bibinfo  {journal} {Phys. Rev. D}\ }\textbf {\bibinfo
  {volume} {93}},\ \bibinfo {pages} {065054} (\bibinfo {year} {2016})},\
  \Eprint {https://arxiv.org/abs/1601.07784} {arXiv:1601.07784 [hep-th]}
  \BibitemShut {NoStop}%
\bibitem [{\citenamefont {Power}\ and\ \citenamefont
  {Thirunamachandran}(1997)}]{power1997}%
  \BibitemOpen
  \bibfield  {author} {\bibinfo {author} {\bibfnamefont {E.~A.}\ \bibnamefont
  {Power}}\ and\ \bibinfo {author} {\bibfnamefont {T.}~\bibnamefont
  {Thirunamachandran}},\ }\bibfield  {title} {\bibinfo {title} {Analysis of the
  causal behavior in energy transfer between atoms},\ }\href
  {https://doi.org/10.1103/physreva.56.3395} {\bibfield  {journal} {\bibinfo
  {journal} {Phys. Rev. A}\ }\textbf {\bibinfo {volume} {56}},\ \bibinfo
  {pages} {3395} (\bibinfo {year} {1997})}\BibitemShut {NoStop}%
\bibitem [{\citenamefont {Milonni}\ \emph {et~al.}(1995)\citenamefont
  {Milonni}, \citenamefont {James},\ and\ \citenamefont {Fearn}}]{milonni1995}%
  \BibitemOpen
  \bibfield  {author} {\bibinfo {author} {\bibfnamefont {P.~W.}\ \bibnamefont
  {Milonni}}, \bibinfo {author} {\bibfnamefont {D.~F.~V.}\ \bibnamefont
  {James}},\ and\ \bibinfo {author} {\bibfnamefont {H.}~\bibnamefont {Fearn}},\
  }\bibfield  {title} {\bibinfo {title} {Photodetection and causality in
  quantum optics},\ }\href {https://doi.org/10.1103/PhysRevA.52.1525}
  {\bibfield  {journal} {\bibinfo  {journal} {Phys. Rev. A}\ }\textbf {\bibinfo
  {volume} {52}},\ \bibinfo {pages} {1525} (\bibinfo {year}
  {1995})}\BibitemShut {NoStop}%
\bibitem [{\citenamefont {Milonni}(1994{\natexlab{b}})}]{milonni1994}%
  \BibitemOpen
  \bibfield  {author} {\bibinfo {author} {\bibfnamefont {P.~W.}\ \bibnamefont
  {Milonni}},\ }\href {https://doi.org/10.1016/C2009-0-21295-5} {\emph
  {\bibinfo {title} {The Quantum Vacuum: An Introduction to Quantum
  Electrodynamics}}}\ (\bibinfo  {publisher} {Academic Press},\ \bibinfo
  {address} {Boston},\ \bibinfo {year} {1994})\ Chap.\ \bibinfo {chapter}
  {4.6}\BibitemShut {NoStop}%
\bibitem [{\citenamefont {Schwartz}(2014)}]{schwartz2014}%
  \BibitemOpen
  \bibfield  {author} {\bibinfo {author} {\bibfnamefont {M.~D.}\ \bibnamefont
  {Schwartz}},\ }\href {https://doi.org/10.1017/9781139540940} {\emph {\bibinfo
  {title} {Quantum Field Theory and the Standard Model}}}\ (\bibinfo
  {publisher} {Cambridge University Press},\ \bibinfo {address} {Cambridge},\
  \bibinfo {year} {2014})\ Chap.\ \bibinfo {chapter} {2.3}\BibitemShut
  {NoStop}%
\bibitem [{Note1()}]{Note1}%
  \BibitemOpen
  \bibinfo {note} {\label {foot:phi_x_0}Note that for some real, scalar quantum
  field given by $\phi (\protect \bm {\protect \mathrm {r}}, t) = \DOTSI \intop
  \ilimits@ _{-\infty }^{\infty } \mathchoice {\protect \hspace
  {-.8em}}{}{}{}\protect \frac {\dd ^3 k}{\mathopen {}\left (2\pi \right )^3}
  \mathchoice {\protect \hspace {.2em}}{\protect \hspace {.2em}}{\protect
  \hspace {.1em}}{\protect \hspace {.06em}}\protect \frac {1}{\protect \sqrt {2
  \omega _{\protect \bm {\protect \mathrm {k}}}}} a_{\protect \bm {\protect
  \mathrm {k}}}^{} e^{i \protect \bm {\protect \mathrm {k}} \cdot \protect \bm
  {\protect \mathrm {r}} - i \omega t} + \protect \text {H.c.}$, the state
  $\left \protect \vert \psi _{\protect \bm {\protect \mathrm {r}}, t} \right
  \protect \rangle = \phi (\protect \bm {\protect \mathrm {r}}, t) \left
  \protect \vert 0 \right \protect \rangle $ gives a nonzero expectation value
  $\left \protect \langle \psi _{\protect \bm {\protect \mathrm {r}}, t} \right
  \protect \vert :\mathchoice {\protect \hspace {.1em}}{\protect \hspace
  {.09em}}{\protect \hspace {.03em}}{}\mathrel {\phi \mathopen {}\left
  (\protect \bm {\protect \mathrm {r}}', t\right )^2}\mathchoice {\protect
  \hspace {.09em}}{\protect \hspace {.06em}}{\protect \hspace {.02em}}{}: \left
  \protect \vert \psi _{\protect \bm {\protect \mathrm {r}}, t} \right \protect
  \rangle $ for all space $\protect \bm {\protect \mathrm {r}}'$ due to the
  factor $1 / \protect \sqrt {2 \omega _{\protect \bm {\protect \mathrm
  {k}}}}$. Hence this state is not localized to any region. It turns out that
  this problem is actually avoidable, as we can instead construct a
  single-particle state $\left \protect \vert \psi _{\protect \bm {\protect
  \mathrm {r}}, t} \right \protect \rangle = \DOTSI \intop \ilimits@ _{-\infty
  }^{\infty } \mathchoice {\protect \hspace {-.8em}}{}{}{}\protect \frac {\dd
  ^3 k}{\mathopen {}\left (2\pi \right )^3} \mathchoice {\protect \hspace
  {.2em}}{\protect \hspace {.2em}}{\protect \hspace {.1em}}{\protect \hspace
  {.06em}}\protect \sqrt {2 \omega _{\protect \bm {\protect \mathrm {k}}}}
  a^\dagger _{\protect \bm {\protect \mathrm {k}}} e^{- i \protect \bm
  {\protect \mathrm {k}} \cdot \protect \bm {\protect \mathrm {r}} + i \omega
  t} \left \protect \vert 0 \right \protect \rangle $, which at time $t$ gives
  an expectation value for $:\mathchoice {\protect \hspace {.1em}}{\protect
  \hspace {.09em}}{\protect \hspace {.03em}}{}\mathrel {\phi (\protect \bm
  {\protect \mathrm {r}}', t)^2}\mathchoice {\protect \hspace {.09em}}{\protect
  \hspace {.06em}}{\protect \hspace {.02em}}{}:$ that is proportional to
  $\delta ^{(3)}\mathopen {}\left (\protect \bm {\protect \mathrm {r}} -
  \protect \bm {\protect \mathrm {r}}'\right )$ as required. However, the real
  localization problem is the absence of negative frequencies in the time
  dependence $e^{i \omega t} = e^{i \protect \sqrt {\protect \bm {\protect
  \mathrm {k}}^2 + m^2} t}$, which makes any single-particle state that is
  localized at time $t$ instantaneously become spread out everywhere in space
  at all times $t' > t$.}\BibitemShut {Stop}%
\bibitem [{\citenamefont {Karpi{\'n}ski}\ \emph {et~al.}(2021)\citenamefont
  {Karpi{\'n}ski}, \citenamefont {Davis}, \citenamefont {So{\' s}nicki},
  \citenamefont {Thiel},\ and\ \citenamefont {Smith}}]{karpinski2021}%
  \BibitemOpen
  \bibfield  {author} {\bibinfo {author} {\bibfnamefont {M.}~\bibnamefont
  {Karpi{\'n}ski}}, \bibinfo {author} {\bibfnamefont {A.~O.~C.}\ \bibnamefont
  {Davis}}, \bibinfo {author} {\bibfnamefont {F.}~\bibnamefont {So{\'
  s}nicki}}, \bibinfo {author} {\bibfnamefont {V.}~\bibnamefont {Thiel}},\ and\
  \bibinfo {author} {\bibfnamefont {B.~J.}\ \bibnamefont {Smith}},\ }\bibfield
  {title} {\bibinfo {title} {Control and measurement of quantum light pulses
  for quantum information science and technology},\ }\href
  {https://doi.org/10.1002/qute.202000150} {\bibfield  {journal} {\bibinfo
  {journal} {Adv. Quantum Technol.}\ }\textbf {\bibinfo {volume} {4}},\
  \bibinfo {pages} {2000150} (\bibinfo {year} {2021})}\BibitemShut {NoStop}%
\bibitem [{\citenamefont {Benincasa}\ \emph {et~al.}(2014)\citenamefont
  {Benincasa}, \citenamefont {Borsten}, \citenamefont {Buck},\ and\
  \citenamefont {Dowker}}]{benincasa2014}%
  \BibitemOpen
  \bibfield  {author} {\bibinfo {author} {\bibfnamefont {D.~M.~T.}\
  \bibnamefont {Benincasa}}, \bibinfo {author} {\bibfnamefont {L.}~\bibnamefont
  {Borsten}}, \bibinfo {author} {\bibfnamefont {M.}~\bibnamefont {Buck}},\ and\
  \bibinfo {author} {\bibfnamefont {F.}~\bibnamefont {Dowker}},\ }\bibfield
  {title} {\bibinfo {title} {Quantum information processing and relativistic
  quantum fields},\ }\href {https://doi.org/10.1088/0264-9381/31/7/075007}
  {\bibfield  {journal} {\bibinfo  {journal} {Class. Quantum Grav.}\ }\textbf
  {\bibinfo {volume} {31}},\ \bibinfo {pages} {075007} (\bibinfo {year}
  {2014})},\ \Eprint {https://arxiv.org/abs/1206.5205} {arXiv:1206.5205
  [quant-ph]} \BibitemShut {NoStop}%
\bibitem [{\citenamefont {Su}\ and\ \citenamefont {Ralph}(2016)}]{su2016a}%
  \BibitemOpen
  \bibfield  {author} {\bibinfo {author} {\bibfnamefont {D.}~\bibnamefont
  {Su}}\ and\ \bibinfo {author} {\bibfnamefont {T.~C.}\ \bibnamefont {Ralph}},\
  }\bibfield  {title} {\bibinfo {title} {Spacetime diamonds},\ }\href
  {https://doi.org/10.1103/physrevd.93.044023} {\bibfield  {journal} {\bibinfo
  {journal} {Phys. Rev. D}\ }\textbf {\bibinfo {volume} {93}},\ \bibinfo
  {pages} {044023} (\bibinfo {year} {2016})},\ \Eprint
  {https://arxiv.org/abs/1507.00423} {arXiv:1507.00423 [quant-ph]} \BibitemShut
  {NoStop}%
\bibitem [{\citenamefont {Foo}\ \emph {et~al.}(2020)\citenamefont {Foo},
  \citenamefont {Onoe}, \citenamefont {Zych},\ and\ \citenamefont
  {Ralph}}]{foo2020}%
  \BibitemOpen
  \bibfield  {author} {\bibinfo {author} {\bibfnamefont {J.}~\bibnamefont
  {Foo}}, \bibinfo {author} {\bibfnamefont {S.}~\bibnamefont {Onoe}}, \bibinfo
  {author} {\bibfnamefont {M.}~\bibnamefont {Zych}},\ and\ \bibinfo {author}
  {\bibfnamefont {T.~C.}\ \bibnamefont {Ralph}},\ }\bibfield  {title} {\bibinfo
  {title} {Generating multi-partite entanglement from the quantum vacuum with a
  finite-lifetime mirror},\ }\href {https://doi.org/10.1088/1367-2630/aba1b2}
  {\bibfield  {journal} {\bibinfo  {journal} {New J. Phys.}\ }\textbf {\bibinfo
  {volume} {22}},\ \bibinfo {pages} {083075} (\bibinfo {year} {2020})},\
  \Eprint {https://arxiv.org/abs/2004.07094} {arXiv:2004.07094 [quant-ph]}
  \BibitemShut {NoStop}%
\bibitem [{\citenamefont {Bykov}\ and\ \citenamefont
  {Tatarskii}(1989)}]{bykov1989}%
  \BibitemOpen
  \bibfield  {author} {\bibinfo {author} {\bibfnamefont {V.~P.}\ \bibnamefont
  {Bykov}}\ and\ \bibinfo {author} {\bibfnamefont {V.~I.}\ \bibnamefont
  {Tatarskii}},\ }\bibfield  {title} {\bibinfo {title} {Causality violation in
  the {G}lauber theory of photodetection},\ }\href
  {https://doi.org/10.1016/0375-9601(89)90682-8} {\bibfield  {journal}
  {\bibinfo  {journal} {Phys. Lett. A}\ }\textbf {\bibinfo {volume} {136}},\
  \bibinfo {pages} {77} (\bibinfo {year} {1989})}\BibitemShut {NoStop}%
\bibitem [{\citenamefont {Plimak}\ and\ \citenamefont
  {Stenholm}(2011)}]{plimak2011}%
  \BibitemOpen
  \bibfield  {author} {\bibinfo {author} {\bibfnamefont {L.~I.}\ \bibnamefont
  {Plimak}}\ and\ \bibinfo {author} {\bibfnamefont {S.~T.}\ \bibnamefont
  {Stenholm}},\ }\bibfield  {title} {\bibinfo {title} {Operator ordering and
  causality beyond the rotating wave approximation},\ }\href
  {https://doi.org/10.1209/0295-5075/96/34002} {\bibfield  {journal} {\bibinfo
  {journal} {{EPL}}\ }\textbf {\bibinfo {volume} {96}},\ \bibinfo {pages}
  {34002} (\bibinfo {year} {2011})}\BibitemShut {NoStop}%
\bibitem [{\citenamefont {Bohr}\ and\ \citenamefont
  {Rosenfeld}(1933)}]{bohr1933}%
  \BibitemOpen
  \bibfield  {author} {\bibinfo {author} {\bibfnamefont {N.}~\bibnamefont
  {Bohr}}\ and\ \bibinfo {author} {\bibfnamefont {L.}~\bibnamefont
  {Rosenfeld}},\ }\bibfield  {title} {\bibinfo {title} {{Zur Frage der
  Messbarkeit der elektromagnetischen Feldgrossen} [{O}n the question of the
  measurability of electromagnetic field sizes]},\ }\href@noop {} {\bibfield
  {journal} {\bibinfo  {journal} {Kgl. Danske Videnskab. Selskab Mat.-Fys.
  Medd.}\ }\textbf {\bibinfo {volume} {12}},\ \bibinfo {pages} {1} (\bibinfo
  {year} {1933})},\ \bibinfo {note} {(in German)}\BibitemShut {NoStop}%
\bibitem [{\citenamefont {Haag}(1959)}]{haag1959a}%
  \BibitemOpen
  \bibfield  {author} {\bibinfo {author} {\bibfnamefont {R.}~\bibnamefont
  {Haag}},\ }\bibfield  {title} {\bibinfo {title} {{Discussion des ``axiomes''
  et des propri{\'e}t{\'e}s asymptotiques d’une th{\'e}orie des champs locale
  avec particules compos{\'e}es} [{D}iscussion of ``axioms'' and asymptotic
  properties of a local field theory with compound particles]},\ }in\
  \href@noop {} {\emph {\bibinfo {booktitle} {{Les Probl{\`e}mes
  Math{\'e}matiques de la Th{\'e}orie Quantique des Champs}, {Colloq. Int. du
  Cent. Natl. Rech. Sci. {LXXV}}, Lille, France, Jun 3--8, 1957}}}\ (\bibinfo
  {publisher} {Centre National de la Recherche Scientifique},\ \bibinfo
  {address} {Paris},\ \bibinfo {year} {1959})\ pp.\ \bibinfo {pages}
  {151--162},\ \bibinfo {note} {(in French)}\BibitemShut {NoStop}%
\bibitem [{\citenamefont {Beckman}\ \emph {et~al.}(2001)\citenamefont
  {Beckman}, \citenamefont {Gottesman}, \citenamefont {Nielsen},\ and\
  \citenamefont {Preskill}}]{beckman2001}%
  \BibitemOpen
  \bibfield  {author} {\bibinfo {author} {\bibfnamefont {D.}~\bibnamefont
  {Beckman}}, \bibinfo {author} {\bibfnamefont {D.}~\bibnamefont {Gottesman}},
  \bibinfo {author} {\bibfnamefont {M.~A.}\ \bibnamefont {Nielsen}},\ and\
  \bibinfo {author} {\bibfnamefont {J.}~\bibnamefont {Preskill}},\ }\bibfield
  {title} {\bibinfo {title} {Causal and localizable quantum operations},\
  }\href {https://doi.org/10.1103/physreva.64.052309} {\bibfield  {journal}
  {\bibinfo  {journal} {Phys. Rev. A}\ }\textbf {\bibinfo {volume} {64}},\
  \bibinfo {pages} {052309} (\bibinfo {year} {2001})},\ \Eprint
  {https://arxiv.org/abs/quant-ph/0102043} {arXiv:quant-ph/0102043}
  \BibitemShut {NoStop}%
\bibitem [{\citenamefont {Borsten}\ \emph {et~al.}(2021)\citenamefont
  {Borsten}, \citenamefont {Jubb},\ and\ \citenamefont {Kells}}]{borsten2021}%
  \BibitemOpen
  \bibfield  {author} {\bibinfo {author} {\bibfnamefont {L.}~\bibnamefont
  {Borsten}}, \bibinfo {author} {\bibfnamefont {I.}~\bibnamefont {Jubb}},\ and\
  \bibinfo {author} {\bibfnamefont {G.}~\bibnamefont {Kells}},\ }\bibfield
  {title} {\bibinfo {title} {Impossible measurements revisited},\ }\href
  {https://doi.org/10.1103/physrevd.104.025012} {\bibfield  {journal} {\bibinfo
   {journal} {Phys. Rev. D}\ }\textbf {\bibinfo {volume} {104}},\ \bibinfo
  {pages} {025012} (\bibinfo {year} {2021})},\ \Eprint
  {https://arxiv.org/abs/1912.06141} {arXiv:1912.06141 [quant-ph]} \BibitemShut
  {NoStop}%
\bibitem [{\citenamefont {Bostelmann}\ \emph {et~al.}(2021)\citenamefont
  {Bostelmann}, \citenamefont {Fewster},\ and\ \citenamefont
  {Ruep}}]{bostelmann2021}%
  \BibitemOpen
  \bibfield  {author} {\bibinfo {author} {\bibfnamefont {H.}~\bibnamefont
  {Bostelmann}}, \bibinfo {author} {\bibfnamefont {C.~J.}\ \bibnamefont
  {Fewster}},\ and\ \bibinfo {author} {\bibfnamefont {M.~H.}\ \bibnamefont
  {Ruep}},\ }\bibfield  {title} {\bibinfo {title} {Impossible measurements
  require impossible apparatus},\ }\href
  {https://doi.org/10.1103/physrevd.103.025017} {\bibfield  {journal} {\bibinfo
   {journal} {Phys. Rev. D}\ }\textbf {\bibinfo {volume} {103}},\ \bibinfo
  {pages} {025017} (\bibinfo {year} {2021})},\ \Eprint
  {https://arxiv.org/abs/2003.04660} {arXiv:2003.04660 [quant-ph]} \BibitemShut
  {NoStop}%
\bibitem [{\citenamefont {de~Ram{\'o}n}\ \emph {et~al.}(2021)\citenamefont
  {de~Ram{\'o}n}, \citenamefont {Papageorgiou},\ and\ \citenamefont
  {Mart{\'\i}n-Mart{\'\i}nez}}]{ramon2021}%
  \BibitemOpen
  \bibfield  {author} {\bibinfo {author} {\bibfnamefont {J.}~\bibnamefont
  {de~Ram{\'o}n}}, \bibinfo {author} {\bibfnamefont {M.}~\bibnamefont
  {Papageorgiou}},\ and\ \bibinfo {author} {\bibfnamefont {E.}~\bibnamefont
  {Mart{\'\i}n-Mart{\'\i}nez}},\ }\bibfield  {title} {\bibinfo {title}
  {Relativistic causality in particle detector models: Faster-than-light
  signaling and impossible measurements},\ }\href
  {https://doi.org/10.1103/physrevd.103.085002} {\bibfield  {journal} {\bibinfo
   {journal} {Phys. Rev. D}\ }\textbf {\bibinfo {volume} {103}},\ \bibinfo
  {pages} {085002} (\bibinfo {year} {2021})},\ \Eprint
  {https://arxiv.org/abs/2102.03408} {arXiv:2102.03408 [quant-ph]} \BibitemShut
  {NoStop}%
\bibitem [{\citenamefont {Jubb}(2022)}]{jubb2022}%
  \BibitemOpen
  \bibfield  {author} {\bibinfo {author} {\bibfnamefont {I.}~\bibnamefont
  {Jubb}},\ }\bibfield  {title} {\bibinfo {title} {Causal state updates in real
  scalar quantum field theory},\ }\href
  {https://doi.org/10.1103/physrevd.105.025003} {\bibfield  {journal} {\bibinfo
   {journal} {Phys. Rev. D}\ }\textbf {\bibinfo {volume} {105}},\ \bibinfo
  {pages} {025003} (\bibinfo {year} {2022})},\ \Eprint
  {https://arxiv.org/abs/2106.09027} {arXiv:2106.09027 [quant-ph]} \BibitemShut
  {NoStop}%
\bibitem [{\citenamefont {Albertini}\ and\ \citenamefont
  {Jubb}(2023)}]{albertini2023}%
  \BibitemOpen
  \bibfield  {author} {\bibinfo {author} {\bibfnamefont {E.}~\bibnamefont
  {Albertini}}\ and\ \bibinfo {author} {\bibfnamefont {I.}~\bibnamefont
  {Jubb}},\ }\href@noop {} {\bibinfo {title} {Are ideal measurements of real
  scalar fields causal?}} (\bibinfo {year} {2023}),\ \Eprint
  {https://arxiv.org/abs/2306.12980} {arXiv:2306.12980 [quant-ph]} \BibitemShut
  {NoStop}%
\bibitem [{\citenamefont {Sorkin}(1993)}]{sorkin1993}%
  \BibitemOpen
  \bibfield  {author} {\bibinfo {author} {\bibfnamefont {R.~D.}\ \bibnamefont
  {Sorkin}},\ }\bibfield  {title} {\bibinfo {title} {Impossible measurements on
  quantum fields},\ }in\ \href {https://doi.org/10.1017/CBO9780511524653.024}
  {\emph {\bibinfo {booktitle} {Directions in General Relativity, Proc. Int.
  Symp. On General Relativity, University of Maryland, MD, USA, May 27--29,
  1993, Vol. 2: Papers in Honor of {D}ieter {B}rill}}},\ \bibinfo {editor}
  {edited by\ \bibinfo {editor} {\bibfnamefont {B.-L.}\ \bibnamefont {Hu}}\
  and\ \bibinfo {editor} {\bibfnamefont {T.~A.}\ \bibnamefont {Jacobson}}}\
  (\bibinfo  {publisher} {Cambridge University Press},\ \bibinfo {address}
  {Cambridge},\ \bibinfo {year} {1993})\ pp.\ \bibinfo {pages} {293--305},\
  \Eprint {https://arxiv.org/abs/gr-qc/9302018} {arXiv:gr-qc/9302018 [gr-qc]}
  \BibitemShut {NoStop}%
\bibitem [{\citenamefont {Beckman}\ \emph {et~al.}(2002)\citenamefont
  {Beckman}, \citenamefont {Gottesman}, \citenamefont {Kitaev},\ and\
  \citenamefont {Preskill}}]{beckman2002}%
  \BibitemOpen
  \bibfield  {author} {\bibinfo {author} {\bibfnamefont {D.}~\bibnamefont
  {Beckman}}, \bibinfo {author} {\bibfnamefont {D.}~\bibnamefont {Gottesman}},
  \bibinfo {author} {\bibfnamefont {A.}~\bibnamefont {Kitaev}},\ and\ \bibinfo
  {author} {\bibfnamefont {J.}~\bibnamefont {Preskill}},\ }\bibfield  {title}
  {\bibinfo {title} {Measurability of {W}ilson loop operators},\ }\href
  {https://doi.org/10.1103/physrevd.65.065022} {\bibfield  {journal} {\bibinfo
  {journal} {Phys. Rev. D}\ }\textbf {\bibinfo {volume} {65}},\ \bibinfo
  {pages} {065022} (\bibinfo {year} {2002})},\ \Eprint
  {https://arxiv.org/abs/hep-th/0110205} {arXiv:hep-th/0110205 [hep-th]}
  \BibitemShut {NoStop}%
\bibitem [{\citenamefont {Unruh}(1976)}]{unruh1976}%
  \BibitemOpen
  \bibfield  {author} {\bibinfo {author} {\bibfnamefont {W.~G.}\ \bibnamefont
  {Unruh}},\ }\bibfield  {title} {\bibinfo {title} {Notes on black-hole
  evaporation},\ }\href {https://doi.org/10.1103/physrevd.14.870} {\bibfield
  {journal} {\bibinfo  {journal} {Phys. Rev. D}\ }\textbf {\bibinfo {volume}
  {14}},\ \bibinfo {pages} {870} (\bibinfo {year} {1976})}\BibitemShut
  {NoStop}%
\bibitem [{\citenamefont {DeWitt}(1979)}]{dewitt1979}%
  \BibitemOpen
  \bibfield  {author} {\bibinfo {author} {\bibfnamefont {B.~S.}\ \bibnamefont
  {DeWitt}},\ }\bibfield  {title} {\bibinfo {title} {Quantum gravity: the new
  synthesis},\ }in\ \href@noop {} {\emph {\bibinfo {booktitle} {General
  Relativity: An {E}instein Centenary Survey}}},\ \bibinfo {editor} {edited by\
  \bibinfo {editor} {\bibfnamefont {S.~W.}\ \bibnamefont {Hawking}}\ and\
  \bibinfo {editor} {\bibfnamefont {W.}~\bibnamefont {Israel}}}\ (\bibinfo
  {publisher} {Cambridge University Press},\ \bibinfo {address} {Cambridge},\
  \bibinfo {year} {1979})\ Chap.~\bibinfo {chapter} {14}, pp.\ \bibinfo {pages}
  {680--745}\BibitemShut {NoStop}%
\bibitem [{\citenamefont
  {Mart{\'\i}n-Mart{\'\i}nez}(2015)}]{martinmartinez2015}%
  \BibitemOpen
  \bibfield  {author} {\bibinfo {author} {\bibfnamefont {E.}~\bibnamefont
  {Mart{\'\i}n-Mart{\'\i}nez}},\ }\bibfield  {title} {\bibinfo {title}
  {Causality issues of particle detector models in {QFT} and quantum optics},\
  }\href {https://doi.org/10.1103/physrevd.92.104019} {\bibfield  {journal}
  {\bibinfo  {journal} {Phys. Rev. D}\ }\textbf {\bibinfo {volume} {92}},\
  \bibinfo {pages} {104019} (\bibinfo {year} {2015})},\ \Eprint
  {https://arxiv.org/abs/1509.07864} {arXiv:1509.07864 [quant-ph]} \BibitemShut
  {NoStop}%
\bibitem [{\citenamefont {Fewster}\ and\ \citenamefont
  {Verch}(2018)}]{fewster2018}%
  \BibitemOpen
  \bibfield  {author} {\bibinfo {author} {\bibfnamefont {C.~J.}\ \bibnamefont
  {Fewster}}\ and\ \bibinfo {author} {\bibfnamefont {R.}~\bibnamefont
  {Verch}},\ }\href@noop {} {\bibinfo {title} {Quantum fields and local
  measurements}} (\bibinfo {year} {2018}),\ \Eprint
  {https://arxiv.org/abs/1810.06512} {arXiv:1810.06512 [math-ph]} \BibitemShut
  {NoStop}%
\bibitem [{\citenamefont {Hall}(2013)}]{hall2013}%
  \BibitemOpen
  \bibfield  {author} {\bibinfo {author} {\bibfnamefont {B.~C.}\ \bibnamefont
  {Hall}},\ }\href {https://doi.org/10.1007/978-1-4614-7116-5} {\emph {\bibinfo
  {title} {Quantum Theory for Mathematicians}}},\ Graduate Texts in Mathematics
  Vol. 267\ (\bibinfo  {publisher} {Springer},\ \bibinfo {address} {New York},\
  \bibinfo {year} {2013})\ Chap.\ \bibinfo {chapter} {10.1}\BibitemShut
  {NoStop}%
\bibitem [{Note2()}]{Note2}%
  \BibitemOpen
  \bibinfo {note} {\label {foot:self_adjoint}There is also a potential
  complication regarding whether $L$ defined in \protect \hyperref
  [{L_G}]{\protect \textup {\hbox {\mathsurround \z@ \protect \normalfont
  (\ignorespaces \protect \ref *{L_G}\unskip \@@italiccorr )}}} is
  self-adjoint, and not just Hermitian, which is needed to use the spectral
  theorem \cite {hall2013} in \protect \hyperref [{E_zeta_spectral}]{\protect
  \textup {\hbox {\mathsurround \z@ \protect \normalfont (\ignorespaces
  \protect \ref *{E_zeta_spectral}\unskip \@@italiccorr )}}}. Here we will not
  treat this question. Instead we will simply assume that $A^\mu (x)$ (and its
  derivatives) is self-adjoint \cite {haag1996}, and that Hermitian
  superpositions of smeared fields defining $L$ \protect \hyperref
  [{L_G}]{\protect \textup {\hbox {\mathsurround \z@ \protect \normalfont
  (\ignorespaces \protect \ref *{L_G}\unskip \@@italiccorr )}}} are
  self-adjoint as well.}\BibitemShut {Stop}%
\bibitem [{Note3()}]{Note3}%
  \BibitemOpen
  \bibinfo {note} {\label {foot:algebra}Our definition of local operators
  \protect \hyperref [{Q_G}]{\protect \textup {\hbox {\mathsurround \z@
  \protect \normalfont (\ignorespaces \protect \ref *{Q_G}\unskip \@@italiccorr
  )}}} matches eq. (1) in \cite {knight1961} and eq. (II.4.1) in \cite
  {haag1996}, and it generates the \protect \emph {polynomial algebra} of the
  field. However, since the (smeared) field is an unbounded operator,
  technically the domains of the operators $Q$ have to be carefully considered.
  To avoid this issue, and to handle the convergence of infinite operator
  sequences, it is convenient to go over to algebras of bounded operators
  instead through the spectral theorem, as discussed in \cite {haag1996}. For
  our purposes however, we will simply use the more straight-forward polynomial
  algebra \protect \hyperref [{Q_G}]{\protect \textup {\hbox {\mathsurround \z@
  \protect \normalfont (\ignorespaces \protect \ref *{Q_G}\unskip \@@italiccorr
  )}}} and ignore the question of the domains of the operators
  $Q$.}\BibitemShut {Stop}%
\bibitem [{\citenamefont {Strocchi}\ and\ \citenamefont
  {Wightman}(1974)}]{strocchi1974}%
  \BibitemOpen
  \bibfield  {author} {\bibinfo {author} {\bibfnamefont {F.}~\bibnamefont
  {Strocchi}}\ and\ \bibinfo {author} {\bibfnamefont {A.~S.}\ \bibnamefont
  {Wightman}},\ }\bibfield  {title} {\bibinfo {title} {Proof of the charge
  superselection rule in local relativistic quantum field theory},\ }\href
  {https://doi.org/10.1063/1.1666601} {\bibfield  {journal} {\bibinfo
  {journal} {J. Math. Phys.}\ }\textbf {\bibinfo {volume} {15}},\ \bibinfo
  {pages} {2198} (\bibinfo {year} {1974})}\BibitemShut {NoStop}%
\bibitem [{\citenamefont {Licht}(1963)}]{licht1963}%
  \BibitemOpen
  \bibfield  {author} {\bibinfo {author} {\bibfnamefont {A.~L.}\ \bibnamefont
  {Licht}},\ }\bibfield  {title} {\bibinfo {title} {Strict localization},\
  }\href {https://doi.org/10.1063/1.1703925} {\bibfield  {journal} {\bibinfo
  {journal} {J. Math. Phys.}\ }\textbf {\bibinfo {volume} {4}},\ \bibinfo
  {pages} {1443} (\bibinfo {year} {1963})}\BibitemShut {NoStop}%
\bibitem [{Note4()}]{Note4}%
  \BibitemOpen
  \bibinfo {note} {\label {foot:partial_trace}Usually the partial trace results
  in a mixed state given by a density matrix. Here it happens to work out so
  that the reduced state is pure, hence we can write it directly as in \protect
  \hyperref [{psi_from_U}]{\protect \textup {\hbox {\mathsurround \z@ \protect
  \normalfont (\ignorespaces \protect \ref *{psi_from_U}\unskip \@@italiccorr
  )}}}.}\BibitemShut {Stop}%
\bibitem [{\citenamefont {Schumaker}\ and\ \citenamefont
  {Caves}(1985)}]{schumaker1985}%
  \BibitemOpen
  \bibfield  {author} {\bibinfo {author} {\bibfnamefont {B.~L.}\ \bibnamefont
  {Schumaker}}\ and\ \bibinfo {author} {\bibfnamefont {C.~M.}\ \bibnamefont
  {Caves}},\ }\bibfield  {title} {\bibinfo {title} {New formalism for
  two-photon quantum optics. {II}. {M}athematical foundation and compact
  notation},\ }\href {https://doi.org/10.1103/physreva.31.3093} {\bibfield
  {journal} {\bibinfo  {journal} {Phys. Rev. A}\ }\textbf {\bibinfo {volume}
  {31}},\ \bibinfo {pages} {3093} (\bibinfo {year} {1985})}\BibitemShut
  {NoStop}%
\bibitem [{Note5()}]{Note5}%
  \BibitemOpen
  \bibinfo {note} {\label {foot:G_w_A_w}For any $g(t)$ in $L^2(0, \infty )$,
  Titchmarsh' theorem \cite {titchmarsh1948} dictates how quickly $G(\omega )$
  falls off for large $\omega $ in the upper half-plane of complex frequency
  $\omega $. The factor $\protect \mathcal {A}(\omega ) \sim 1/\protect \sqrt
  {\omega }$ only strengthens this convergence, but there is a possibility of
  divergence at the origin. This only happens for very special choices of seed
  functions $g(t)$ that have ``almost-diverging'' norm at $\omega = 0$, which
  we must exclude.}\BibitemShut {Stop}%
\bibitem [{\citenamefont {Rudin}(1973)}]{rudin1973}%
  \BibitemOpen
  \bibfield  {author} {\bibinfo {author} {\bibfnamefont {W.}~\bibnamefont
  {Rudin}},\ }\href@noop {} {\emph {\bibinfo {title} {Functional Analysis}}},\
  McGraw-Hill Series in Higher Mathematics\ (\bibinfo  {publisher}
  {McGraw-Hill},\ \bibinfo {address} {New York},\ \bibinfo {year} {1973})\
  Chap.~\bibinfo {chapter} {13}\BibitemShut {NoStop}%
\bibitem [{\citenamefont {Indritz}(1961)}]{indritz1961}%
  \BibitemOpen
  \bibfield  {author} {\bibinfo {author} {\bibfnamefont {J.}~\bibnamefont
  {Indritz}},\ }\bibfield  {title} {\bibinfo {title} {An inequality for
  {H}ermite polynomials},\ }\href
  {https://doi.org/10.1090/s0002-9939-1961-0132852-2} {\bibfield  {journal}
  {\bibinfo  {journal} {Proc. {A}m. Math. Soc.}\ }\textbf {\bibinfo {volume}
  {12}},\ \bibinfo {pages} {981} (\bibinfo {year} {1961})}\BibitemShut
  {NoStop}%
\bibitem [{\citenamefont {Sansone}\ \emph {et~al.}(2006)\citenamefont
  {Sansone}, \citenamefont {Benedetti}, \citenamefont {Calegari}, \citenamefont
  {Vozzi}, \citenamefont {Avaldi}, \citenamefont {Flammini}, \citenamefont
  {Poletto}, \citenamefont {Villoresi}, \citenamefont {Altucci}, \citenamefont
  {Velotta}, \citenamefont {Stagira}, \citenamefont {Silvestri},\ and\
  \citenamefont {Nisoli}}]{sansone2006}%
  \BibitemOpen
  \bibfield  {author} {\bibinfo {author} {\bibfnamefont {G.}~\bibnamefont
  {Sansone}}, \bibinfo {author} {\bibfnamefont {E.}~\bibnamefont {Benedetti}},
  \bibinfo {author} {\bibfnamefont {F.}~\bibnamefont {Calegari}}, \bibinfo
  {author} {\bibfnamefont {C.}~\bibnamefont {Vozzi}}, \bibinfo {author}
  {\bibfnamefont {L.}~\bibnamefont {Avaldi}}, \bibinfo {author} {\bibfnamefont
  {R.}~\bibnamefont {Flammini}}, \bibinfo {author} {\bibfnamefont
  {L.}~\bibnamefont {Poletto}}, \bibinfo {author} {\bibfnamefont
  {P.}~\bibnamefont {Villoresi}}, \bibinfo {author} {\bibfnamefont
  {C.}~\bibnamefont {Altucci}}, \bibinfo {author} {\bibfnamefont
  {R.}~\bibnamefont {Velotta}}, \bibinfo {author} {\bibfnamefont
  {S.}~\bibnamefont {Stagira}}, \bibinfo {author} {\bibfnamefont {S.~D.}\
  \bibnamefont {Silvestri}},\ and\ \bibinfo {author} {\bibfnamefont
  {M.}~\bibnamefont {Nisoli}},\ }\bibfield  {title} {\bibinfo {title} {Isolated
  single-cycle attosecond pulses},\ }\href
  {https://doi.org/10.1126/science.1132838} {\bibfield  {journal} {\bibinfo
  {journal} {Science}\ }\textbf {\bibinfo {volume} {314}},\ \bibinfo {pages}
  {443} (\bibinfo {year} {2006})}\BibitemShut {NoStop}%
\bibitem [{\citenamefont {Su}\ \emph {et~al.}(2016)\citenamefont {Su},
  \citenamefont {Chinnarasu}, \citenamefont {Kuo},\ and\ \citenamefont
  {Chuu}}]{su2016}%
  \BibitemOpen
  \bibfield  {author} {\bibinfo {author} {\bibfnamefont {W.-M.}\ \bibnamefont
  {Su}}, \bibinfo {author} {\bibfnamefont {R.}~\bibnamefont {Chinnarasu}},
  \bibinfo {author} {\bibfnamefont {C.-H.}\ \bibnamefont {Kuo}},\ and\ \bibinfo
  {author} {\bibfnamefont {C.-S.}\ \bibnamefont {Chuu}},\ }\bibfield  {title}
  {\bibinfo {title} {Shaping single photons and biphotons by inherent losses},\
  }\href {https://doi.org/10.1103/PhysRevA.94.033805} {\bibfield  {journal}
  {\bibinfo  {journal} {Phys. Rev. A}\ }\textbf {\bibinfo {volume} {94}},\
  \bibinfo {pages} {033805} (\bibinfo {year} {2016})},\ \Eprint
  {https://arxiv.org/abs/1609.00761} {arXiv:1609.00761 [quant-ph]} \BibitemShut
  {NoStop}%
\bibitem [{\citenamefont {Ryen}\ \emph {et~al.}(2022)\citenamefont {Ryen},
  \citenamefont {Gulla},\ and\ \citenamefont {Skaar}}]{ryen2022}%
  \BibitemOpen
  \bibfield  {author} {\bibinfo {author} {\bibfnamefont {K.}~\bibnamefont
  {Ryen}}, \bibinfo {author} {\bibfnamefont {J.}~\bibnamefont {Gulla}},\ and\
  \bibinfo {author} {\bibfnamefont {J.}~\bibnamefont {Skaar}},\ }\bibfield
  {title} {\bibinfo {title} {Strictly localized three-dimensional states close
  to single photons},\ }\href {https://doi.org/10.1007/s10773-022-05133-7}
  {\bibfield  {journal} {\bibinfo  {journal} {Int. J. Theor. Phys.}\ }\textbf
  {\bibinfo {volume} {61}},\ \bibinfo {pages} {143} (\bibinfo {year} {2022})},\
  \Eprint {https://arxiv.org/abs/2109.07998} {arXiv:2109.07998 [quant-ph]}
  \BibitemShut {NoStop}%
\bibitem [{\citenamefont {Bruschi}\ \emph {et~al.}(2010)\citenamefont
  {Bruschi}, \citenamefont {Louko}, \citenamefont {Mart{\'\i}n-Mart{\'\i}nez},
  \citenamefont {Dragan},\ and\ \citenamefont {Fuentes}}]{bruschi2010}%
  \BibitemOpen
  \bibfield  {author} {\bibinfo {author} {\bibfnamefont {D.~E.}\ \bibnamefont
  {Bruschi}}, \bibinfo {author} {\bibfnamefont {J.}~\bibnamefont {Louko}},
  \bibinfo {author} {\bibfnamefont {E.}~\bibnamefont
  {Mart{\'\i}n-Mart{\'\i}nez}}, \bibinfo {author} {\bibfnamefont
  {A.}~\bibnamefont {Dragan}},\ and\ \bibinfo {author} {\bibfnamefont
  {I.}~\bibnamefont {Fuentes}},\ }\bibfield  {title} {\bibinfo {title} {{U}nruh
  effect in quantum information beyond the single-mode approximation},\ }\href
  {https://doi.org/10.1103/physreva.82.042332} {\bibfield  {journal} {\bibinfo
  {journal} {Phys. Rev. A}\ }\textbf {\bibinfo {volume} {82}},\ \bibinfo
  {pages} {042332} (\bibinfo {year} {2010})},\ \Eprint
  {https://arxiv.org/abs/1007.4670} {arXiv:1007.4670 [quant-ph]} \BibitemShut
  {NoStop}%
\bibitem [{\citenamefont {Birrell}\ and\ \citenamefont
  {Davies}(1984)}]{birrell1984}%
  \BibitemOpen
  \bibfield  {author} {\bibinfo {author} {\bibfnamefont {N.~D.}\ \bibnamefont
  {Birrell}}\ and\ \bibinfo {author} {\bibfnamefont {P.~C.~W.}\ \bibnamefont
  {Davies}},\ }\href@noop {} {\emph {\bibinfo {title} {Quantum Fields in Curved
  Space}}}\ (\bibinfo  {publisher} {Cambridge University Press},\ \bibinfo
  {address} {Cambridge},\ \bibinfo {year} {1984})\ Chap.\ \bibinfo {chapter}
  {3.2}\BibitemShut {NoStop}%
\bibitem [{\citenamefont {Barnett}\ and\ \citenamefont
  {Radmore}(2002)}]{barnett2002}%
  \BibitemOpen
  \bibfield  {author} {\bibinfo {author} {\bibfnamefont {S.~M.}\ \bibnamefont
  {Barnett}}\ and\ \bibinfo {author} {\bibfnamefont {P.~M.}\ \bibnamefont
  {Radmore}},\ }\href
  {https://doi.org/10.1093/acprof:oso/9780198563617.001.0001} {\emph {\bibinfo
  {title} {Methods in Theoretical Quantum Optics}}},\ Oxford Series in Optical
  and Imaging Sciences 15\ (\bibinfo  {publisher} {Clarendon Press},\ \bibinfo
  {address} {Oxford},\ \bibinfo {year} {2002})\ Chap.\ \bibinfo {chapter} {App.
  4}\BibitemShut {NoStop}%
\end{thebibliography}

%

\onecolumngrid
\end{document}